\documentclass[reprint,superscriptaddress,nofootinbib,amsmath,amssymb,aps,pra,longbibliography]{revtex4-1}
\usepackage[utf8x]{inputenc}
\usepackage{amsmath,amsthm}
\usepackage{dcolumn}
\usepackage{bm}
\usepackage[colorlinks=true,citecolor=blue,urlcolor=blue]{hyperref}
\usepackage{graphicx}
\usepackage{subcaption}
\usepackage{braket}
\usepackage{color}
\usepackage[T1]{fontenc}
\usepackage[usenames,dvipsnames,svgnames,table]{xcolor}
\usepackage{lipsum}
\allowdisplaybreaks

\newcommand{\be}{\begin{equation}}
\newcommand{\ee}{\end{equation}}
\newcommand{\ba}{\begin{eqnarray}}
\newcommand{\ea}{\end{eqnarray}}

\newtheorem{thm}{Theorem}

\begin{document}
 \newwrite\bibnotes
    \def\bibnotesext{Notes.bib}
    \immediate\openout\bibnotes=\jobname\bibnotesext
    \immediate\write\bibnotes{@CONTROL{REVTEX41Control}}
    \immediate\write\bibnotes{@CONTROL{%
    apsrev41Control,author="08",editor="1",pages="1",title="0",year="1"}}
     \if@filesw
     \immediate\write\@auxout{\string\citation{apsrev41Control}}%
    \fi

\title{ Distillation of genuine tripartite Einstein-Podolsky-Rosen steering}

\author{Shashank Gupta}
\email{shashankg687@bose.res.in}
\affiliation{S.   N.  Bose  National Centre  for Basic  Sciences, Salt
 Lake,    Kolkata    700    106,   India}  
 
  \author{Debarshi Das}
\email{dasdebarshi90@gmail.com}
\affiliation{S.   N.  Bose  National Centre  for Basic  Sciences, Salt
 Lake,    Kolkata    700    106,   India}

\author{A. S. Majumdar}
\email{archan@bose.res.in}
 \affiliation{S. N. Bose National Centre for
  Basic Sciences, Salt Lake, Kolkata 700 106, India}

\date{\today}

\begin{abstract}

We show that  a perfectly genuine tripartite steerable assemblage can be distilled from partially genuine tripartite EPR steerable assemblages. In particular, we consider two types of hybrid scenarios: one-sided device-independent (1SDI) scenario (where one observer is untrusted, and other two observers are trusted)  and two-sided device-independent (2SDI) scenario (where two observers are untrusted, and one observer is trusted). In both the scenarios, we show distillation of perfectly genuine steerable assemblage of three-qubit Greenberger–Horne–Zeilinger (GHZ) states or three-qubit W states from many copies of initial partially genuine steerable assemblages of the corresponding  states. In each of these cases, we demonstrate that at least one copy of a perfectly genuine steerable assemblage can be distilled with certainty from infinitely many copies of initial assemblages. In case of practical scenarios employing finite copies, we show that the efficiency of our distillation protocols reaches near perfect levels using only a few number of initial assemblages.
 
\end{abstract}


\maketitle

\section{Introduction}

Einstein-Podolsky-Rosen (EPR) steering \cite{Wiseman07,Jones07,Costa20}, originally proposed by Schrödinger  \cite{Schrodinger35}, is a form of quantum inseparability that lies  between entanglement \cite{Horo09} and Bell-nonlocality \cite{Brunner14}. Unlike entanglement and Bell-nonlocality, the scenario of EPR steering is asymmetric in the sense that one observer is considered to be trusted while the other one is not.  This asymmetric scenario is often referred to as  semi-device-independent (SDI)  scenario, where the trusted observer has full knowledge of her/his measuring devices, while the untrusted observer does not have any knowledge of it, thus operating on black-box devices.

Importantly, EPR steering certifies the presence of entanglement in a SDI scenario, but it is not equivalent to entanglement \cite{Wiseman07}. Apart from its fundamental relevance, EPR steering has numerous information theoretic applications in the SDI scenario ranging from quantum key distribution \cite{Cyril12}, advantages in sub-channel discrimination \cite{Piani15}, secure quantum teleportation \cite{Reid13,He15}, quantum communication \cite{Reid13}, detecting bound entanglement \cite{Moroder14}, randomness generation \cite{Law2014,Passaro2015,Skrzypczyk18,Coyle2018}, self-testing of pure entangled states \cite{supic2016,Gheorghiu2017,Goswami18,Bian20,shrotriya20}.
 
Quantum networks composed of multiple observers sharing multipartite quantum states are emerging out to be significant more than ever in quantum communication. In practical scenarios, it is natural to expect hybrid quantum networks where some observers have more knowledge of their measuring devices than the others. Genuine EPR steering \cite{Cavalcanti11,He13,Li15,Cavalcanti15,Cavalcanti16,Bihalan18,Riccardi18} is a form of genuine multipartite quantum correlation that is present in such hybrid quantum network having more than two observers sharing genuine entanglement \cite{GUHNE91}. This kind of hybrid quantum networks where some of the observers are considered to be trusted while other observers are not, act as the most natural building blocks of quantum internet \cite{Kimble2008}. However, similar to the case of other quantum correlations, genuine multipartite EPR steering also faces environmental interactions which inevitably deteriorate the quality of the associated hybrid quantum network. 
 
Genuine EPR steering has a wide range of applications in quantum information processing protocols, such as in quantum metrology \cite{Giovannetti2011,Li15}, multipartite secret sharing in a generic SDI scenario \cite{Armstrong2015}, commercial quantum key distributions and commercial random number generations where the general consumers  may not want to trust their providers \cite{Cavalcanti15}. Due to its foundational importance and the kaleidoscopic range of information theoretic applications, detection of genuine EPR steering has recently gained much attention \cite{Armstrong2015,Mattar2017,Liu2020}. 
Further, genuine EPR steering certifies genuine entanglement in SDI scenarios, in a way which is less experimentally demanding  than the standard fully device-independent approach, requiring fewer assumptions or resources than the standard cases of quantum state tomography or genuine entanglement witnesses. 
 
Perfect quantum resources (in the sense that they maximize some quantifier of the relevant correlation) are the most desirable in any quantum information processing task. In experiments, the environmental interactions and experimental imprecision obstruct one from realizing the  perfect quantum correlation, thus, degrading the performance of the implemented task. One strategy to overcome this practical drawback is distillation which concentrates the imperfect or partial resources contained in multiple copies into a perfect quantum resource. In the bipartite context, distillation protocols exist for entanglement \cite{Bennet96,Horo98,Horo99,Horod99,Horo01}, EPR steering \cite{Nery20,Liu20} and Bell nonlocality \cite{Forster09,Forster11,Peter10,Brunner11,Wu2013,Hyer2013}. For multipartite states, distillation schemes have also been proposed for entanglement \cite{Huang14,Ben08} and Bell nonlocality \cite{Wu10,Ye12,Ebbe2013,Pan2015}. However, no such distillation protocol has hitherto been formulated for EPR steering involving more than two parties.  
 
With the above motivation, in this work we take the first step towards distillation of EPR steering in multipartite networks. Specifically, we propose certain protocols to distill genuine EPR steering in the tripartite scenario. Due to their monogamous character, quantum correlations possess special attributes for tripartite systems that are not shared by bipartite ones \cite{Dey13,Jeba18,DAS201855,Gupta18,Maity20,Gupta21}. The inherent asymmetry of EPR steering adds an extra flavour, thereby making the question of tripartite steering distillation even more interesting. Here we consider the two possible hybrid networks in the tripartite scenario, namely, one-sided device-independent (1SDI) scenario (where one observer is untrusted, and the other two observers are trusted)  and two-sided device-independent (2SDI) scenario (where two observers are untrusted, and one observer is trusted). In each of these cases, we propose distillation strategies to extract a perfectly genuine steerable assemblage of the three-qubit Greenberger–Horne–Zeilinger (GHZ) state or the three-qubit W state from many copies of initial partially genuine steerable assemblages of the corresponding states. 

 In the multipartite context, the idea of maximally entangled state or `perfectly genuine tripartite EPR steerable assemblage’ is  scenario dependent. For example, in the tripartite case, there are two SLOCC (stochastic local operations and classical communication) inequivalent classes of states (GHZ class and W class) \cite{Dur2000}. The GHZ state and the W state can be taken as two extreme members in the family of maximally entangled states \cite{Tamaryan09}. Each of these two states maximizes some entanglement measure \cite{Tamaryan09} . For example, the W state maximizes the relative entropy of entanglement  \cite{Plenio2001}. On the other hand, the GHZ state maximizes the negativity  across any bipartition \cite{Karol1998}. Importantly, the GHZ state and W state  serve as the most resourceful states in separate information theoretic tasks. Moreover, the assemblages produced from the GHZ state and the W state by applying particular orthogonal von Neumann measurements of rank-1 by each of the untrusted parties lead to the optimal quantum violations of different genuine EPR steering inequalities \cite{Cavalcanti15}. Therefore, in the steering scenario, it is sensible to consider these assemblages derived from the GHZ state as well as from the W state as perfectly genuine tripartite EPR steerable assemblages. Following this spirit, we also consider the assemblages that produce
sub-optimal violations of the above-mentioned inequalities \cite{Cavalcanti15} as partially steerable assemblages.

When one starts with infinitely many copies of the initial partially steerable assemblage,  we show that at least one copy of a perfectly genuine steerable assemblage can be distilled with certainty in 1SDI  and 2SDI scenarios. On the other hand, when one starts with finite copies of the initial assemblage, we quantify the performance of our distillation protocols using the notion of assemblage fidelity which quantifies the closeness between a perfectly genuine steerable assemblage and an assemblage obtained on average, when the distillation protocol is applied on finite copies of the initial assemblages. Through the concept of assemblage fidelity we demonstrate the efficacy of our distillation protocols in realistic scenarios involving only few copies of the initial assemblages.

The paper is organised as follows. In Sec. \ref{section2}, we briefly outline the essential features of genuine tripartite steering and assemblage fidelity. In Sec. \ref{section3}, we demonstrate our steering distillation protocols in the two types of hybrid tripartite networks. Concluding remarks  are presented in Sec. \ref{section4}.

\section{Preliminaries}\label{section2}

Consider a tripartite network where three spatially separated parties, say Alice, Bob and Charlie, share an unknown quantum system $\rho^{\text{ABC}} \in \mathcal{B}(\mathcal{H}_A \otimes \mathcal{H}_B \otimes \mathcal{H}_C)$.
Here, $\mathcal{B}(\mathcal{H}_A \otimes \mathcal{H}_B \otimes \mathcal{H}_C)$ stands for the set of all density operators acting on the Hilbert space $\mathcal{H}_A \otimes \mathcal{H}_B \otimes \mathcal{H}_C$. We begin by recapitulating the definitions of genuine tripartite EPR steering \cite{Cavalcanti15,Cavalcanti16,Bihalan18}.

\subsection{Genuine Tripartite EPR Steering}

Let us consider that a tripartite state $\rho^{\text{ABC}}$ is shared among three observers, say, Alice, Bob and Charlie. In the context of SDI tripartite network, there can be two scenarios based on the number of trusted or untrusted party: 1) one-sided device-independent (1SDI) scenario and 2) two-sided device-independent (2SDI) scenario.


For the 1SDI scenario, the local Hilbert space dimension of Alice's subsystem (untrusted party) is arbitrary and the local Hilbert space dimensions of Bob's and Charlie's subsystems (trusted parties) are fixed. Alice performs black-box measurements, Bob and Charlie perform characterized measurements. Alice's measurements are denoted by $X_x$ with outcomes $a$. Here, $x$ $\in$ $\{0,1, \cdots, n_A-1\}$ denotes the measurement choices of Alice and $a$ $\in$ $\{0, 1, \cdots, d_A-1\}$. Here,  $n_A$ and $d_A$ are natural numbers. The POVM elements associated with Alice's measurements are $\{M^A_{a|x}\}_{a,x}$ (where $M^A_{a|x} \geq 0$ $\forall a, x$; and $\sum_{a} M^A_{a|x} = \openone$ $\forall x$). Since, Bob's and Charlie's system are characterized/trusted, they can perform state tomography. The steering scenario is characterised by the assemblage  $\{\sigma^{\text{BC}}_{a|x}\}_{a,x}$ which is the set of unnormalized conditional states on Bob's and Charlie's sides with
\begin{equation}
	\sigma^{\text{BC}}_{a|x} = \text{tr}_A\Big[(\text{M}_{a|x}^{A} \otimes \openone_B \otimes \openone_C) \rho^{\text{ABC}} \Big]. 
	\label{assemblage1}
\end{equation}
Each element in the assemblage  is given by $\sigma^{BC}_{a|x}=p(a|x)\varrho_{BC}(a, x)$,  where $p(a|x)$ is the conditional probability of getting the outcome $a$ when Alice performs the measurement $x$; $\varrho_{BC}(a, x)$ is the normalized conditional state on Bob's and Charlie's end.

If the state $\rho^{\text{ABC}}$ contains no genuine entanglement then it can be decomposed in the bi-separable form as following 
\begin{eqnarray} \nonumber
	\rho^{\text{ABC}} &=& \sum_\lambda p_\lambda^{A:BC}\rho_\lambda^A \otimes \rho_\lambda^{BC} 
						 + \sum_\mu p_\mu^{B:AC}\rho_\mu^B\otimes \rho_\mu^{AC}\\ \nonumber
						 &+& \sum_\nu p_\nu^{AB:C} \rho_{\nu}^{AB}\otimes \rho_\nu^C,\\ 
\label{bisep}
\end{eqnarray} 
where $p_\lambda^{A:BC}, p_\mu^{B:AC}$ and $p_\nu^{AB:C}$ are probability distributions.
In this case, the assemblage (\ref{assemblage1}) has the following form \cite{Cavalcanti15}
\begin{eqnarray} 
	\sigma^{\text{BC}}_{a|x} &=& \sum_\lambda p_\lambda^{A:BC}p_\lambda(a|x)\rho_\lambda^{BC}
							\label{1SA:BC} \\
						 &+& \sum_\mu p_\mu^{B:AC}\rho_\mu^B\otimes \sigma_{a|x\mu}^{C} 
						        \label{1SB:AC}. \\ 
						 &+& \sum_\nu p_\nu^{AB:C} \sigma_{a|x\nu}^{B}\otimes \rho_\nu^C. \label{1SAB:C} \\
						 \nonumber 
\end{eqnarray} 
The bi-separable form of the state (\ref{bisep}) imposes constraints on the observed assemblage. For instance, (\ref{1SA:BC}) is an unsteerable assemblage from Alice to Bob-Charlie. The assemblage in (\ref{1SB:AC}) has two features: (i) It is unsteerable from Alice to Bob, but not necessarily from Alice to Charlie; (ii) It is separable. Similarly, the assemblage in (\ref{1SAB:C}) has two features: (i) it is unsteerable from Alice to Charlie, but not necessarily from Alice to Bob; (ii) It is separable. 

When each element of an assemblage $\{\sigma^{\text{BC}}_{a|x}\}_{a,x}$ can be written in the above form, then the assemblage does not demonstrate genuine EPR steering in 1SDI scenario, otherwise it demonstrates genuine EPR steering in 1SDI scenario. 


In 2SDI scenario, the local Hilbert space dimension of both Alice's and Bob's subsystems (untrusted parties) are arbitrary and the local Hilbert space dimension of Charlie's subsystem (trusted party) is fixed. Here, Alice and Bob perform black-box measurements, Charlie performs characterized measurements. Similar to 1SDI scenario, Alice's measurements are denoted by $X_x$ with outcomes $a$. On the other hand, Bob's measurements are denoted by  $Y_y$  with outcomes $b$. Here, $y$ $\in$ $\{0,1, \cdots, n_B-1\}$ denotes the measurement choices of Bob and $b$ $\in$ $\{0, 1, \cdots, d_B-1\}$. Here, $n_B$ and $d_B$ are natural numbers. The POVM elements associated with Bob's measurements are $\{M^B_{b|{y}}\}_{b,y}$ (where $M^B_{b|{y}} \geq 0$ $\forall b, y$; and $\sum_{b} M^B_{b|{y}} = \openone$ $\forall y$). These local measurements by Alice and Bob prepare the assemblage $\{\sigma^{C}_{a,b|x,y}\}_{a,b,x,y}$, which are the set of unnormalized conditional states on Charlie's side with 
\begin{equation}
	\sigma^{\text{C}}_{a,b|x,y} = \text{tr}_{\text{AB}} \Big[(\text{M}_{a|x}^{A} \otimes \text{M}_{b|y}^{B} \otimes \openone_C) \rho^{\text{ABC}} \Big].
	\label{assemblage2}
\end{equation}
Each element in the assemblage  is given by $\sigma^{C}_{a,b|x,y}=p(a, b|x,y)\varrho_{C}(a, b, x, y)$,  where $p(a, b|x, y)$ is the conditional probability of getting the outcome $a$ and $b$ when Alice performs the measurement $x$ and Bob performs measurement $y$ respectively; $\varrho_{C}(a, b, x, y)$ is the normalized conditional state on Charlie's end. 

If the state $\rho^{\text{ABC}}$ contains no genuine entanglement (i.e., it is bi-separable (\ref{bisep})), then the assemblage (\ref{assemblage2}) has the following form \cite{Cavalcanti15}
\begin{eqnarray} 
	\sigma^{\text{C}}_{a,b|x,y} &=& \sum_\lambda p_\lambda^{A:BC}p_\lambda(a|x)\sigma_{b|y\lambda}^{C}
		\label{2SA:BC}\\ 
						 &+& \sum_\mu p_\mu^{B:AC}p_\mu(b|y) \sigma_{a|x\mu}^{C}
						 \label{2SB:AC}\\ 
						 &+& \sum_\nu p_\nu^{AB:C} p_{\nu}(a,b|x,y) \rho_\nu^C .
						 \label{2SAB:C} \\ \nonumber
\end{eqnarray} 
The fact that the state $\rho^{\text{ABC}}$ is bi-separable imposes constraints on the  observed assemblage. For instance, the assemblage (\ref{2SA:BC}) is an unsteerable assemblage from Alice to Charlie, but not necessarily from Bob to Charlie. Similarly, the assemblage (\ref{2SB:AC}) is unsteerable from Bob to Charlie, but not necessarily from Alice to Charlie. The assemblage (\ref{2SAB:C}) has two features: (i) It is unsteerable from Alice-Bob to  Charlie,  (ii) The probability distribution $p_{\nu}(a,b|x, y)$ arises due to local measurements performed on a possibly entangled state, it may contain nonlocal quantum correlations. 

When each element of an assemblage $\{\sigma^{\text{C}}_{a,b|x, y}\}_{a,b,x,y}$ can be written in the above form, then the assemblage does not demonstrate genuine EPR steering in 2SDI scenario, otherwise it demonstrates genuine EPR steering in 2SDI scenario.   


\subsubsection{Genuine tripartite EPR steering inequalities}

Cavalcanti \textit{et al.} designed several inequalities \cite{Cavalcanti15} which detect genuine entanglement of GHZ state given by, $\frac{1}{\sqrt{2}}(|000\rangle + |111 \rangle)$ and W state given by, $\frac{1}{\sqrt{3}}(|001\rangle + |010 \rangle + |100\rangle)$ in the two scenarios mentioned above. These inequalities are nothing but genuine EPR steering inequalities \cite{Cavalcanti15}. For GHZ state in 1SDI scenario, the inequality has the following form: 
\begin{align}
G_1 &= 1 + 0.1547 \langle Z_BZ_C\rangle - \frac{1}{3} ( \langle A_3Z_B \rangle + \langle A_3Z_C \rangle  \nonumber \\
& + \langle A_1X_BX_C\rangle - \langle A_1Y_BY_C\rangle - \langle A_2X_BY_C\rangle \nonumber \\
& - \langle A_2Y_BX_C\rangle )\geq 0,
\label{GHZ1}
\end{align}
with $A_i$ for $i=1, 2, 3$, being the observables associated with Alice's uncharacterized measurements with outcomes $\pm 1$ and $X$, $Y$ and $Z$ represent Pauli operators. The GHZ state violates the inequality by $-0.845$ when Alice's measurements are $X$, $Y$ and $Z$, which numerical optimisation suggests are the optimal choices for Alice. 

 In this 1SDI scenario, we will consider the assemblage produced from the GHZ state contingent upon using the measurements of $X$, $Y$ and $Z$ by Alice as a perfectly genuine tripartite steerable assemblage because of the following reasons:

$\bullet$ This assemblage is produced from the GHZ state, which is one extreme member in the family of maximally entangled three-qubit states.

$\bullet$ It is produced by applying orthogonal von Neumann measurements of rank-1.

$\bullet$ It maximally violates the genuine tripartite EPR steering inequality (\ref{GHZ1}).

For GHZ state in 2SDI scenario, the inequality has the following form: 
\begin{align}
G_2 &= 1 - \alpha (\langle A_3B_3\rangle +  \langle A_3Z \rangle + \langle B_3Z \rangle) - \beta( \langle A_1B_1X \rangle \nonumber \\
& -  \langle A_1B_2Y\rangle -  \langle A_2B_1Y\rangle -   \langle A_2B_2X\rangle )\geq 0,
\label{GHZ2}
\end{align}
where $\alpha = 0.183$, $\beta = 0.258$, $A_i$ and $B_i$ with $i=1, 2, 3$ represent the observables associated with Alice and Bob's uncharacterized measurements, respectively, with outcomes $\pm 1$. The optimal quantum violation of this inequality is $-0.582$. This is achieved by GHZ state when Alice and Bob both perform $X$, $Y$ and $Z$ measurements. 

 Since, the assemblage produced from the GHZ state contingent upon using the orthogonal von Neumann measurements of $X$, $Y$ and $Z$ by Alice as well as by Bob in the above 2SDI scenario violates the genuine steering inequality (\ref{GHZ2}) maximally, this assemblage will be considered as a perfectly genuine tripartite steerable assemblage in the 2SDI scenario.

Similar inequalities for W-state are given for both the scenarios. For W state in 1SDI scenario, the inequality has the following form: 
\begin{align}
W_1 &= 1 + 0.4405(\langle Z_B \rangle + \langle Z_C \rangle) - 0.0037  \langle Z_BZ_C\rangle   \nonumber \\
& - 0.1570 ( \langle X_B X_C \rangle + \langle Y_B Y_C \rangle + \langle A_3 X_B X_C \rangle \nonumber \\
&+ \langle A_3 Y_B Y_C \rangle )   + 0.2424 ( \langle A_3 \rangle +   \langle A_3Z_BZ_C \rangle ) \nonumber \\
&+ 0.1848 ( \langle A_3Z_B\rangle + \langle A_3Z_C \rangle )  - 0.2533( \langle A_1X_B\rangle \nonumber \\
&+ \langle A_1X_C\rangle + \langle A_2Y_B\rangle + \langle A_2Y_C\rangle   + \langle A_1X_BZ_C\rangle \nonumber \\
&+ \langle A_1Z_BX_C\rangle ) + \langle A_2Y_BZ_C\rangle + \langle A_2Z_BY_C\rangle )  \geq 0,
\label{W1}
\end{align}
with the  W state achieving the optimal quantum violation $-0.759$. 

Similar to the case of the GHZ state, the assemblage derived from the W state by applying the orthogonal von Neumann measurements of $X$, $Y$ and $Z$ on Alice's subsystem in the above 1SDI scenario violates the genuine steering inequality (\ref{W1}) maximally. This is why this assemblage will be considered as a perfectly genuine tripartite steerable assemblage in the 1SDI scenario.

For W state in 2SDI scenario, the inequality has the following form: 
\begin{align}
W_2 &= 1 + 0.2517(\langle A_3 \rangle + \langle B_3 \rangle) + 0.3520  \langle Z\rangle \nonumber \\
&- 0.1112 ( \langle A_1 X \rangle + \langle A_2 Y \rangle + \langle B_1 X  \rangle + \langle B_2 Y  \rangle ) \nonumber \\
&+ 0.1296 ( \langle A_3 Z \rangle + \langle B_3 Z \rangle)  - 0.1943 ( \langle A_1B_1\rangle \nonumber \\
&+ \langle A_2B_2 \rangle ) + 0.2277  \langle A_3B_3\rangle  - 0.1590( \langle A_1B_1Z\rangle \nonumber \\
&+ \langle A_2B_2Z\rangle ) + 0.2228 \langle A_3B_3Z\rangle   -0.2298 (\langle A_1B_3X\rangle \nonumber \\
&+ \langle A_2B_3Y\rangle + \langle A_3B_1X\rangle + \langle A_3B_2 Y\rangle )  \geq 0,
\label{W2}
\end{align}
 with the  W state achieving the optimal quantum violation $-0.480$. 

In the 2SDI case,  the assemblage derived from the W state by performing the orthogonal von Neumann measurements of $X$, $Y$ and $Z$ on Alice's subsystem as well as on Bob's subsystem  violates the genuine steering inequality (\ref{W2}) maximally. This assemblage will thus be considered as a perfectly genuine tripartite steerable assemblage in the 2SDI scenario.

\subsection{Assemblage fidelity}

Assemblage is a set of unnormalized conditional states (known as assemblage elements). The property of an assemblage depends on the properties of all of its elements. In a particular steering scenario, two assemblages are inequivalent if at least one element of the first assemblage is different from the corresponding element of the second assemblage. In a steering distillation task, we start with $N$ copies of partially genuine steerable assemblage. In the asymptotically many copies, $N \rightarrow \infty$ limit, any distillation protocol must guarantee extraction of at least one copy of the target assemblage. The target assemblage is derived from a pure maximally entangled state by applying orthogonal Von Neumann measurements of rank-1 by the untrusted parties and gives the optimal quantum violation of the appropriate genuine EPR steering inequality mentioned earlier. That is why, we consider these assemblages as perfectly genuine steerable assemblages. However, it needs further investigation to address whether these target assemblages are indeed perfect assemblages. For this, we need to formulate proper resource theory of genuine tripartite EPR steering, proper quantifier of genuine steering, which are beyond the scope of the present paper. On the other hand, any assemblage not giving the optimal quantum violation is defined as the partially genuine steerable assemblage. For a limited number of copies, perfect extraction may not be achievable. Hence, we use the notion of assemblage fidelity \cite{Nery20} to capture the equivalence between two assemblages and have a figure of merit of the protocol for the finite number of copies $N$. 

\textit{1SDI scenario:} Let $\{\sigma_{a|x}^{dist} \}_{a,x}$ be the assemblage after the application of distillation protocol and  $\{\sigma_{a|x}^{t} \}_{a,x}$ be the target assemblage. Here, $\{\sigma_{a|x}^{dist} \}_{a,x}$ and $\{\sigma_{a|x}^{t} \}_{a,x}$ are generic assemblages with same number of inputs and outputs and with components acting on the same Hilbert space. The assemblage fidelity between $\{\sigma_{a|x}^{dist} \}_{a,x}$ and $\{\sigma_{a|x}^{t} \}_{a,x}$ is defined as \cite{Nery20}
\begin{equation}
	\mathcal{F}_A \Big(\{\sigma_{a|x}^{dist} \}_{a,x} , \{\sigma_{a|x}^{t} \}_{a,x} \Big):= \underset{x}{\mathrm{min}}\sum_{a} \mathcal{F}(\sigma_{a|x}^{dist}, \sigma_{a|x}^{t}) ,
\end{equation}
where $\mathcal{F}(A ,B) = \text{Tr}[\sqrt{\sqrt{A}B \sqrt{A}}]$ is the fidelity applied on two bounded positive semidefinite operators $A$ and $B$. $\mathcal{F}_A$ is non-negative and $\mathcal{F}_A \Big(\{\sigma_{a|x}^{dist} \}_{a,x} , \{\sigma_{a|x}^{t} \}_{a,x} \Big) \leq 1$, with the equality holding iff $\{\sigma_{a|x}^{dist} \}_{a,x} = \{\sigma_{a|x}^{t} \}_{a,x}$ \cite{Nery20}.  The minimum is taken in the definition of the assemblage fidelity in order to capture the equivalence between two assemblages by characterizing the minimum overlap between the elements of two assemblages under consideration.

\textit{2SDI scenario:} Let $\{\sigma_{a,b|x, y}^{dist}\}_{a,b,x,y}$ be the assemblage after the application of distillation protocol and $\{\sigma_{a,b|x,y}^{t}\}_{a,b,x,y}$ is the target assemblage. Here, $\{\sigma_{a,b|x, y}^{dist}\}_{a,b,x,y}$ and $\{\sigma_{a,b|x,y}^{t}\}_{a,b,x,y}$ are generic assemblages with same number of inputs and outputs and with components acting on the same Hilbert space. The assemblage fidelity between $\{\sigma_{a,b|x, y}^{dist}\}_{a,b,x,y}$ and $\{\sigma_{a,b|x,y}^{t}\}_{a,b,x,y}$ is defined as 

\begin{align}
	& \mathcal{F}_A \Big(\{\sigma_{a,b|x, y}^{dist}\}_{a,b,x,y}, \{\sigma_{a,b|x,y}^{t}\}_{a,b,x,y} \Big) \nonumber \\
	&:= \underset{x,y}{\mathrm{min}}\sum_{a,b} \mathcal{F}(\sigma_{a,b|x,y}^{dist}, \sigma_{a,b|x,y}^{t}),
\end{align}

with $\mathcal{F}_A$ having the same properties as mentioned in 1SDI scenario. 

\subsection{Genuine tripartite steering versus genuine tripartite entanglement distillation}
In a genuine tripartite steering distillation protocol, only the trusted parties can do local quantum operations, whereas, in a genuine tripartite entanglement distillation protocol, all parties generally perform local quantum operations. The advantage of designing a genuine tripartite EPR steering distillation protocol is that the same protocol can be adopted for genuine tripartite entanglement distillation with lesser resource (fewer parties perform quantum  operations). However, a genuine tripartite entanglement distillation protocol does not always serve as a genuine tripartite steering distillation protocol because free operations in the context of genuine entanglement are not always the free operations in case of genuine tripartite steering. For example, following our distillation protocol, one can even distill genuine entanglement in generalized GHZ or W state. On the other hand, following the approach of distilling genuine tripartite entanglement of GHZ state or W state (which involves local quantum measurements by all parties that is not a free operation in genuine tripartite steering scenario), one does not obtain a distillation protocol of perfectly genuine steerable assemblage derived from the GHZ state or W state. Next, we will discuss the distillation protocols in the two aforementioned genuine tripartite steering scenarios.


\section{Genuine tripartite steering distillation}\label{section3}

Consider a hybrid quantum tripartite network constituted by three spatially separated parties, Alice, Bob and Charlie, sharing a three-qubit state. Alice's subsystem is always uncharacterized/untrusted and Charlie's subsystem is always fully characterized/trusted. Depending on whether the scenario is 1SDI or 2SDI, Bob's subsystem is characterized or uncharacterized respectively. 

Given $N \geq 2$ copies of the partly genuine steerable assemblages, the task of the genuine steering distillation is to create $M$ $(M < N )$ copies of perfectly genuine steerable target assemblage using free operations (that cannot create genuine steerable assemblage from assemblage not demonstrating genuine steering) only. Now, we  discuss our distillation protocol for 1SDI and 2SDI scenarios.

$\bullet$ {\bf Distillation protocol in 1SDI scenario:}  Here the assemblages at Bob-Charlie's end are produced due to uncharacterized measurements by Alice. These assemblages are unnormalized two-qubit states. In our protocol, we start with $N \geq 2$ copies of initial partially genuine steerable assemblages $\sigma_{a|x}^{BC} =p(a|x)\varrho_{BC}(a, x)$. Each of the two trusted parties (Bob and Charlie) perform a dichotomic qubit POVM  $\textbf{G}^{i}:=\{G_{0}^{i},G_{1}^{i}\}$, satisfying $G_{o^u_i}^{i} \ge 0$ $\forall$ $o^u_i \in \{0,1\}$ and $G_0^{i} + G_1^{i} = \openone$ on $u^{\text{th}}$ copy of the initial assemblage (with $u \in \{1, 2, \cdots, N-1\}$) and gets an outcome $o^u_i \in \{ 0, 1\}$. Here $i=B$ for Bob's POVM and $i=C$ for Charlie's POVM. If any of the trusted parties gets $o^u_i=1$ for all $u \in \{1, 2, \cdots, N-1\}$, then that trusted party sets $o^N_i=0$ for the $N^{\text{th}}$ copy, otherwise sets $o^N_i=1$ for the $N^{\text{th}}$ copy. Next, that trusted party sends the string $\textbf{o}_{i}:=\{o^1_i, o^2_i, o^3_i, \cdots, o^{N}_i\}$ to all the other parties (trusted as well as untrusted) in the multipartite network through a classical channel. Finally, all parties discard the $u^{\text{th}}$ copy ($u \in \{1, 2, \cdots, N\}$) for which $o^u_B=1$ or $o^u_C=1$. The output of the protocol is given by the remaining  copies of assemblages post Bob's and Charlie's measurements.

Next, we introduce the notation $K_{o^u_i}^{i} = \sqrt{G_{o^u_i}^{i}}$ such that $G_{o^u_i}^{i} = K_{o^u_i}^{i^{\dagger}} K_{o^u_i}^{i}$. In the above protocol, when Bob gets the outcome $o^u_B$ and Charlie gets the outcome $o^u_C$ contingent upon performing the above POVMs on any of the $u^{\text{th}}$ copy (with $u \in \{1, 2, \cdots, N-1\}$), the assemblage’s components are updated by,
\begin{align}
    \widetilde{\sigma}_{a|x}^{BC} = p(a|x, o^u_B, o^u_C) \, \varrho_{BC}^{o^u_B, o^u_C}(a, x) \, \forall \, a,x,
    \label{een1}
\end{align}
where the effects of the POVMs on Alice’s black box are considered (as the final output $a$ depends on $o^u_B$ and $o^u_C$) and
\begin{align}
 \varrho_{BC}^{o^u_B, o^u_C}(a, x) &=    \frac{\Big(K_{o^u_B}^{B} \otimes K_{o^u_C}^{C}\Big) \varrho_{BC}(a, x) \Big(K_{o^u_B}^{B^{\dagger}} \otimes K_{o^u_C}^{C^{\dagger}} \Big)}{\text{Tr} \Big[ \Big(K_{o^u_B}^{B} \otimes K_{o^u_C}^{C} \Big) \varrho_{BC}(a, x) \Big(K_{o^u_B}^{B^{\dagger}} \otimes K_{o^u_C}^{C^{\dagger}} \Big) \Big]} \nonumber \\
 &=    \frac{\Big(K_{o^u_B}^{B} \otimes K_{o^u_C}^{C}\Big) \varrho_{BC}(a, x) \Big(K_{o^u_B}^{B^{\dagger}} \otimes K_{o^u_C}^{C^{\dagger}} \Big)}{p(o^u_B, o^u_C|a,x)}.
 \label{een2}
\end{align}
Next, using Bayes' rule, we have 
\begin{align}
 p(a|x, o^u_B, o^u_C) = \dfrac{p(a|x) \, p(o^u_B, o^u_C|a,x)}{p(o^u_B, o^u_C|x)}  , 
 \label{een3}
\end{align}
where
\begin{align}
 p(o^u_B, o^u_C|x) &= p(o^u_B, o^u_C) \nonumber \\
 &= \text{Tr} \Big[ \Big(K_{o^u_B}^{B} \otimes K_{o^u_C}^{C} \Big) \rho^{BC} \Big(K_{o^u_B}^{B^{\dagger}} \otimes K_{o^u_C}^{C^{\dagger}} \Big) \Big].
 \label{een4}
\end{align}
Here, $p(o^u_B, o^u_C)$ denotes the probability that Bob gets the outcome $o^u_B$ and Charlie gets the outcome $o^u_C$ contingent upon performing the above POVMs on any of the $u^{\text{th}}$ copy (with $u \in \{1, 2, \cdots, N-1\}$) and is independent of Alice's input $x$; $\rho^{BC} = \text{Tr}_{A}[\rho^{ABC}] = \sum_{a} \sigma_{a|x}^{BC} $. Hence, using Eqs.(\ref{een1}), (\ref{een2}), (\ref{een3}), (\ref{een4}), we get
\begin{align}
    \widetilde{\sigma}_{a|x}^{BC} = \frac{\Big(K_{o^u_B}^{B} \otimes K_{o^u_C}^{C}\Big) \sigma_{a|x}^{BC} \Big(K_{o^u_B}^{B^{\dagger}} \otimes K_{o^u_C}^{C^{\dagger}} \Big)}{\text{Tr} \Big[ \Big(K_{o^u_B}^{B} \otimes K_{o^u_C}^{C} \Big) \rho^{BC} \Big(K_{o^u_B}^{B^{\dagger}} \otimes K_{o^u_C}^{C^{\dagger}} \Big) \Big]} \, \forall \, a,x.
    \label{finalass1}
\end{align}

In an alternative protocol, any one of the two trusted parties, say Bob, performs a dichotomic qubit POVM  $\textbf{G}^B:=\{G_{0}^B,G_{1}^B \}$, satisfying $G^B_{o^u_B} \ge 0$ $\forall$ $o^u_B \in \{0,1\}$ and $G^B_0 + G^B_1 = \openone$ on $u^{\text{th}}$ copy of the initial assemblage (with $u \in \{1, 2, \cdots, N-1\}$) and gets an outcome $o^u_B \in \{ 0, 1\}$. The other trusted party (Charlie) does nothing. If Bob gets $o^u_B=1$ for all $u \in \{1, 2, \cdots, N-1\}$, then he sets $o^N_B=0$ for the $N^{\text{th}}$ copy, otherwise sets $o^N_B=1$ for the $N^{\text{th}}$ copy. Then Bob sends the string $\textbf{o}_B:=\{o^1_B, o^2_B, o^3_B, \cdots, o^{N}_B\}$ to Alice and Charlie. Finally, all parties discard $u^{\text{th}}$ copy ($u \in \{1, 2, \cdots, N\}$) for which $o^u_B = 1$. The output of this protocol is given by the remaining  copies of assemblages post Bob's measurements. In this protocol, when Bob gets the outcome $o^u_B$ contingent upon performing the above POVM on any of the $u^{\text{th}}$ copy (with $u \in \{1, 2, \cdots, N-1\}$), then the assemblage’s components are updated by 
\begin{align}
    \widetilde{\sigma}_{a|x}^{BC} = \frac{\Big(K_{o^u_B}^B \otimes \openone \Big) \sigma_{a|x}^{BC} \Big(K_{o^u_B}^{B^{\dagger}} \otimes \openone \Big)}{\text{Tr} \Big[ \Big(K_{o^u_B}^B \otimes \openone \Big) \rho^{BC} \Big(K_{o^u_B}^{B^{\dagger}} \otimes \openone \Big) \Big]} \, \forall \, a,x.
    \label{finalass2}
\end{align}
Here, $K_{o^u_B}^B = \sqrt{G_{o^u_B}^B}$ and $p(o^u_B) = \text{Tr} \Big[ \Big(K_{o^u_B}^B \otimes \openone \Big) \rho^{BC} \Big(K_{o^u_B}^{B^{\dagger}} \otimes \openone \Big) \Big]$ is the probability that Bob gets the outcome $o^u_B$.

For any single copy of the initial assemblage, the above operations consist of local quantum operations by each of the trusted parties and classical communications from each of the trusted parties to other (trusted and untrusted) parties. In Appendix \ref{app1}, we show that this operation is a free operation of genuine tripartite EPR steering in 1SDI scenario.

$\bullet$ {\bf Distillation protocol in 2SDI scenario:}  Here the assemblages at Charlie's end are produced due to uncharacterized local measurements by Alice and Bob. These assemblages are unnormalized qubit states. In our protocol, we start with $N \geq 2$ copies of initial partly genuine steerable assemblages $\sigma_{a,b|x,y}^{C} =p(a,b|x,y)\varrho_{C}(a,b, x,y)$. The trusted party (Charlie) performs a dichotomic qubit POVM  $\textbf{G}^{C}:=\{G_{0}^{C},G_{1}^{C}\}$, satisfying $G_{o^u_C}^{C} \ge 0$ $\forall$ $o^u_C \in \{0,1\}$ and $G_0^{C} + G_1^{C} = \openone$ on the $u^{\text{th}}$ copy of the initial assemblage (with $u \in \{1, 2, \cdots, N-1\}$) and gets an outcome $o^u_C \in \{ 0, 1\}$. If Charlie gets $o^u_C=1$ for all $u \in \{1, 2, \cdots, N-1\}$, then he sets $o^N_C=0$ for the $N^{\text{th}}$ copy, otherwise sets $o^N_C=1$ for the $N^{\text{th}}$ copy. Charlie then sends the string $\textbf{o}_{i}:=\{o^1_i, o^2_i, o^3_i, \cdots, o^{N}_i\}$ to Alice and Bob. Finally, all parties discard $u^{\text{th}}$ copy ($u \in \{1, 2, \cdots, N\}$) for which $o^u_C=1$. The output of the protocol is given by the remaining  copies of assemblages post Charlie's measurements. In this protocol, when Charlie gets the outcome $o^u_C$ contingent upon performing the above POVM on any of the $u^{\text{th}}$ copy (with $u \in \{1, 2, \cdots, N-1\}$), then the assemblage’s components are updated by 
\begin{align}
    \widetilde{\sigma}_{a,b|x,y}^{C} = \frac{ K_{o^u_C}^C \, \sigma_{a,b|x,y}^{C} \, K_{o^u_C}^{C^{\dagger}} }{\text{Tr} \Big[  K_{o^u_C}^C \, \rho^{C} \, K_{o^u_C}^{C^{\dagger}} \Big]} \, \forall \, a,b,x,y,
    \label{finalass3}
\end{align}
where $\rho^{C} = \text{Tr}_{AB}[\rho^{ABC}] = \sum_{a,b} \sigma_{a,b|x,y}^{C}$ $\forall$ $x,y$, $K_{o^u_C}^C = \sqrt{G_{o^u_C}^C}$ and $p(o^u_C) = \text{Tr} \Big[  K_{o^u_C}^C \, \rho^{C} \, K_{o^u_C}^{C^{\dagger}} \Big]$ is the probability that Charlie gets the outcome $o^u_C$.

For any single copy of the initial assemblage, the above operations consist of local quantum operations by  the trusted party and classical communications from  the trusted party to the other untrusted parties. In Appendix \ref{app2}, we show that this operation is a free operation of genuine tripartite EPR steering in 2SDI scenario. 

Next, we will demonstrate different distillation protocols for different initial assemblages in 1SDI and 2SDI scenarios.

\subsection{Distillation of genuine steerable assemblage of GHZ state}\label{subsection1}

$\bullet$ {\bf 1SDI scenario:} Consider a tripartite steering scenario where Alice, Bob and Charlie initially  share  $N \geq 2$ copies of the assemblage $\{\sigma_{a|x}^{BC^{\text{GGHZ}}}\}_{a,x}$ obtained from the three-qubit generalized GHZ (GGHZ) state,
\begin{equation}
	|\psi_{\text{GGHZ}} \rangle = \cos \theta \ket{000} + \sin \theta \ket{111}, \quad 0 < \theta < \frac{\pi}{4},
	\label{GGHZ}
\end{equation}
through the measurements of observables $X_0= \sigma_x$, $X_1=\sigma_y$ and $X_2= \sigma_z$ by Alice. The components of the assemblages are given by,
\begin{eqnarray}
	\sigma_{0|0}^{BC^{\text{GGHZ}}} &=& \frac{1}{2} \Big| \theta_+^0 \Big\rangle \Big\langle \theta_+^0 \Big|,\quad \hspace{0.45cm} \sigma_{1|0}^{BC^{\text{GGHZ}}} = \frac{1}{2} \Big| \theta_-^0 \Big\rangle \Big\langle \theta_-^0 \Big| \nonumber \\ 
    \sigma_{0|1}^{BC^{\text{GGHZ}}} &=& \frac{1}{2} \Big| \theta_-^1 \Big\rangle \Big\langle \theta_-^1 \Big|,\quad \hspace{0.45cm} \sigma_{1|1}^{BC^{\text{GGHZ}}} = \frac{1}{2} \Big| \theta_+^1 \Big\rangle \Big\langle \theta_+^1 \Big| \nonumber \\ 
	\sigma_{0|2}^{BC^{\text{GGHZ}}} &=& \cos^2\theta \ket{00}\bra{00},\nonumber \\
	\sigma_{1|2}^{BC^{\text{GGHZ}}} &=& \sin^2\theta \ket{11}\bra{11},
	\label{GGHZassemblage}
\end{eqnarray}
where $\Big| \theta_\pm^0 \Big\rangle = \cos\theta \ket{00} \pm \sin\theta \ket{11}$ and $\Big| \theta_\pm^1 \Big\rangle = \cos\theta \ket{00} \pm i \, \sin\theta \ket{11}$. 

When these assemblages are subjected to the measurements  $\sigma_x$,  $\sigma_y$,  $\sigma_z$ by Bob and  Charlie as specified in the inequality (\ref{GHZ1}), they cannot give optimum quantum violation of the inequality (\ref{GHZ1}). Hence,  these initial assemblages are considered as partially genuine steerable assemblages in 1SDI scenario. Moreover, the magnitude of quantum violation of inequality (\ref{GHZ1}) by the above assemblages is a monotonic function of $\theta$ in the range $0 < \theta < \frac{\pi}{4}$.
Importantly, the inequality (\ref{GHZ1}) is violated by the GGHZ states (\ref{GGHZ}) for $\theta \in ( 0.185, \frac{\pi}{4})$ and hence it ensures genuine steering of the initial assemblage in this range.

Now, consider the  assemblage $\{\sigma_{a|x}^{{BC}^{\text{GHZ}}}\}_{a,x}$, produced from the GHZ state when Alice performs the measurements $X_0=\sigma_x$, $X_1=\sigma_y$ and $X_2=\sigma_z$, with components
\begin{eqnarray}
	\sigma_{0|0}^{{BC}^{\text{GHZ}}} &=& \frac{1}{2} \Bigg| \frac{\pi}{4}_+^0 \Bigg\rangle \Bigg\langle \frac{\pi}{4}_+^0 \Bigg|,\quad 
	\sigma_{1|0}^{{BC}^{\text{GHZ}}} = \frac{1}{2} \Bigg| \frac{\pi}{4}_-^0 \Bigg\rangle \Bigg\langle \frac{\pi}{4}_-^0 \Bigg| \nonumber \\ 
\sigma_{0|1}^{{BC}^{\text{GHZ}}} &=& \frac{1}{2} \Bigg| \frac{\pi}{4}_-^1 \Bigg\rangle \Bigg\langle \frac{\pi}{4}_-^1 \Bigg|,\quad 
\sigma_{1|1}^{{BC}^{\text{GHZ}}} = \frac{1}{2} \Bigg| \frac{\pi}{4}_+^1 \Bigg\rangle \Bigg\langle \frac{\pi}{4}_+^1 \Bigg| \nonumber \\ 
	\sigma_{0|2}^{{BC}^{\text{GHZ}}} &=& \frac{1}{2} \ket{00}\bra{00},\quad  \hspace{0.5cm}
	\sigma_{1|2}^{{BC}^{\text{GHZ}}} = \frac{1}{2} \ket{11}\bra{11},
	\label{GHZassemblage}
\end{eqnarray}
where $\Bigg| \dfrac{\pi}{4}_{\pm}^0 \Bigg\rangle:= \dfrac{1}{\sqrt{2}}(\ket{00} \pm \ket{11})$ and  $\Bigg| \dfrac{\pi}{4}_{\pm}^1 \Bigg\rangle:= \dfrac{1}{\sqrt{2}}( \ket{00} \pm i  \ket{11})$. The assemblage is derived from the GHZ state by applying orthogonal Von Neumann measurements of rank-1 by the untrusted parties. In literature, there are several information theoretic tasks where GHZ state serve as the most resourceful state. Also, when this assemblage is subjected to the measurements $\sigma_x$,  $\sigma_y$, $\sigma_z$ by Bob, Charlie,  the produced correlations give the optimum quantum violation of the inequality (\ref{GHZ1}). Hence,  this assemblage (\ref{GHZassemblage}) is a perfectly genuine steerable assemblage for the generalized GHZ states in 1SDI scenario. In our distillation protocol for the generalized GHZ states, this perfectly genuine steerable assemblage is the target assemblage.

Next, Bob and Charlie perform any of the two distillation strategies described below on the initial assemblage (\ref{GGHZassemblage}).

$\textit{Strategy 1: "Equal participation"}$ Both Bob and Charlie participate and perform quantum measurements (POVM) with the following measurement operators on the $u^{\text{th}}$ copy (for all $u \in \{1, 2, \cdots, N-1\}$) of the initial assemblage $\{\sigma_{a|x}^{BC^{\text{GGHZ}}}\}_{a,x}$:
\begin{eqnarray}
	K_{0}^{B} &=& \left(
\begin{array}{cc}
\sqrt{ \tan \theta } & 0 \\
 0 & 1 \\
\end{array}
\right), \quad K_{1}^{B} =  \left(
\begin{array}{cc}
 \sqrt{1-\tan \theta } & 0 \\
 0 & 0 \\
\end{array}
\right), \nonumber \\
K_{0}^{C} &=& \left(
\begin{array}{cc}
\sqrt{ \tan \theta } & 0 \\
 0 & 1 \\
\end{array}
\right), \quad K_{1}^{C} =  \left(
\begin{array}{cc}
 \sqrt{1-\tan \theta } & 0 \\
 0 & 0 \\
\end{array}
\right),
\label{GHZsteer1}
\end{eqnarray}
 where these matrices and all matrices henceforth are  in the computational basis $\{|0\rangle, |1\rangle\}$.

 As discussed earlier, all parties will not discard those copies (post Bob and Charlie's POVMs) for which both Bob and Charlie have obtained the outcome $0$. Hence, the probability with which each copy is not discarded is given by, $\tilde{P}_{\text{ND}}^{\text{GHZ}_1^1}$ = $p(o^u_B=0, o^u_C=0)$ = $\text{Tr} \Big[ \Big(K_{0}^{B} \otimes K_{0}^{C} \Big) \rho_{\text{GGHZ}}^{BC} \Big(K_{0}^{B^{\dagger}} \otimes K_{0}^{C^{\dagger}} \Big) \Big]$ = $2 \sin^2 \theta$. Here, $\rho_{\text{GGHZ}}^{BC} = \text{Tr}_{A}\Big[ |\psi_{\text{GGHZ}} \rangle \langle \psi_{\text{GGHZ}}|\Big]$. Note that $0 < 2 \sin^2 \theta < 1$ for all $\theta \in ( 0.185, \frac{\pi}{4})$. The updated assemblage after these POVMs by Bob and Charlie, when both of them get the outcome $0$, is nothing but our target assemblage (\ref{GHZassemblage}). This can be checked using Eq.(\ref{finalass1}).
 
$\textit{Strategy 2: "Single party participation"}$ Here, either Bob or Charlie performs quantum measurement with the following measurement operators on the $u^{\text{th}}$ copy (for all $u \in \{1, 2, \cdots, N-1\}$) of the initial assemblage $\{\sigma_{a|x}^{BC^{\text{GGHZ}}}\}_{a,x}$:
\begin{equation}
	K_{0}^{B/C} = \left(
\begin{array}{cc}
 \tan \theta  & 0 \\
 0 & 1 \\
\end{array}
\right), \quad K_{1}^{B/C} =  \left(
\begin{array}{cc}
 \sqrt{1 - \tan^2 \theta}  & 0 \\
 0 & 0 \\
\end{array}
\right).
\label{GHZsteer2}
\end{equation}

 All parties will not discard those copies (post Bob or Charlie's POVM) for which the outcome $0$ is obtained by Bob or Charlie (who performs the above measurement). Hence, when Bob performs POVM and Charlie performs nothing, the probability with which each copy is not discarded is given by, $\tilde{P}_{\text{ND}}^{\text{GHZ}_1^2}$ = $p(o^u_B=0)$ = $\text{Tr} \Big[ \Big(K_{0}^{B} \otimes \openone \Big) \rho^{BC} \Big(K_{0}^{B^{\dagger}} \otimes \openone \Big) \Big]$ = $2 \sin^2 \theta$. On the other hand, when Charlie performs POVM and Bob performs nothing, the probability with which each copy is not discarded is given by, $\tilde{P}_{\text{ND}}^{\text{GHZ}_1^2}$ = $p(o^u_C=0)$ = $\text{Tr} \Big[ \Big( \openone \otimes K_{0}^{C} \Big) \rho_{\text{GGHZ}}^{BC} \Big( \openone \otimes K_{0}^{C^{\dagger}} \Big) \Big]$ = $2 \sin^2 \theta$. The updated assemblage after the above quantum measurement by Bob or Charlie, when the outcome $0$ is obtained, is the target assemblage (\ref{GHZassemblage}). When Bob performs the POVM, then the updated assemblage is calculated using Eq.(\ref{finalass2}). On the other hand, when Charlie performs the POVM, then the updated assemblage is calculated using a similar equation.

 In case of the $\textit{Strategy 1}$, when either Bob or Charlie or both of them get the outcome $1$ contingent upon performing the POVMs on the $u^{\text{th}}$ copy for all $u \in \{1, 2, \cdots, N-1\}$, then all copies except the $N^{\text{th}}$ copy are discarded after the classical communications mentioned earlier. Hence, in this case the output of the distillation protocol is the $N^{\text{th}}$ copy on which no POVM is performed. This implies failure of the protocol as the $N^{\text{th}}$ copy in this case is nothing but the initial assemblage. Now, since the local quantum measurements are performed independently on each of $N-1$ copies, the  probability with which the distillation protocol fails is given by,
\begin{align}
 P^{\text{GHZ}_1}_{\text{fail}} &= \big(1- \tilde{P}_{\text{ND}}^{\text{GHZ}_1^i} \big)^{N-1}   \quad \text{with} \quad i=1 \nonumber \\
 &=(1-2\sin^2\theta)^{N-1}.
\end{align} 

Following similar arguments, it can be shown that the  probability with which the distillation protocol associated with $\textit{Strategy 2}$ fails is given by,
\begin{align}
 P^{\text{GHZ}_1}_{\text{fail}} &= \big(1- \tilde{P}_{\text{ND}}^{\text{GHZ}_1^i} \big)^{N-1}  \quad \text{with} \quad i=2 \nonumber \\
 &=(1-2\sin^2\theta)^{N-1},
\end{align} 
where $0 < P^{\text{GHZ}_1}_{\text{fail}} < 1$ $\forall$ $\theta \in ( 0.185, \frac{\pi}{4})$.

To summarize, in our distillation protocol associated with any of the two strategies, after the aforementioned classical communications, the parties manage to either keep at least one successfully distilled target assemblage (\ref{GHZassemblage})  with probability $P^{\text{GHZ}_1}_{\text{success}} = 1-(1-2\sin^2\theta)^{N-1}$, or the last copy of the initial assemblage  given by Eq.(\ref{GGHZassemblage}) with the probability $P^{\text{GHZ}_1}_{\text{fail}}$.

As number of copies $N$ tends to infinity, we get $\lim\limits_{N\to\infty} P^{\text{GHZ}_1}_{\text{success}} \rightarrow 1$ and, hence, the protocol ensures distillation of at least one copy of the target assemblage in the asymptotic regime.

 In the regime of finite $N$ copies with $N \geq 2$, a single assemblage can be extracted  $\{\sigma_{a|x}^{\text{dist}^{\text{GHZ}}} \}_{a,x}$ as a convex combination of the initial assemblage $\{\sigma_{a|x}^{BC^{\text{GGHZ}}}\}_{a,x}$ with components given in Eq.(\ref{GGHZassemblage}) and the target assemblage $\{\sigma_{a|x}^{{BC}^{\text{GHZ}}}\}_{a,x}$ with components given in Eq.(\ref{GHZassemblage}). The components of $\{\sigma_{a|x}^{\text{dist}^{\text{GHZ}}} \}_{a,x}$ are given by,
\begin{eqnarray}
	\sigma_{0|0}^{\text{dist}^{\text{GHZ}}} &=& \frac{1}{2} \Bigg( P^{\text{GHZ}_1}_{\text{success}}\Bigg| \frac{\pi}{4}_+^0 \Bigg\rangle \Bigg\langle \frac{\pi}{4}_+^0 \Bigg| + P^{\text{GHZ}_1}_{\text{fail}} \Big| \theta_+^0 \Big\rangle \Big\langle \theta_+^0 \Big| \Bigg),\nonumber \\ \sigma_{1|0}^{\text{dist}^{\text{GHZ}}} &=& \frac{1}{2} \Bigg( P^{\text{GHZ}_1}_{\text{success}} \Bigg| \frac{\pi}{4}_-^0 \Bigg\rangle \Bigg\langle \frac{\pi}{4}_-^0 \Bigg| + P^{\text{GHZ}_1}_{\text{fail}} \Big| \theta_-^0 \Big\rangle \Big\langle \theta_-^0 \Big| \Bigg) \nonumber\\
	\sigma_{0|1}^{\text{dist}^{\text{GHZ}}} &=& \frac{1}{2} \Bigg( P^{\text{GHZ}_1}_{\text{success}} \Bigg| \frac{\pi}{4}_-^1 \Bigg\rangle \Bigg\langle \frac{\pi}{4}_-^1 \Bigg| + P^{\text{GHZ}_1}_{\text{fail}} \Big| \theta_-^1 \Big\rangle \Big\langle \theta_-^1 \Big| \Bigg),\nonumber \\ \sigma_{1|1}^{\text{dist}^{\text{GHZ}}} &=& \frac{1}{2} \Bigg( P^{\text{GHZ}_1}_{\text{success}} \Bigg| \frac{\pi}{4}_+^1 \Bigg\rangle \Bigg\langle \frac{\pi}{4}_+^1 \Bigg| + P^{\text{GHZ}_1}_{\text{fail}}\Big| \theta_+^1 \Big\rangle \Big\langle \theta_+^1 \Big| \Bigg) \nonumber \\
	\sigma_{0|2}^{\text{dist}^{\text{GHZ}}} &=& \Bigg( \frac{1}{2} P^{\text{GHZ}_1}_{\text{success}} + \cos^2\theta \, P^{\text{GHZ}_1}_{\text{fail}} \Bigg) \ket{00}\bra{00},\nonumber \\ \sigma_{1|2}^{\text{dist}^{\text{GHZ}}} &=& \Bigg(\frac{1}{2} P^{\text{GHZ}_1}_{\text{success}} + \sin^2\theta \, P^{\text{GHZ}_1}_{\text{fail}} \Bigg)\ket{11}\bra{11}.
	\label{GGHZdistill}
\end{eqnarray}

\begin{figure}[t!]
\resizebox{8cm}{7cm}{\includegraphics{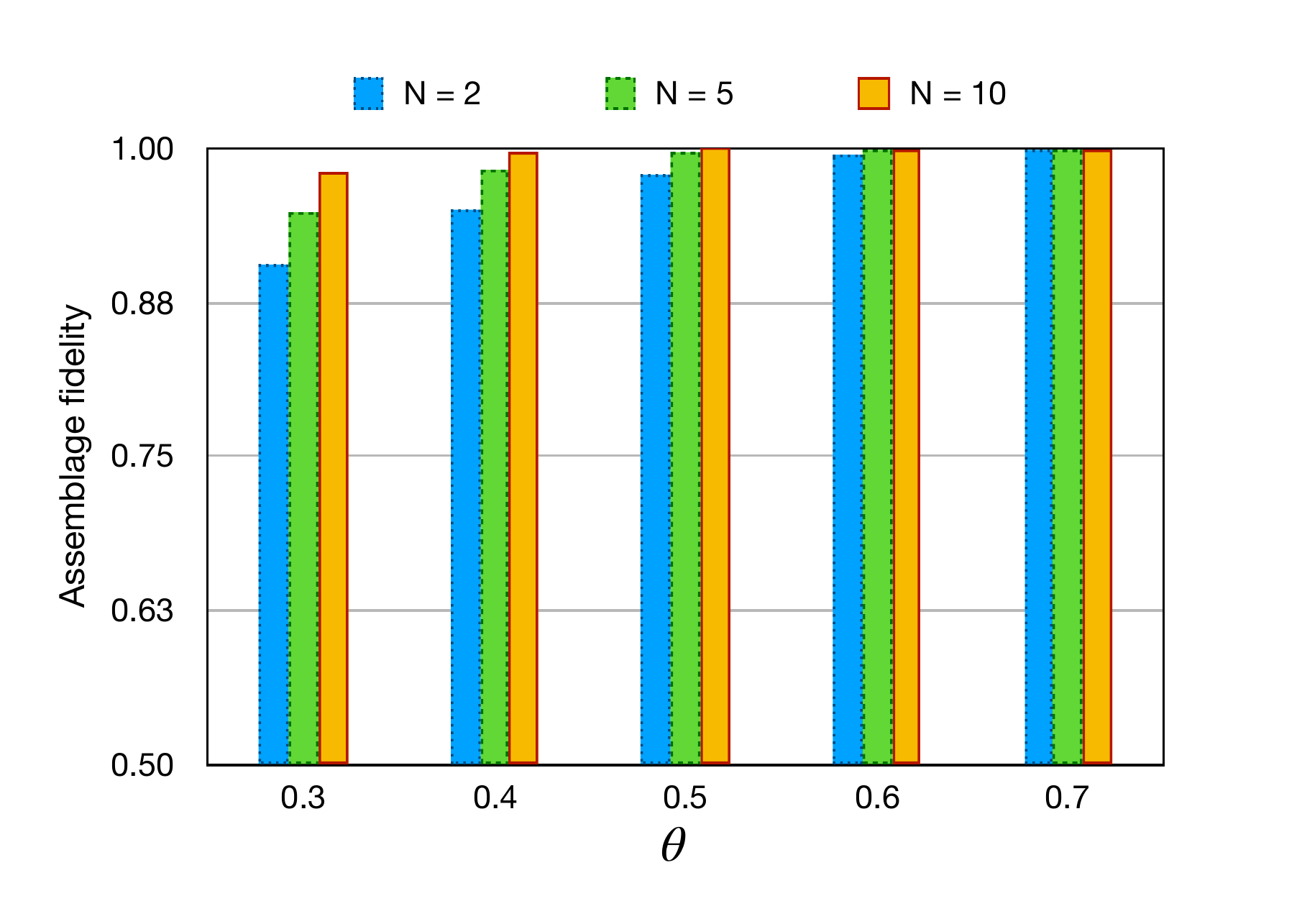}}
\caption{\footnotesize (Coloronline)  Variation of the assemblage fidelity versus the state parameter $\theta$ (radian) in our 1SDI distillation protocol for GGHZ states. The quantity in the vertical axis has no units.}  
\label{fig1}
\end{figure}

 The assemblage fidelity between the assemblage (\ref{GGHZdistill}) and our target assemblage (\ref{GHZassemblage}) corresponding to GHZ state  is given by, 
\begin{align}
    &\mathcal{F}_{A} \Big( \{\sigma_{a|x}^{\text{dist}^{\text{GHZ}}} \}_{a,x}, \{\sigma_{a|x}^{BC^{\text{GHZ}}}\}_{a,x} \Big) \nonumber \\
     &=\sqrt{1 - \frac{1}{2} (1 - \sin 2 \theta ) (\cos 2 \theta )^{N-1}}.
     \label{assfidghz1}
\end{align}
The detailed derivation of the above expression (\ref{assfidghz1}) is presented in Appendix \ref{app3}.

Hence, we can present the following theorem:
\begin{thm}
In the regime of finite $N \geq 2$ copies of initial assemblage given by Eq.(\ref{GGHZassemblage}) of GGHZ state in 1SDI scenario, an assemblage can be obtained on average which is close to the target assemblage given in Eq.(\ref{GHZassemblage}) in 1SDI scenario with assemblage fidelity $\sqrt{1 - \frac{1}{2} (1 - \sin 2 \theta ) (\cos 2 \theta )^{N-1}}$.
\end{thm}

Let us now consider how efficient our distillation protocol is in this case when only a few copies of the initial assemblages are taken. Consider, for example, that the distillation protocol is performed on $N=7$ copies of the initial assemblages produced from the GGHZ state (\ref{GGHZ}) with $\theta = 0.25$. In this case, the assemblage fidelity $\mathcal{F}_{A} \Big( \{\sigma_{a|x}^{\text{dist}^{\text{GHZ}}} \}_{a,x}, \{\sigma_{a|x}^{BC^{\text{GHZ}}}\}_{a,x} \Big)$ = $0.939$. If one starts with $N=5$ copies of the initial assemblages produced from the GGHZ state (\ref{GGHZ}) with $\theta = 0.50$, then the above assemblage fidelity  turns out to be $0.997$. In Fig. \ref{fig1}, we plot the above assemblage fidelity (\ref{assfidghz1}) for different values of the state parameter $\theta$ of the GGHZ state (\ref{GGHZ}) taking a few number  of copies of the initial assemblage  given in Eq.(\ref{GGHZassemblage}). 

\begin{figure}[t!]
\resizebox{7.5cm}{6.5cm}{\includegraphics{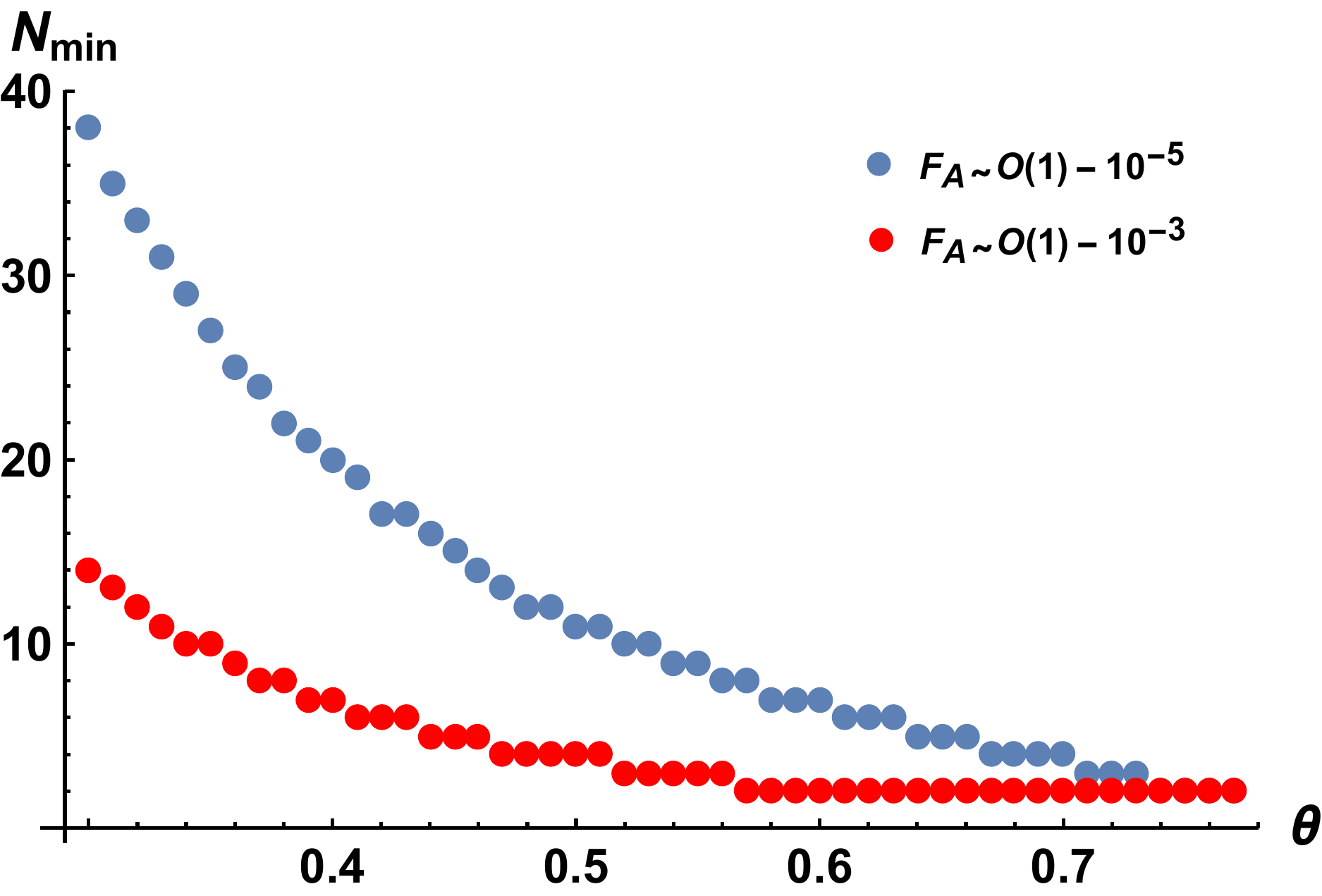}}
\caption{\footnotesize (Coloronline)  Variation of the minimum number of copies ($N_{\text{min}}$)  with the state parameter $\theta$ (radian) required to achieve near perfect assemblage fidelity in our 1SDI scenario  for GGHZ states. Blue dotted upper curve is for $\mathcal{F}_{A} \sim O(1) - 10^{-5}$ and red dotted bottom curve is for $\mathcal{F}_{A} \sim O(1) - 10^{-3}$. The quantity in the vertical axis has no units. }   
\label{N_min_GGHZ}
\end{figure}

We can further evaluate the minimum number of copies ($N_{\text{min}}$) of the initial assemblages required to achieve an approximately perfect assemblage fidelity  (in particular, $\mathcal{F}_{A} \sim O(1) - 10^{-5}$). For example, consider the initial assemblages produced from the GGHZ state (\ref{GGHZ}) with $\theta = 0.25$.  In this case, $N_{\text{min}} = 56$. On the other hand, when $\theta = 0.50$, then $N_{\text{min}} =11 $. In Fig. \ref{N_min_GGHZ}, $N_{\text{min}}$ is plotted for different values of the state parameter $\theta$ of the GGHZ state (\ref{GGHZ}). From this figure, it is evident that $N_{\text{min}}$ decreases with increasing values of $\theta$. These examples signify the efficacy of our distillation protocol for realistic scenarios.

\vspace{1cm}
$\bullet$ {\bf 2SDI scenario:}  Consider a tripartite steering scenario where Alice, Bob and Charlie  initially share $N \geq 2$ copies of the assemblage $\{\sigma_{a,b|x,y}^{C^{\text{GGHZ}}}\}_{a,b,x,y}$ obtained from the three qubit GGHZ state (\ref{GGHZ}) with $0 < \theta < \frac{\pi}{4}$ through the measurements of observables $X_0= \sigma_x$, $X_1= \sigma_y$ and $X_2=\sigma_z$ by Alice and $Y_0= \sigma_x$, $Y_1= \sigma_y$ and $Y_2=\sigma_z$ by Bob. The components of these initial assemblages are given in the Appendix \ref{2SAGGHZ}.

When these assemblages are subjected to the measurements $ \sigma_x,  \sigma_y,  \sigma_z$ by Charlie as specified in the inequality (\ref{GHZ2}), the produced correlations cannot give the optimum quantum violation of the inequality (\ref{GHZ2}). Hence,   these initial assemblages $\{\sigma_{a,b|x,y}^{C^{\text{GGHZ}}}\}_{a,b,x,y}$ are partially genuine steerable assemblages in 2SDI scenario. The left hand side of the inequality (\ref{GHZ2}) for the GGHZ state is a monotonic function of $\theta$ for  $0 < \theta < \frac{\pi}{4}$.  The inequality (\ref{GHZ2}) is violated  by the GGHZ states for $\theta \in ( 0.22, \frac{\pi}{4})$ and, hence, it ensures 2SDI genuine steering of the initial assemblage in this range of $\theta$. 

Next, consider the assemblage $\{ \sigma_{a,b|x,y}^{C^{\text{GHZ}}} \}_{a,b,x,y}$, produced from the GHZ state when Alice performs measurements of the observables $X_0= \sigma_x$, $X_1= \sigma_y$ and $X_2=\sigma_z$ and Bob performs measurements of the observables $Y_0= \sigma_x$, $Y_1= \sigma_y$ and $Y_2=\sigma_z$, with components given in the Appendix \ref{2SAGHZ}.  This assemblage is derived from the GHZ state by applying orthogonal Von Neumann measurements of rank-1 by the untrusted parties. Also, when this assemblage is subjected to the measurements $ \sigma_x, \sigma_y$ and $ \sigma_z$ by Charlie, the produced correlations give the optimum quantum violation of the inequality (\ref{GHZ2}). Hence,  this assemblage $\{ \sigma_{a,b|x,y}^{C^{\text{GHZ}}} \}_{a,b,x,y}$ is a perfectly genuine steerable assemblage of the generalized GHZ states in 2SDI scenario. In this case, this $\{ \sigma_{a,b|x,y}^{C^{\text{GHZ}}} \}_{a,b,x,y}$  is our target assemblage of the distillation protocol.

As discussed earlier, on the $u^{\text{th}}$ copy of the initial assemblage $\{\sigma_{a,b|x,y}^{C^{\text{GGHZ}}}\}_{a,b,x,y}$ (for all $u \in \{1, 2, \cdots, N-1\}$), Charlie performs quantum measurements (POVM) with the following measurement operators,
\begin{equation}
	K_0^{C} = \left(
\begin{array}{cc}
 \tan \theta  & 0 \\
 0 & 1 \\
\end{array}
\right), \quad K_{1}^{C} =  \left(
\begin{array}{cc}
 \sqrt{1 - \tan^2 \theta} & 0 \\
 0 & 0 \\
\end{array}
\right),
\label{GHZ2Ssteer2}
\end{equation}
 and sends the string of outputs to Alice and Bob.  As mentioned earlier, all parties will not discard those copies (post Charlie's POVM) for which Charlie has obtained the outcome $0$. Hence, the probability with which each copy is not discarded is given by, $\tilde{P}^{\text{GHZ}_2}_{\text{ND}} = p(o_C^u = 0) = \text{Tr} \Big[  K_{o^u_C}^C \, \rho_{\text{GGHZ}}^{C} \, K_{o^u_C}^{C^{\dagger}} \Big]$ = $2\sin^2\theta$. Here, $\rho_{\text{GGHZ}}^C = \text{Tr}_{AB}\Big[ |\psi_{\text{GGHZ}} \rangle \langle \psi_{\text{GGHZ}}|\Big]$. The modified assemblage after this POVM by Charlie, when he gets the outcome $0$, is nothing but our target assemblage  $\{ \sigma_{a,b|x,y}^{C^{\text{GHZ}}} \}_{a,b,x,y}$. This can be checked using Eq.(\ref{finalass3}).
 
On the other hand, when Charlie gets the outcome $1$ contingent upon performing the POVM on the $u^{\text{th}}$ copy for all $u \in \{ 1, 2, ..., N-1\}$, then all copies except the $N^{\text{th}}$ copy are discarded. Hence, in this case the output of the distillation protocol is the $N^{\text{th}}$ copy of the initial assemblage. Now, since the local quantum measurements are performed independently on each of the $N-1$ copies, the probability with which the distillation protocol fails is given by,
\begin{equation}
    P^{\text{GHZ}_2}_{\text{fail}} = (1 - \tilde{P}^{\text{GHZ}_2}_{\text{ND}})^{N-1} = (1 - 2\sin^2\theta)^{N-1}
\end{equation}
This means, the parties manage to either keep at least one successfully distilled target assemblage  $\{ \sigma_{a,b|x,y}^{C^{\text{GHZ}}} \}_{a,b,x,y}$ with probability $P^{\text{GHZ}_2}_{\text{success}} = 1 - (1-2\sin^2\theta)^{N-1}$, or the last copy of the initial assemblage $\{\sigma_{a,b|x,y}^{C^{\text{GGHZ}}}\}_{a,b,x,y}$  with the probability $P^{\text{GHZ}_2}_{\text{fail}}$.
In the asymptotic limit as the number of copies tends to infinity, we get $\underset{N \rightarrow \infty}{\text{lim}} P^{\text{GHZ}_2}_{\text{success}} \rightarrow 1$ and, hence, the protocol ensures distillation of at least one copy of target assemblage as $N$ tends to infinity.

 Let $\{\sigma_{a,b|x,y}^{\text{dist}^{\text{GHZ}}} \}_{a,b,x,y}$ denote the assemblage  obtained, on average, after the distillation protocol in the regime of finite $N$ copies with $N \geq 2$. Then the assemblage fidelity between the assemblage $\{\sigma_{a,b|x,y}^{\text{dist}^{\text{GHZ}}}\}_{a,b,x,y}$ and the target assemblage $\{\sigma_{a,b|x,y}^{C^{\text{GHZ}}}\}_{a,b,x,y}$ is given by,
\begin{align}
    &\mathcal{F}_A \Big(\{\sigma_{a,b|x,y}^{\text{dist}^{\text{GHZ}}}\}_{a,b,x,y}, \{\sigma_{a,b|x,y}^{C^{\text{GHZ}}}\}_{a,b,x,y} \Big)  \nonumber \\
    &=\sqrt{1 - \frac{1}{2} (1 - \sin 2 \theta ) (\cos 2 \theta )^{N-1}}.
     \label{assfidghz2sdi}
\end{align}
The details of this calculation are provided in Appendix \ref{app6}. Hence, we can present the following theorem:
\begin{thm}
In the regime of finite $N \geq 2$ copies of initial assemblage $\{\sigma_{a,b|x,y}^{C^{\text{GGHZ}}}\}_{a,b,x,y}$ mentioned in Appendix \ref{2SAGGHZ} of GGHZ state in 2SDI scenario, an assemblage can be obtained on average which is close to the target assemblage $\{ \sigma_{a,b|x,y}^{C^{\text{GHZ}}} \}_{a,b,x,y}$ in 2SDI scenario mentioned in Appendix \ref{2SAGHZ} with assemblage fidelity $\sqrt{1 - \frac{1}{2} (1 - \sin 2 \theta ) (\cos 2 \theta )^{N-1}}$.
\end{thm}

Since the above expression of assemblage fidelity is exactly similar to the expression (\ref{assfidghz1}), the characteristics of the above assemblage fidelity (\ref{assfidghz2sdi}) are similar to that of (\ref{assfidghz1}) in 1SDI scenario.

\subsection{Distillation of genuine steerable assemblage of W state}\label{subsection2}

$\bullet$ {\bf 1SDI scenario:} We consider the following generalized pure three-qubit W states,

\begin{equation}
	|\psi_\text{GW} \rangle = c_0\ket{001} + c_1\ket{010} + \sqrt{1-c_0^2-c_1^2} \ket{100} ,
\label{wclass}
\end{equation}
with $c_0$, $c_1$ being real and $0 < c_0 \leq \dfrac{1}{\sqrt{3}}$, $0 < c_1 < \dfrac{1}{\sqrt{3}}$. We, therefore, have $\dfrac{1}{\sqrt{3}} < \sqrt{1-c_0^2-c_1^2} < 1$. 
 Let $\mathcal{C}_{0,1}$ = $\{c_0, c_1|0 < c_0 \leq \frac{1}{\sqrt{3}}, 0 < c_1 < \frac{1}{\sqrt{3}} \}$. Note that the above states do not represent the most general W-class pure states, rather a subset of them \cite{ALS1}.

  Consider a tripartite steering scenario where Alice, Bob and Charlie initially  share $N \geq 2$ copies of the assemblage $\{\sigma_{a|x}^{BC^{\text{GW}}}\}_{a,x}$ obtained from the three-qubit  generalized W states given by Eq.(\ref{wclass}) through the measurements of the observables $X_0=\sigma_x$, $X_1=\sigma_y$ and $X_2=\sigma_z$ by Alice. The components of the assemblages are given in the Appendix \ref{W_assemb_1SDI}.

When these assemblages are subjected to the measurements $ \sigma_x,  \sigma_y,  \sigma_z$ by Bob and  Charlie as specified in the inequality (\ref{W1}), they cannot give optimum violation of the inequality (\ref{W1}). Hence,  these initial assemblages are partially genuine steerable assemblages in 1SDI scenario.

Note that the inequality (\ref{W1}) is violated by the generalized W states (\ref{wclass}) for specific ranges of $c_0$ and $c_1$  and, hence, it ensures 1SDI genuine steering in that range only. We will denote the set of values of $c_0$ and $c_1$, for which the inequality (\ref{W1}) is violated by the generalized W states (\ref{wclass}) with $c_0$, $c_1$ $\in$ $\mathcal{C}_{0,1}$, by the notation $\mathcal{C}^{\text{GW}_1}_{0,1}$. Here, $\mathcal{C}^{\text{GW}_1}_{0,1}$ $\subset$ $\mathcal{C}_{0,1}$. Henceforth, we will only consider $c_0$, $c_1$ $\in$ $\mathcal{C}^{\text{GW}_1}_{0,1}$. For other values of $c_0$, $c_1$, the initial assemblage may or may not be genuine steerable in 1SDI scenario. But we will not consider those values.

Next, consider the assemblage $\{\sigma_{a|x}^{BC^{\text{W}}}\}_{a,x}$, produced from the W state- $\dfrac{1}{\sqrt{3}}(\ket{001} + \ket{010} + \ket{100})$ when Alice performs the orthogonal von Neumann measurements of the observables $X_0=\sigma_x$, $X_1= \sigma_y$ and $X_2=\sigma_z$,  with components given in the Appendix \ref{W_assemb_1SDI}.

When this assemblage is subjected to the  measurements of $\sigma_x,  \sigma_y, \sigma_z$ by Bob and Charlie,  the produced correlations give the optimum quantum violation of the inequality (\ref{W1}). Hence,  this assemblage is taken as the target assemblage for this case.

Our distillation protocol proceeds as follows. At first, Bob and Charlie perform quantum measurements (POVM) with the following measurement operators on the $u^{\text{th}}$ copy of the initial assemblage $\{\sigma_{a|x}^{BC^{\text{GW}}}\}_{a,x}$ (for all $u \in \{1, 2, \cdots, N-1\}$):
\begin{eqnarray}
	K_{0}^{B} &=& \left(
\begin{array}{cc}
 \frac{c_1}{\sqrt{1-c_0^2-c_1^2}} & 0 \\
 0 & 1 \\
\end{array}
\right), \quad K_{1}^{B} =  \left(
\begin{array}{cc}
\sqrt{\frac{1-c_0^2-2c_1^2}{1-c_0^2-c_1^2}} & 0 \\
 0 & 0 \\
\end{array}
\right), \nonumber  \\
K_{0}^{C} &=& \left(
\begin{array}{cc}
 \frac{c_0}{\sqrt{1-c_0^2-c_1^2}} & 0 \\
 0 & 1 \\
\end{array}
\right), \quad K_{1}^{C} =  \left(
\begin{array}{cc}
 \sqrt{\frac{1-2c_0^2-c_1^2}{1-c_0^2-c_1^2}} & 0 \\
 0 & 0 \\
\end{array}
\right) .
\label{Wsteer1}
\end{eqnarray}
As stated earlier, all parties do not discard those copies (after Bob and Charlie's POVMs) for which both Bob and Charlie have obtained the outcome $0$. Hence, the probability with which each copy is not discarded is given by, $\tilde{P}^{\text{W}_1}_{\text{ND}} = p( o_B^u = 0, o_C^u = 0)$ = $\text{Tr} \Big[ \Big(K_{0}^{B} \otimes K_{0}^{C} \Big) \rho_{\text{GW}}^{BC} \Big(K_{0}^{B^{\dagger}} \otimes K_{0}^{C^{\dagger}} \Big) \Big]$ = $\dfrac{3 c_0^2 c_1^2}{ 1-c_0^2-c_1^2}$. Here, $\rho_{\text{GW}}^{BC} = \text{Tr}_{A}\Big[ |\psi_{\text{GW}} \rangle \langle \psi_{\text{GW}}|\Big]$. It can be checked that $0 < \dfrac{3 c_0^2 c_1^2}{ 1-c_0^2-c_1^2} < 1$ for all $c_0$, $c_1$ $\in$ $\mathcal{C}^{\text{GW}_1}_{0,1}$. The updated assemblage after these POVMs by Bob and Charlie, when both of them get the outcome $0$, is nothing but our target assemblage $\{\sigma_{a|x}^{BC^{\text{W}}}\}_{a,x}$. This can be checked using Eq.(\ref{finalass1}).

In case, if either Bob or Charlie or both of them get the outcome $1$ contingent upon performing the POVMs on the $u^{th}$ copy for all $u \in \{1, 2, ..., N-1\}$, then all copies, except the $N^{th}$ copy, are discarded and the distillation protocol fails. The probability with which it fails is given by,

\begin{equation}
    P^{\text{W}_1}_{\text{fail}} = \Big(1-\tilde{P}^{\text{W}_1}_{\text{ND}}\Big)^{N-1} = \Bigg(1-\dfrac{3 c_0^2 c_1^2}{ 1-c_0^2-c_1^2}\Bigg)^{N-1},
    \label{failurew1sdi}
\end{equation}
where $0 < P^{\text{W}_1}_{\text{fail}} < 1$ for all $c_0$, $c_1$ $\in$ $\mathcal{C}^{\text{GW}_1}_{0,1}$. 

Hence, the parties manage to either keep at least one successfully distilled target assemblage $\{\sigma_{a|x}^{BC^{\text{W}}}\}_{a,x}$ with the probability 
\begin{equation}
   P^{\text{W}_1}_{\text{success}} = \Bigg[1-\Bigg(1-\dfrac{3 c_0^2 c_1^2}{ 1-c_0^2-c_1^2}\Bigg)^{N-1}\Bigg], 
\end{equation} 
or the last copy of the initial assemblage $\{\sigma_{a|x}^{BC^{\text{GW}}}\}_{a,x}$ with the probability $P^{\text{W}_1}_{\text{fail}}$.

In the asymptotic limit as $N$ tends to infinity, we get $\underset{N \rightarrow \infty}{\text{lim}} P^{\text{W}_1}_{\text{success}} \rightarrow 1$. This is because $0 < \dfrac{3 c_0^2 c_1^2}{ 1-c_0^2-c_1^2} < 1$ for all $0 < c_0 \leq \dfrac{1}{\sqrt{3}}$, $0 < c_1 < \dfrac{1}{\sqrt{3}}$ (and, hence, for all $c_0$, $c_1$ $\in$ $\mathcal{C}^{\text{GW}_1}_{0,1}$). Therefore, the protocol ensures distillation of at least one copy of target assemblage $\{\sigma_{a|x}^{BC^{\text{W}}}\}_{a,x}$ with certainty.

\begin{figure*}
   \begin{subfigure}{6cm}
   \centering\includegraphics[width=6cm]{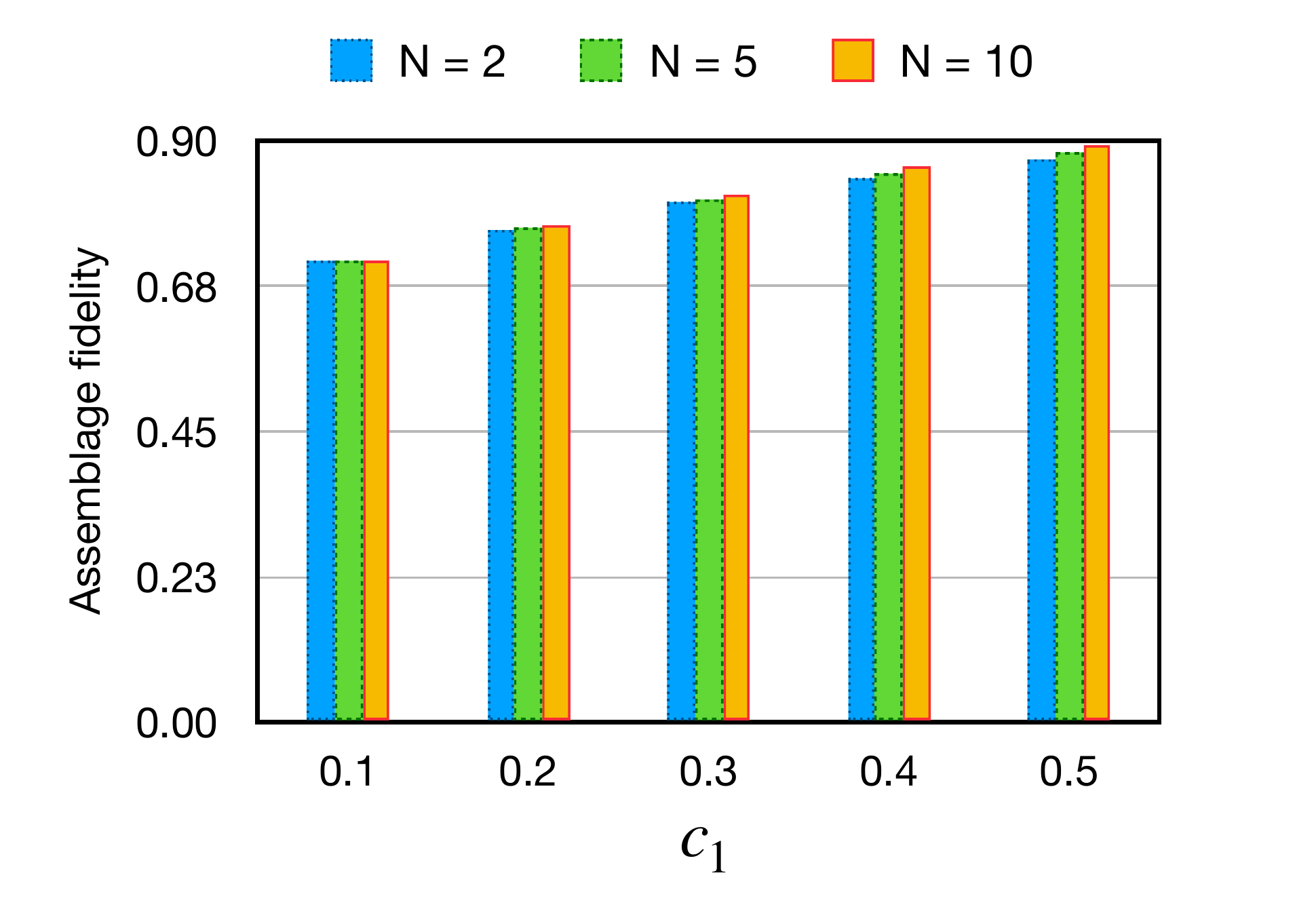}
    \caption{\footnotesize $c_0=0.15$.}
    \label{AFc015}
  \end{subfigure}%
  \begin{subfigure}{6cm}
    \centering\includegraphics[width=6cm]{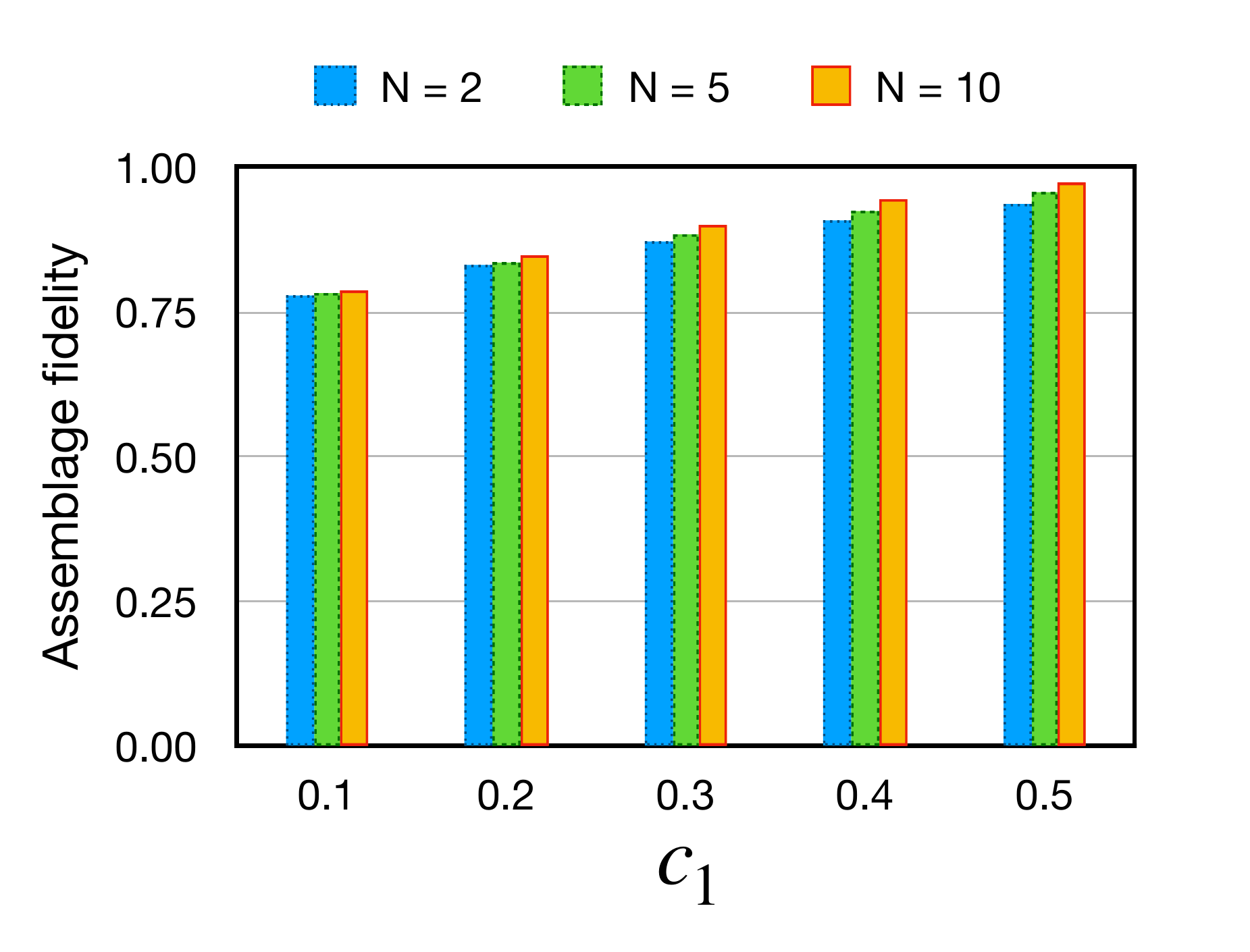}
    \caption{\footnotesize $c_0=0.30$.}
    \label{AFc30}
  \end{subfigure}%
   \begin{subfigure}{6cm}
    \centering\includegraphics[width=6cm]{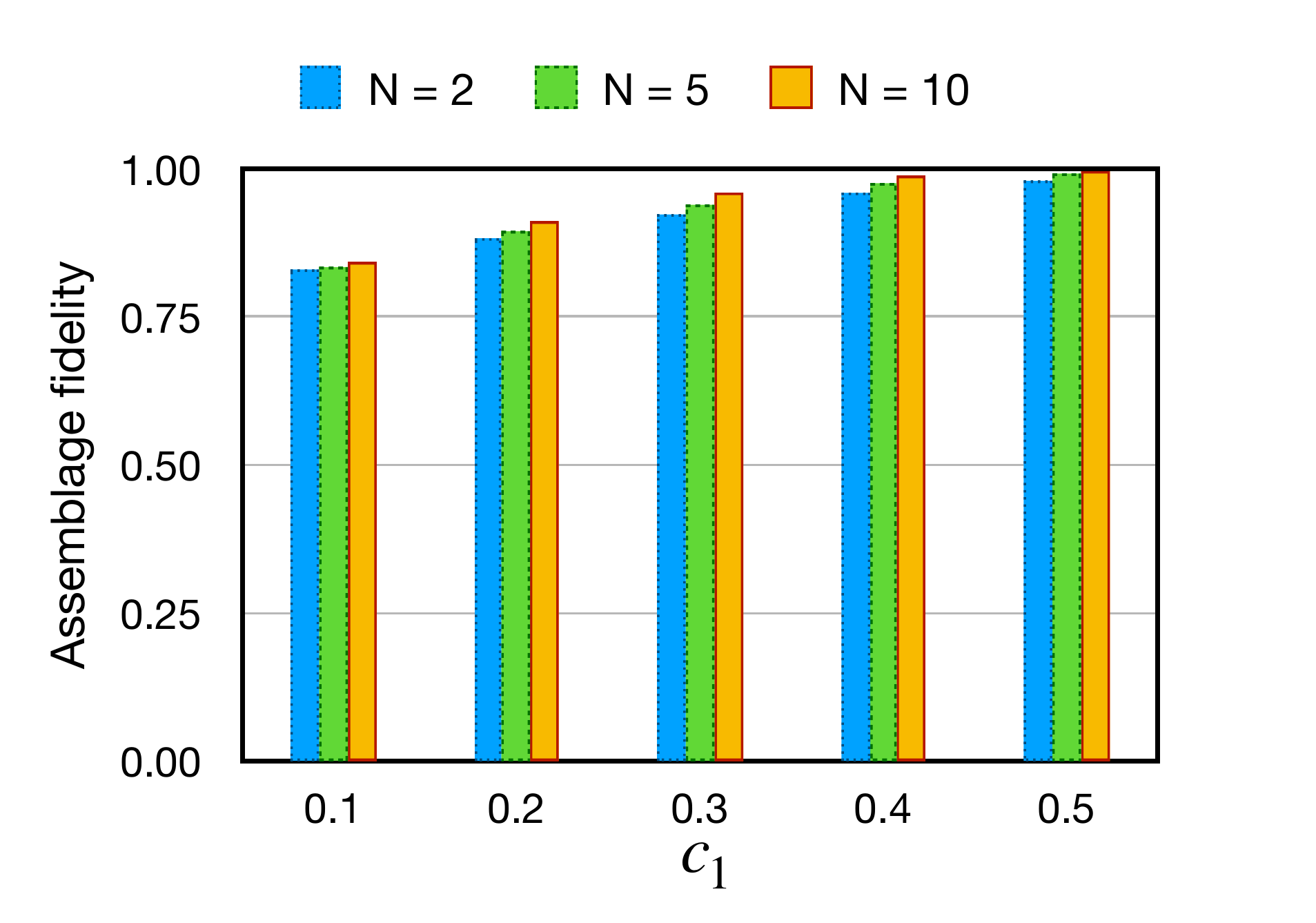}
    \caption{\footnotesize $c_0=0.45$.}
    \label{AFc45}
  \end{subfigure}%
\caption{\footnotesize (Coloronline) Variations of the assemblage fidelity of our distilation protocol for W states in the 1SDI scenario.  Variation is shown with respect
to $c_1$ with $c_0 = 0.15$ in Fig. \ref{AFc015}, $c_0 = 0.30$ in Fig. \ref{AFc30},
and $c_0 = 0.45$ in Fig. \ref{AFc45}. The quantities in the horizontal and vertical axes are numbers and have no unit.}
\label{AF_w}
\end{figure*}

 Let $\{\sigma_{a|x}^{\text{dist}^{\text{W}}}\}_{a,x}$ denote the assemblage obtained, on average, after the distillation protocol in the regime of finite N copies with $N \geq 2$. The assemblage $\{\sigma_{a|x}^{\text{dist}^{\text{W}}}\}_{a,x}$ can be written as a convex combination of the initial assemblage $\{\sigma_{a|x}^{BC^{\text{GW}}}\}_{a,x}$ and the target assemblage of W state $\{\sigma_{a|x}^{BC^{\text{W}}}\}_{a,x}$.

The assemblage fidelity between the assemblage $\{\sigma_{a|x}^{\text{dist}^{\text{W}}}\}_{a,x}$ and our target assemblage $\{\sigma_{a|x}^{BC^{\text{W}}}\}_{a,x}$ corresponding to W state  is given by,
\begin{align}
    &\mathcal{F}_{A} \Big( \{\sigma_{a|x}^{\text{dist}^{\text{W}}} \}_{a,x}, \{\sigma_{a|x}^{BC^{\text{W}}}\}_{a,x} \Big) \nonumber \\
    &= \frac{1}{\sqrt{3}} \sqrt{ X^{N-1}  \Big[ (c_0 + c_1 + \sqrt{1- c_0^2 - c_1^2})^2 -3 \Big] + 3},
     \label{assfidw1app}
\end{align}
where $X$ = $1-\dfrac{3 c_0^2 c_1^2}{ 1-c_0^2-c_1^2}$. The details of this derivation are provided in Appendix \ref{appnew}.

\begin{figure*}
   \begin{subfigure}{6cm}
   \centering\includegraphics[width=6cm]{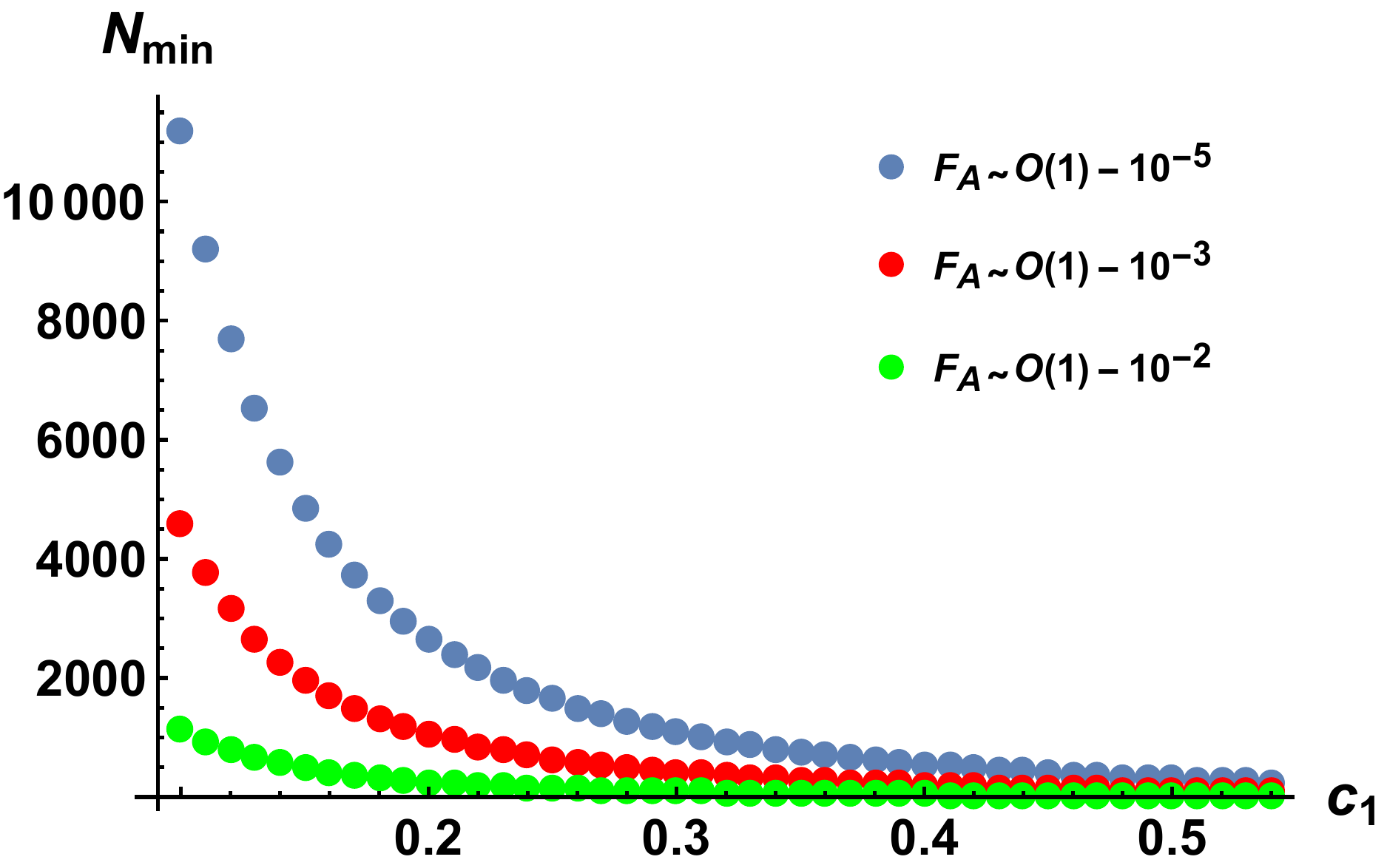}
    \caption{\footnotesize $c_0=0.15$.}
    \label{c015}
  \end{subfigure}%
  \begin{subfigure}{6cm}
    \centering\includegraphics[width=6cm]{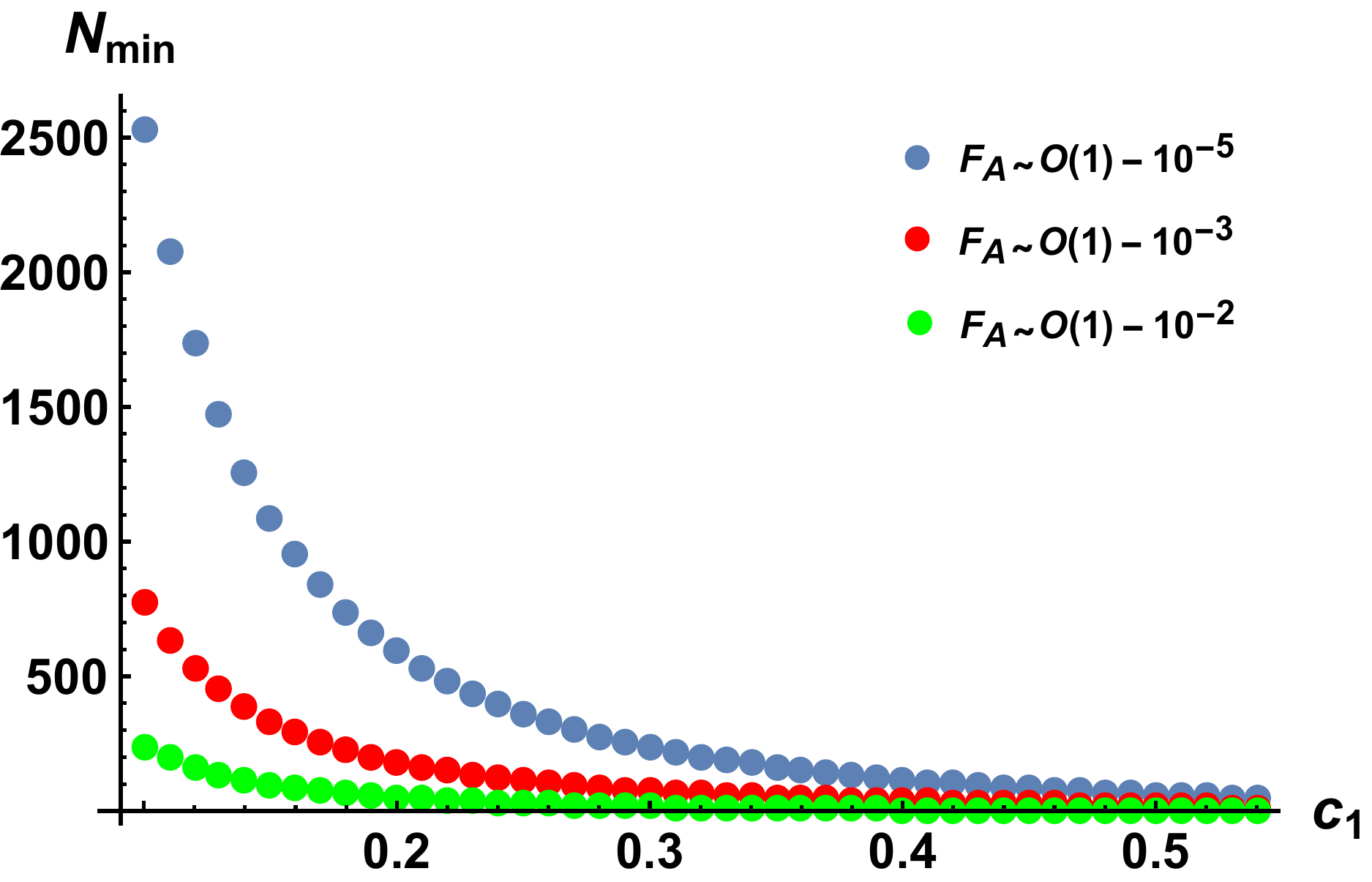}
    \caption{\footnotesize $c_0=0.30$.}
    \label{c30}
  \end{subfigure}%
   \begin{subfigure}{6cm}
    \centering\includegraphics[width=6cm]{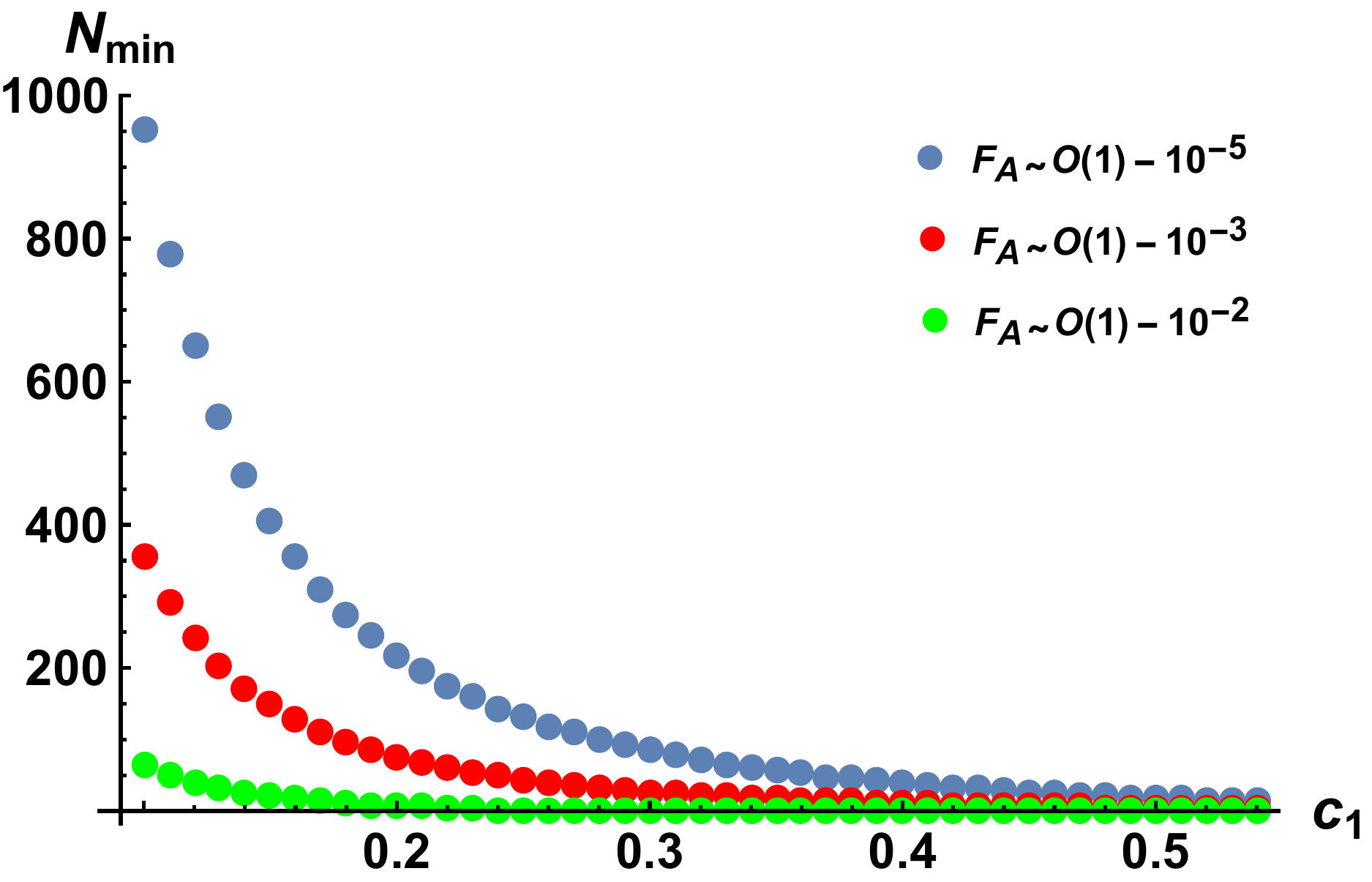}
    \caption{\footnotesize $c_0=0.45$.}
    \label{c45}
  \end{subfigure}%
\caption{\footnotesize (Coloronline) Variation of the minimum number of copies $(N_{min})$ required to achieve near perfect assemblage fidelity in our 1SDI scenario for W states.
 The variation is shown with respect to $c_1$ with $c_0 = 0.15$ in Fig. \ref{AFc015}, $c_0 = 0.30$ in Fig. \ref{AFc30},
and $c_0 = 0.45$ in Fig. \ref{AFc45} for different accuracy of the assemblage fidelity. The quantities in the horizontal and vertical axes are numbers and have no unit.}
\label{N_min_w}
\end{figure*}

We can, therefore, state the following theorem
\begin{thm}
In the regime of finite $N \geq 2$ copies of initial assemblage $\{\sigma_{a|x}^{BC^{\text{GW}}}\}_{a,x}$ of generalized W pure states in 1SDI scenario, an assemblage can
be obtained on average which is close to the target assemblage $\{\sigma_{a|x}^{BC^{\text{W}}}\}_{a,x}$ derived from the W state in 1SDI scenario with assemblage fidelity $\frac{1}{\sqrt{3}} \sqrt{X^{N-1}  \Big[ (c_0 + c_1 + \sqrt{1- c_0^2 - c_1^2})^2 -3 \Big] + 3}$, where $X$ = $1-\dfrac{3 c_0^2 c_1^2}{ 1-c_0^2-c_1^2}$.
\end{thm}

In Fig. \ref{AF_w}, we  demonstrate the variation of the above assemblage fidelity (\ref{assfidw1app}) for different values of the state parameters $c_0$, $c_1$ of the  generalized W states (\ref{wclass}) and for different number $N \geq 2$ of copies of the initial assemblage  $\{\sigma_{a|x}^{BC^{\text{GW}}}\}_{a,x}$.

Let us now evaluate the  minimum number of copies ($N_{\text{min}}$) of the initial assemblages required to achieve an approximately perfect assemblage fidelity  (in particular, $\mathcal{F}_{A} \sim O(1) - 10^{-5}$).   
For example, the initial assemblages produced from the generalized W pure states (\ref{wclass}) with  $c_0 = 0.15$ and $c_1 = 0.25$, leads to $N_{\text{min}} = 1792$. On the other hand, when  $c_0 = 0.45$, $c_1 = 0.50$, we get $N_{\text{min}} = 18$.
In Fig. \ref{N_min_w}, we  plot $N_{\text{min}}$ for different values of the state parameters $c_0$, $c_1$.
 It can be seen that our distillation protocol is efficient  in realistic scenarios where one cannot have infinitely many copies of the initial assemblages.

\vspace{1cm}

$\bullet$ {\bf 2SDI scenario:} Here, we consider the following one parameter family of pure three-qubit  generalized W states, 
\begin{equation}
	\ket{\widetilde{\psi}_{\text{GW}}} = d_0 \ket{001} + \sqrt{\frac{1-d_0^2}{2}} \ket{010} + \sqrt{\frac{1-d_0^2}{2}} \ket{100},
	\label{owclass}
\end{equation}
with $d_0$ being real and $0<d_0<\dfrac{1}{\sqrt{3}}$. Hence, we have $\dfrac{1}{\sqrt{3}} < \sqrt{\dfrac{1-d_0^2}{2}} < \dfrac{1}{\sqrt{2}}$.  It is evident that the above states do not represent the most general W-class pure states, rather a subset of them \cite{ALS1}.

Consider a tripartite steering scenario where Alice, Bob and Charlie  initially share $N \geq 2$ copies of the assemblage $\{\sigma_{a,b|x,y}^{C^{\text{GW}}}\}_{a,b,x,y}$ obtained from the one parameter pure three-qubit generalized W states given by Eq.(\ref{owclass}) with $0 < d_0 < \dfrac{1}{\sqrt{3}}$ through the measurements of the observables $X_0=\sigma_x$, $X_1=\sigma_y$, $X_2=\sigma_z$ by Alice and $Y_0=\sigma_x$, $Y_1=\sigma_y$, $Y_2=\sigma_z$ by Bob. The components of the assemblages $\{\sigma_{a,b|x,y}^{C^{\text{GW}}} \}_{a,b,x,y}$ on Charlie's side are  explicitly mentioned in the Appendix \ref{OW2Sassemblage}.

When these assemblages are subjected to the measurements $ \sigma_x,  \sigma_y,  \sigma_z$ by Charlie as specified in the inequality (\ref{W2}), then the produced correlations cannot give the optimum violation of the inequality (\ref{W2}). Hence,  these assemblages are taken to be partially genuine steerable assemblage in 2SDI scenario.

Note that the inequality (\ref{W2}) is violated by the one parameter generalized W states (\ref{owclass}) when $d_0$ $\in$ $\Bigg( \dfrac{3}{25}, \dfrac{1}{\sqrt{3}} \Bigg)$ and, hence, it ensures 2SDI genuine steering in this range. Henceforth, we will only consider $d_0$ $\in$ $\Bigg( \dfrac{3}{25}, \dfrac{1}{\sqrt{3}} \Bigg)$. For other values of $d_0$, the initial assemblage may or may not be genuine steerable in our 2SDI scenario.

Consider the assemblage $\{\sigma_{a,b|x,y}^{C^{\text{W}}} \}$, produced from W state- $\frac{1}{\sqrt{3}}(\ket{001}+\ket{010}+\ket{100})$ when Alice performs the  orthogonal von Neumann measurements of the observables $X_0=\sigma_x$,  $X_1=\sigma_y$,  $X_2=\sigma_z$ and Bob performs the  orthogonal von Neumann measurements of the observables $Y_0=\sigma_x$,  $Y_1=\sigma_y$,  $Y_2=\sigma_z$, with components given in Appendix \ref{W2Sassemblage}. When this assemblage is subjected to the  measurements of $ \sigma_x,  \sigma_y,  \sigma_z$ by Charlie,  the produced correlations give the optimum quantum violation of the inequality (\ref{W2}). Hence,  this assemblage is considered as the target assemblage for this case.
    
Our distillation protocol for this case proceeds as follows. At first, Charlie performs quantum measurements (POVM) with the following measurement operators on the $u^{\text{th}}$ copy of the initial assemblage $\{\sigma_{a,b|x,y}^{C^{\text{OW}}} \}_{a,b,x,y}$ (for all $u \in \{1, 2, . . . , N-1\}$):
\begin{equation}
	K_0^C = \left(
\begin{array}{cc}
 \frac{\sqrt{2} d_0}{\sqrt{1-d_0^2}} & 0 \\
 0 & 1 \\
\end{array}
\right), \quad K_1^C = \left(
\begin{array}{cc}
 \sqrt{1- \frac{2 d_0^2}{1-d_0^2}} & 0 \\
 0 & 0 \\
\end{array}
\right)
\label{POVM2SW}
\end{equation}
In this case also all parties do not discard those copies (after Charlie's POVM measurements) for which he has obtained the outcome $0$. Hence, the probability with which each copy is not discarded is given by, $\tilde{P}_{\text{ND}}^{W_2} = p(o_C^u = 0) = \text{Tr}\big[ \big(K_0^C\big)\rho_{\text{GW}}^C\big(K_0^{C^{\dagger}}\big)\big] = 3d_0^2$ $< 1$ for all $d_0$ $\in$ $\Bigg( \dfrac{3}{25}, \dfrac{1}{\sqrt{3}} \Bigg)$. Here, $\rho_{\text{GW}}^C = \text{Tr}_{\text{AB}} \big[\ket{\widetilde{\psi}_{\text{GW}}}\bra{\widetilde{\psi}_{\text{GW}}}\big]$. The updated assemblage after this POVM by Charlie, when he gets the outcome $0$, is nothing but our target assemblage $\{\sigma_{a,b|x,y}^{C^{\text{W}}}\}_{a,b,x,y}$. 

When Charlie gets the outcome $1$ after performing the POVM (\ref{POVM2SW}) on the $u^{\text{th}}$ copy for all $u \in \{1, 2, . . . , N-1\}$,  all copies except the $N^{\text{th}}$ copy are discarded. Hence, in this case the output of the distillation protocol is the $N^{\text{th}}$ copy of the initial assemblage. Now, since the local quantum measurements are performed independently on each of the $N-1$ copies, the probability with which the distillation protocol fails is given by,
\begin{equation}
	P_{\text{fail}}^{\text{W}_2} = (1- \tilde{P}_{\text{ND}}^{\text{W}_2})^{N-1} = (1-3d_0^2)^{N-1}
	\label{PfailW2}
\end{equation} 
This means that the parties manage to either keep at least one successfully distilled target assemblage $\{\sigma_{a,b|x,y}^{C^{\text{W}}}\}_{a,b,x,y}$ with the probability, 
\begin{align}
P_{\text{success}}^{\text{W}_2} = 1-(1-3d_0^2)^{N-1},
\end{align}
or the last copy of the initial assemblage $\{\sigma_{a,b|x,y}^{C^{\text{GW}}}\}_{a,b,x,y}$ with the probability $P_{\text{fail}}^{\text{W}_2}$.
In the asymptotic limit as the number of copies tends to infinity, we get $\underset{N \rightarrow \infty}{\text{lim}} P_{\text{success}}^{\text{W}_2} \rightarrow 1$ and hence, the protocol ensures distillation of at least one copy of the target assemblage of the W state as $N$ tends to infinity.

\begin{figure}[!t]
\resizebox{8cm}{7cm}{\includegraphics{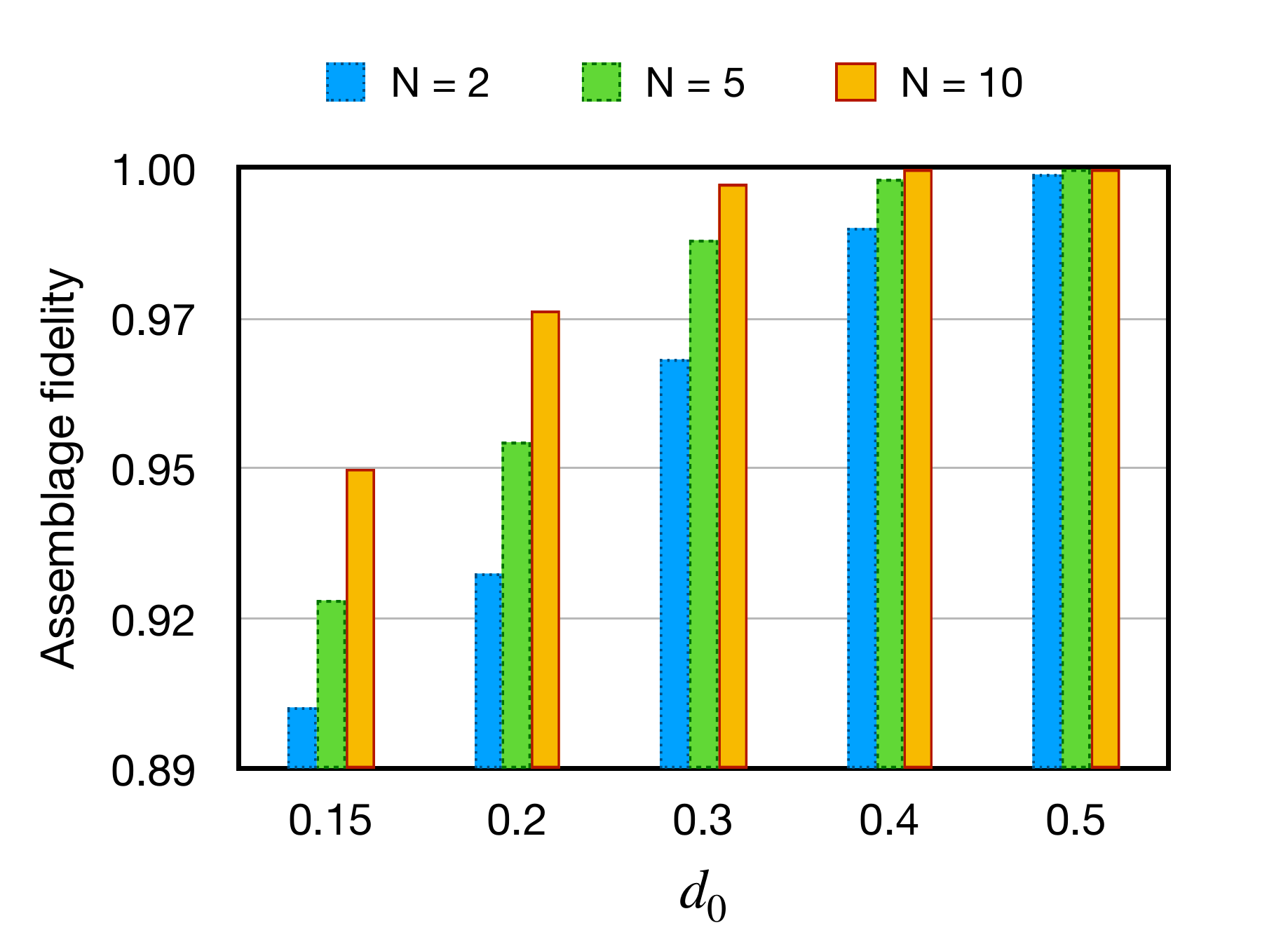}}
\caption{\footnotesize (Coloronline) Variation of the assemblage fidelity with the state parameter $d_0$ of generalized W pure states in our distillation protocol in the 2SDI scenario. The quantities in the horizontal and vertical axes are numbers and have no unit.} 
\label{W2SAF_fig}
\end{figure}

On the other hand, in the regime of finite $N$ copies with $N \geq 2$, let $\{\sigma_{a,b|x,y}^{\text{dist}^{\text{W}}}\}_{a,b,x,y}$ denote the assemblage obtained, on average, after the distillation protocol. Then the assemblage fidelity between the assemblage $\{\sigma_{a,b|x,y}^{\text{dist}^{\text{W}}}\}_{a,b,x,y}$ and the target assemblage $\{\sigma_{a,b|x,y}^{C^{\text{W}}}\}_{a,b,x,y}$ is given in by,

\begin{align}
    &\mathcal{F}_A \Big(\{\sigma_{a,b|x,y}^{\text{dist}^{\text{W}}}\}_{a,b,x,y}, \{\sigma_{a,b|x,y}^{C^{\text{W}}}\}_{a,b,x,y} \Big)  \nonumber \\
    &=\dfrac{1}{\sqrt{3}}\sqrt{3 - 2 \Bigg(\sqrt{\frac{1-d_0^2}{2}} - d_0 \Bigg)^2 \Big( 1-3d_0^2 \Big)^{N-1}}.
    \label{AF2SDIW}
\end{align}
The details of the derivation of the above expression are provided in Appendix \ref{app10}. Hence, we can present the following theorem:
\begin{thm} 
In the regime of finite $N \geq 2$ copies of initial assemblage $\{\sigma_{a,b|x, y}^{C^{\text{GW}}}\}_{a,b,x,y}$ of the one parameter generalized W states (\ref{owclass}) in our 2SDI scenario, an assemblage can be obtained on average, which is close to the target assemblage $\{\sigma_{a,b|x, y}^{C^{\text{W}}}\}_{a,b,x,y}$ in our 2SDI scenario  with assemblage fidelity $\dfrac{1}{\sqrt{3}}\sqrt{3 - 2 \Bigg(\sqrt{\dfrac{1-d_0^2}{2}} - d_0 \Bigg)^2 \Big( 1-3d_0^2 \Big)^{N-1}}$.        
\end{thm}

In Fig. \ref{W2SAF_fig} we demonstrate the variation of the assemblage fidelity (\ref{AF2SDIW}) for different values of the state parameter $d_0$ of the one-parameter generalized W three-qubit pure states (\ref{owclass}) and for different number $N \geq 2$ of copies of the initial assemblage $\{\sigma_{a,b|x,y}^{C^{\text{GW}}}\}_{a,b,x,y}$ mentioned in Appendix \ref{OW2Sassemblage}. From the figure it is clear that our distillation protocol ensures significant amount of the assemblage fidelity even for the small number of copies of the initial assemblage.

\begin{figure}[!t]
\resizebox{7.5cm}{6.5cm}{\includegraphics{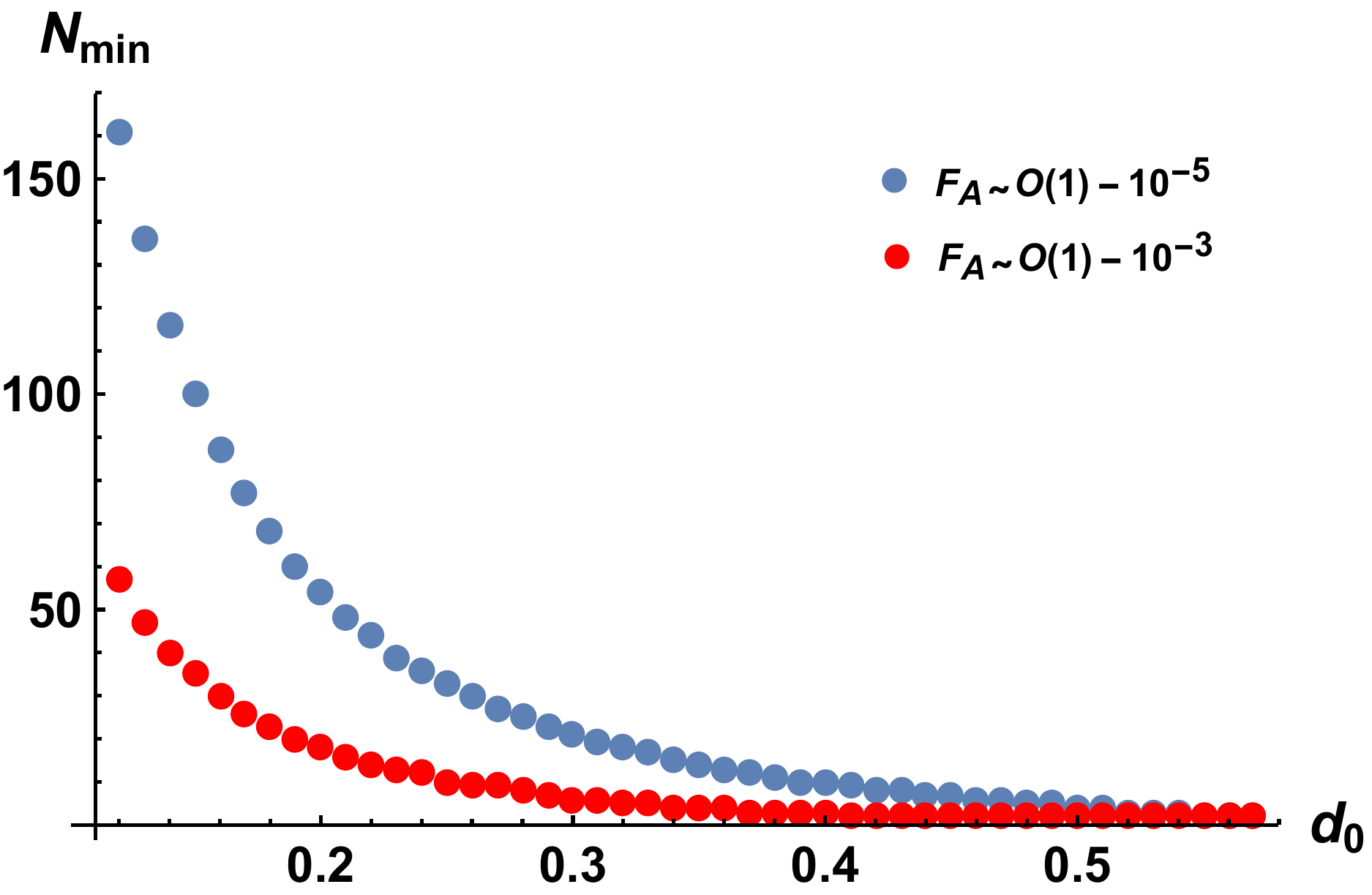}}
\caption{\footnotesize (Coloronline) Variation of the minimum number of copies ($N_{\text{min}}$) required to achieve near perfect assemblage fidelity of our
distillation protocol in the 2SDI scenario for generalized W states with state parameter $d_0$. Blue dotted upper curve is for $\mathcal{F}_{A} \sim O(1) - 10^{-5}$ and red dotted bottom curve is for $\mathcal{F}_{A} \sim O(1) - 10^{-3}$. The quantities in the horizontal and vertical axes are numbers and have no unit.}   
\label{N_min_W2S}
\end{figure}

We evaluate the minimum number of copies ($N_{\text{min}}$) of the initial assemblages required to have  a near perfect  assemblage fidelity ($\mathcal{F}_{A} \sim O(1) - 10^{-5}$). In Fig. \ref{N_min_W2S}, we plot the variation of $N_{\text{min}}$  for different values of the state parameter $d_0$ of the one parameter generalized W three-qubit pure states (\ref{owclass}). This figure shows that $N_{\text{min}}$ decreases with increasing values of $d_0$. It is clear that our distillation protocol works significantly well in realistic scenarios where one cannot have infinitely many copies of the initial assemblages.

\section{Summary and outlook}\label{section4}

Multipartite quantum correlations are the key resources for   information processing tasks in  quantum networks. Depending on the degree of trust/characterization of the measuring devices of the observers, there can be three nonequivalent forms of multipartite quantum correlations- multipartite entanglement, multipartite EPR steering and multipartite Bell-nonlocality. In practical quantum information processing, distillation of resources plays an important role in the presence of ubiquitous decoherence effects. Several distillation protocols are known for  multipartite entanglement \cite{Huang14,Ben08} and  multipartite Bell-nonlocality \cite{Wu10,Ye12,Ebbe2013,Pan2015}. However, no distillation protocol has been proposed till date for multipartite EPR steering. In the present paper, we have taken the first step towards filling this gap. 

Here, we have considered the two types of hybrid quantum networks that are possible in the tripartite scenario: 1SDI and 2SDI networks. In each of these cases, we have proposed distillation protocols of genuine tripartite EPR steering for the two SLOCC inequivalent classes of genuine tripartite entangled states, {\it viz.}, the GHZ  state and W  state. We have taken the assemblages obtained from these states, when untrusted parties perform some particular orthogonal von Neumann measurements of rank-1, as  perfectly genuine steerable assemblages. These perfectly genuine tripartite steerable assemblages have been considered as the target assemblages of our distillation protocols. In each of these cases, we have derived the exact filtering operation and have given the exact analytical expression of the assemblage fidelity  that would be very useful in finding the optimal number of copies of the initial assemblage upto the desired accuracy of the assemblage fidelity.

In particular, we have demonstrated the  possibility of perfect distillation in the limit of infinitely many copies of the initial assemblages. On the other hand, in practical scenarios  when one starts with a finite number of copies of the initial assemblages, our protocols have been shown to have high efficiency. For example, in case of GGHZ state with $\theta = 0.3$, the assemblage fidelity is 0.906, 0.948 and 0.98 for 2, 5 and 10 copies respectively. These are reasonable values of fidelities with realistic experimental devices taking into account their intrinsic tolerance levels. For the W states, distillation in the the 2SDI network outperforms that in the 1SDI network, achieving an accuracy comparable with the GHZ state distillation. 

Designing a genuine tripartite steering distillation protocol is always advantageous in the sense that the same protocol can also be used for genuine tripartite entanglement distillation with smaller number of parties (only the trusted parties)  performing the local quantum operations. However, genuine tripartite entanglement distillation protocol cannot always be used for genuine tripartite EPR steering distillation as the free operations in the context of genuine tripartite entanglement are generally not the free operations in the genuine tripartite EPR steering scenario. Following our protocol, it is possible to distill genuine tripartite entanglement of three-qubit generalized GHZ or W state. 

Before concluding, it is worth mentioning certain possible offshoots of our work. Our study opens up new avenues of research for the genuine steering distillation protocols for other states and in multipartite settings that will also be relevant for the distillation of genuine multipartite entanglement with lesser resource, {\it i.e.}, with less number of observers performing quantum operations, in such scenarios. Extensions of the present analysis could be performed to devise distillation protocols  for EPR steering of $n$-qubit or higher dimensional quantum states. It is worth exploring other pre-processing operations or to devise optimal filtering technique in such distillation protocols. Finally, it would be interesting to investigate whether  the concept of  bound genuine multipartite steering exists analogous to bound entanglement \cite{Horo98}. \\
  
\section*{Acknowledgement} 
 S.G. acknowledges the S. N. Bose National Centre for Basic Sciences, Kolkata for financial support. D.D. acknowledges the Science and Engineering Research Board (SERB), Government of India for financial support through the National Post Doctoral Fellowship (NPDF) (File No. PDF/2020/001358).  A.S.M. acknowledges Project No. DST/ICPS/QuEST/2018/98
from the Department of Science and Technology, Government of India. The authors would like to thank C. Jebarathinam and Arup Roy for fruitful comments. 
\bibliography{Gsteer} 
\begin{widetext}
\appendix

 \section{Free operations of genuine tripartite steering distillation in 1SDI scenario}\label{app1}
 Here we  show that for a single copy of assemblage, the operations involved in our distillation protocol in the 1SDI scenario are free operations of 1SDI genuine tripartite steering (i.e., they cannot create genuine steerable assemblage from an assemblage which does not demonstrate genuine steering). 
 
 Our distillation protocol consists of the following operations:
 
 (A1) Local quantum measurements (i.e., local completely positive non-trace preserving maps) by the two trusted parties, Bob and Charlie.
 
 (B1) Classical communication from each of the trusted parties to other (trusted and untrusted) parties. In other words, Bob can send classical information to Alice and Charlie, Charlie can send classical information to Alice and Bob, but Alice does not send any classical information.
 
 (C1) The untrusted party, Alice, can post-process her output depending on her input and the classical informations received from Bob and Charlie.
 
  An assemblage $\{ \sigma^{\text{BC}}_{a|x} \}_{a,x}$ is genuine steerable in 1SDI scenario  if it is not the case that \cite{Cavalcanti15}
\begin{eqnarray} 
	\sigma^{\text{BC}}_{a|x} &=& \sum_\lambda p_\lambda^{A:BC}p_\lambda(a|x)\rho_\lambda^{BC}
							+ \sum_\mu p_\mu^{B:AC}\rho_\mu^B\otimes \sigma_{a|x\mu}^{C} 
						        + \sum_\nu p_\nu^{AB:C} \sigma_{a|x\nu}^{B}\otimes \rho_\nu^C \quad \forall a, x \nonumber \\
						        &=& \Pi^1_{a|x} + \Pi^2_{a|x} + \Pi^3_{a|x} \quad \forall a, x,
						 \label{1SAB:Cnew} 
\end{eqnarray} 
where
\begin{align}
 \Pi^1_{a|x} &=  \sum_\lambda p_\lambda^{A:BC}p_\lambda(a|x)\rho_\lambda^{BC}, \nonumber \\  
 \Pi^2_{a|x} &=  \sum_\mu p_\mu^{B:AC}\rho_\mu^B\otimes \sigma_{a|x\mu}^{C},  \nonumber \\  
 \Pi^3_{a|x} &=  \sum_\nu p_\nu^{AB:C} \sigma_{a|x\nu}^{B}\otimes \rho_\nu^C. \nonumber 						        
\end{align}
Now, we  show that the above operations (A1), (B1), (C1) cannot create 1SDI genuine steerable assemblage from an assemblage having the form in Eq.(\ref{1SAB:Cnew}). In other words,  the characteristics of the assemblage (\ref{1SAB:Cnew}) remain the same under the above-mentioned operations.

The first term $\Pi^1_{a|x}$ in Eq.(\ref{1SAB:Cnew}) is an unsteerable assemblage from  Alice to Bob-Charlie. This remains unsteerable from Alice to Bob-Charlie even after applying the above operations  \cite{Nery20}, as local operations assisted by one-way classical communication (1W-LOCCs) from the quantum part (Bob-Charlie) to the black box (Alice) cannot create steering from Alice to Bob-Charlie \cite{Rodrigo15}.

The second term $\Pi^2_{a|x}$ in Eq.(\ref{1SAB:Cnew}) has two features: (i) it is unsteerable from Alice to Bob, but not necessarily from Alice to Charlie; (ii)  it is separable across the bipartition- Bob versus Charlie. Now, we want to check whether this assemblage $\Pi^2_{a|x}$ remains unsteerable from Alice to Bob after the above operations  (A1), (B1), (C1). Note that, Alice's black-box distribution after the above operations involved in our distillation protocol is $p(a|x, o^u_B, o^u_C)$ [see Eq.(\ref{een1})]. Since, this is an arbitrary black-box distribution, we can write it as $\tilde{p}(a|x, o^u_B)$ = $p(a|x, o^u_B, o^u_C)$, where $\tilde{p}(a|x, o^u_B)$ can be interpreted as a black-box distribution of $a$ conditioned on $x$ and $o^u_B$. Now, neglecting Charlie's part, the above operations  (A1), (B1), (C1) involve  1W-LOCCs from Bob to Alice, and they cannot create steering from Alice to Bob \cite{Rodrigo15}. Moreover, this assemblage remains separable across the bipartition- Bob versus Charlie after the above operations, as local operations (by Bob and Charlie) and classical communications (between Bob and Charlie) cannot create entanglement between Bob and Charlie \cite{Horo09}. Hence, the characteristics of the second term $\Pi^2_{a|x}$ remains unchanged after performing the operations  (A1), (B1), (C1) mentioned above.

The third term $\Pi^3_{a|x}$ in Eq.(\ref{1SAB:Cnew}) has two features:  (i) it is unsteerable from Alice to Charlie, but not necessarily from Alice to Bob; (ii) it is separable across the bipartition- Bob versus Charlie. Following the same approach mentioned in the case of the second term $\Pi^2_{a|x}$, it can be shown that the characteristics of the third term $\Pi^3_{a|x}$ remains unchanged after performing the operations  (A1), (B1), (C1) mentioned above.

Since, the characteristics of the assemblage given by Eq.(\ref{1SAB:Cnew}) remain the same under the above-mentioned operations (A1), (B1), (C1), we can conclude that these operations constitute free operations of genuine tripartite steering in the 1SDI scenario.
Here, it may be noted that the above operations (A), (B), (C) may not be the most general free operations of genuine tripartite steering in the 1SDI scenario.

 \section{Free operations of genuine tripartite steering distillation in 2SDI scenario}\label{app2}
 
 We  show that, for a single copy of assemblage, the operations involved in our distillation protocol in the 2SDI scenario are free operations of the 2SDI genuine tripartite steering. Our distillation protocol in the 2SDI scenario consists of the following operations:
 
 (A2) Local quantum measurements (i.e., local completely positive non-trace preserving maps) by the trusted party, Charlie.
 
 (B2) Classical communication from the trusted party (Charlie) to the untrusted  parties (Alice and Bob). In other words, Charlie sends classical information to Alice and Bob, but Alice and Bob do not send any classical information.
 
 (C2) The untrusted parties, Alice and Bob, can locally post-process their output depending on their input and the classical information received from Charlie.
 
  An assemblage $\{ \sigma^{\text{C}}_{a,b|x,y} \}_{a,b,x,y}$ is genuine steerable in the 2SDI scenario if it is not the case that \cite{Cavalcanti15}
 \begin{eqnarray} 
	\sigma^{\text{C}}_{a,b|x,y} &=& \sum_\lambda p_\lambda^{A:BC}p_\lambda(a|x)\sigma_{b|y\lambda}^{C}
	+ \sum_\mu p_\mu^{B:AC}p_\mu(b|y) \sigma_{a|x\mu}^{C}
	+ \sum_\nu p_\nu^{AB:C}  p_{\nu}(a,b|x,y) \rho_\nu^C  \quad \forall a,b,x,y \nonumber \\
&=& \Pi^1_{a,b|x,y} + \Pi^2_{a,b|x,y} + \Pi^3_{a,b|x,y} \quad \forall a, b,x,y,
						 \label{2SAB:Cnew} 
\end{eqnarray} 
where
\begin{align}
 \Pi^1_{a,b|x,y} &=  \sum_\lambda p_\lambda^{A:BC}p_\lambda(a|x)\sigma_{b|y\lambda}^{C},\nonumber \\  
 \Pi^2_{a,b|x,y} &=  \sum_\mu p_\mu^{B:AC}p_\mu(b|y) \sigma_{a|x\mu}^{C}, \nonumber \\  
 \Pi^3_{a,b|x,y} &=  \sum_\nu p_\nu^{AB:C} p_{\nu}(a,b|x,y) \rho_\nu^C . \nonumber 						        
\end{align}
We show that the above operations (A2), (B2), (C2) cannot create 2SDI genuine steerable assemblage from an assemblage having the form given by Eq.(\ref{2SAB:Cnew}). 


The first term $\Pi^1_{a,b|x,y}$ in Eq.(\ref{2SAB:Cnew}) is unsteerable from Alice to Charlie. This remains unsteerable from Alice to Charlie even after applying the above operations (A2), (B2), (C2) \cite{Nery20} as 1W-LOCCs from the quantum part (Charlie) to the black box (Alice) cannot create steering from Alice to Charlie \cite{Rodrigo15}. Here, Charlie sends classical communication to Bob as well. But, since no communication from Bob to Alice or Bob to Charlie is involved in our distillation protocol, we can completely neglect Bob's subsystem while checking unsteerability from Alice and Charlie.

The second term $\Pi^2_{a,b|x,y}$ in Eq.(\ref{2SAB:Cnew}) is unsteerable from Bob to Charlie. Following the above approach, it can be shown that the operations (A2), (B2), (C2) cannot create steering from Bob to Charlie.

The third term $\Pi^3_{a,b|x,y}$ in Eq.(\ref{2SAB:Cnew}) has two features:  (i) it is unsteerable from Alice-Bob to  Charlie, (ii) the probability distribution $p_{\nu}(a,b|x, y)$ arises due to local measurements performed on a possibly entangled state, it may contain nonlocal quantum correlations. Now, the operations  (A2), (B2), (C2) are nothing but 1W-LOCCs from Charlie to Alice-Bob and it cannot create steering from Alice-Bob to Charlie \cite{Rodrigo15, Nery20}. Moreover, the local post-processing  by the untrusted parties (Alice and Bob) involves discarding or not discarding the assemblage. Hence, it will not effect the characteristic (ii) mentioned above on a single copy of the assemblage  \cite{Nery20}.

Since, the characteristics of the assemblage Eq.(\ref{2SAB:Cnew}) remain the same under the above-mentioned operations (A2), (B2), (C2), we can conclude that these operations are free operations of genuine tripartite steering in 2SDI scenario.
Here too, the above operations (A2), (B2), (C2) may not be the most general free operations of genuine tripartite steering in 2SDI scenario.

\section{Derivation of the assemblage fidelity of GHZ states in 1SDI scenario}\label{app3}

Since, our target assemblage $\{\sigma_{a|x}^{BC^{\text{GHZ}}} \}_{a,x}$ given by  Eq.(\ref{GHZassemblage}) is pure, we can write $\sigma_{a|x}^{BC^{\text{GHZ}}}$ = $\dfrac{1}{2} \Big| \phi_{BC}^{\text{GHZ}} (a, x) \Big\rangle \Big\langle \phi_{BC}^{\text{GHZ}} (a, x) \Big|$ $\forall$ $a,x$. Hence, the assemblage fidelity of the assemblage (\ref{GGHZdistill}) and our target assemblage (\ref{GHZassemblage}) corresponding to GHZ state  can be written as: 
\begin{align}
    &\mathcal{F}_{A} \Big( \{\sigma_{a|x}^{\text{dist}^{\text{GHZ}}} \}_{a,x}, \{\sigma_{a|x}^{BC^{\text{GHZ}}}\}_{a,x} \Big) = \underset{x}{\mathrm{min}} \sum_a \sqrt{\frac{1}{2} \Big\langle \phi_{BC}^{\text{GHZ}} (a, x) \Big| \sigma_{a|x}^{\text{dist}^{\text{GHZ}}} \Big| \phi_{BC}^{\text{GHZ}} (a, x)} \Big\rangle.
     \label{assfidghz1app}
\end{align}
Next, we have,
\begin{align}
 &\sum_a \sqrt{\frac{1}{2} \Big\langle \phi_{BC}^{\text{GHZ}} (a, 0) \Big| \sigma_{a|0}^{\text{dist}^{\text{GHZ}}} \Big| \phi_{BC}^{\text{GHZ}} (a, 0)} \Big\rangle =\sqrt{1 - \frac{1}{2} (1 - \sin 2 \theta ) (\cos 2 \theta )^{N-1}}, \nonumber \\
 &\sum_a \sqrt{\frac{1}{2} \Big\langle \phi_{BC}^{\text{GHZ}} (a, 1) \Big| \sigma_{a|1}^{\text{dist}^{\text{GHZ}}} \Big| \phi_{BC}^{\text{GHZ}} (a, 1)} \Big\rangle =\sqrt{1 - \frac{1}{2} (1 - \sin 2 \theta ) (\cos 2 \theta )^{N-1}}, \nonumber \\
 &\sum_a \sqrt{\frac{1}{2} \Big\langle \phi_{BC}^{\text{GHZ}} (a, 2) \Big| \sigma_{a|2}^{\text{dist}^{\text{GHZ}}} \Big| \phi_{BC}^{\text{GHZ}} (a, 2)} \Big\rangle =\frac{1}{2} \left(\sqrt{1- (\cos 2 \theta )^N}+\sqrt{ 1+ (\cos 2 \theta )^N }\right). \nonumber
\end{align}
Hence, we can write
\begin{align}
    &\mathcal{F}_{A} \Big( \{\sigma_{a|x}^{\text{dist}^{\text{GHZ}}} \}_{a,x}, \{\sigma_{a|x}^{BC^{\text{GHZ}}}\}_{a,x} \Big) = \mathrm{min} \Bigg[\sqrt{1 - \frac{1}{2} (1 - \sin 2 \theta ) (\cos 2 \theta )^{N-1}},  \frac{1}{2} \left(\sqrt{1- (\cos 2 \theta )^N}+\sqrt{ 1+ (\cos 2 \theta )^N }\right) \Bigg].
    \label{app3eq3}
\end{align}
Let us now define $f_0$ and $f_1$ as
\begin{align}
    f_0 =& \sqrt{1 - \frac{1}{2} (1 - \sin 2 \theta ) (\cos 2 \theta )^{N-1}} ,\nonumber \\
    f_1 =& \frac{1}{2} \left(\sqrt{1- (\cos 2 \theta )^N}+\sqrt{ 1+ (\cos 2 \theta )^N }\right).
\end{align}
Since, by definition of fidelity, $ \sqrt{ \Big\langle \phi_{BC}^{\text{GHZ}} (a, x) \Big| \sigma_{a|x}^{\text{dist}^{\text{GHZ}}} \Big| \phi_{BC}^{\text{GHZ}} (a, x)} \Big\rangle \geq 0$ $\forall$ $x,a$, we have $f_0 \geq 0$ and $f_1 \geq 0$ $\forall$ $\theta$ $\in$ $( 0.185, \frac{\pi}{4})$ and $\forall$ $N \geq 2$ $\in$ $\mathbb{Z}$ (where $\mathbb{Z}$ denotes the set of positive integers). 

Let also $u_0 = 2f_0^2-1$ and $u_1 = 2f_1^2-1$. Hence, we can write
\begin{align}
    u_0 &= 1 -  (1 - \sin 2 \theta ) (\cos 2 \theta )^{N-1} > 0 \quad \forall \quad \theta \in ( 0.185, \frac{\pi}{4}), \, N \geq 2 \, \in \, \mathbb{Z}, \nonumber \\
    u_1 &= \sqrt{1- (\cos 2 \theta )^{2N}}   > 0 \quad \forall \quad \theta \in ( 0.185, \frac{\pi}{4}), \, N \geq 2 \, \in \, \mathbb{Z},
    \label{app3eq1}
\end{align}
where the last inequality is obtained from the definition of fidelity \cite{Nery20}: $\sqrt{\frac{1}{2} \Big\langle \phi_{BC}^{\text{GHZ}} (1, 2) \Big| \sigma_{1|2}^{\text{dist}^{\text{GHZ}}} \Big| \phi_{BC}^{\text{GHZ}} (1, 2)} \Big\rangle$ = $\frac{1}{2}\sqrt{1- (\cos 2 \theta )^{2N}}$ $>$ $0$. Next, an algebraic manipulation leads to the expression,
\begin{align}
u_0^2 - u_1^2 = - 2 (\cos 2 \theta )^{N-1} (1 - \sin 2 \theta ) [1 - (\cos 2 \theta )^{N-1}]   < 0 \quad \forall \quad \theta \in ( 0.185, \frac{\pi}{4}), \, N \geq 2 \, \in \, \mathbb{Z}.
\label{app3eq2}
\end{align}
Hence, from Eqs.(\ref{app3eq2}) and (\ref{app3eq1}), we have $u_1 \geq u_0$ $\forall$ $\theta$ $\in$ $( 0.185, \frac{\pi}{4})$ and $\forall$  $N$ $\geq$ $2$ $\in$ $\mathbb{Z}$. Now, using the facts $f_0 \geq 0$ and $f_1 \geq 0$, we can conclude the following,
\begin{align}
   &f_0 = \sqrt{1 - \frac{1}{2} (1 - \sin 2 \theta ) (\cos 2 \theta )^{N-1}} < \frac{1}{2} \left(\sqrt{1- (\cos 2 \theta )^N}+\sqrt{ 1+ (\cos 2 \theta )^N }\right) = f_1 \quad \forall \quad \theta \in ( 0.185, \frac{\pi}{4}), \, N \geq 2 \, \in \, \mathbb{Z}.
\end{align}

Therefore, from Eq.(\ref{app3eq3}), we can write
\begin{align}
    &\mathcal{F}_{A} \Big( \{\sigma_{a|x}^{\text{dist}^{\text{GHZ}}} \}_{a,x}, \{\sigma_{a|x}^{BC^{\text{GHZ}}}\}_{a,x} \Big) = \sqrt{1 - \frac{1}{2} (1 - \sin 2 \theta ) (\cos 2 \theta )^{N-1}} \quad \forall \quad \theta \in ( 0.185, \frac{\pi}{4}), \, N \geq 2 \, \in \, \mathbb{Z}.
\end{align}

\section{Components of the assemblages produced from the GGHZ states in our 2SDI scenario}\label{2SAGGHZ}
The components of the assemblages $\{\sigma_{a,b|x,y}^{C^{\text{GGHZ}}}\}_{a,b,x,y}$ (the initial assemblages of our distillation protocol), obtained from the three qubit GGHZ state (\ref{GGHZ}) through measurements of the observables $X_0= \sigma_x$, $X_1=\sigma_y$ and $X_2=\sigma_z$ by Alice and $Y_0= \sigma_x$, $Y_1=\sigma_y$ and $Y_2=\sigma_z$ by Bob, are given by,
\begin{align}
	\sigma_{0,0|0,0}^{C^{\text{GGHZ}}} &= \sigma_{1,1|0,0}^{C^{\text{GGHZ}}} = \sigma_{0,1|1,1}^{C^{\text{GGHZ}}} = \sigma_{1,0|1,1}^{C^{\text{GGHZ}}} = \frac{1}{4} \Big| \theta_+^2 \Big\rangle \Big\langle \theta_+^2 \Big|, \nonumber \\ 
	\sigma_{0,1|0,0}^{C^{\text{GGHZ}}} &= \sigma_{1,0|0,0}^{C^{\text{GGHZ}}} = \sigma_{0,0|1,1}^{C^{\text{GGHZ}}} = \sigma_{1,1|1,1}^{C^{\text{GGHZ}}} = \frac{1}{4} \Big| \theta_-^2 \Big\rangle \Big\langle \theta_-^2 \Big|, \nonumber \\ 
    \sigma_{0,0|0,1}^{C^{\text{GGHZ}}} &= \sigma_{1,1|0,1}^{C^{\text{GGHZ}}} = \sigma_{0,0|1,0}^{C^{\text{GGHZ}}} = \sigma_{1,1|1,0}^{C^{\text{GGHZ}}} = \frac{1}{4} \Big| \theta_-^3 \Big\rangle \Big\langle \theta_-^3 \Big|, \nonumber \\ 
    \sigma_{0,1|0,1}^{C^{\text{GGHZ}}} &= \sigma_{1,0|0,1}^{C^{\text{GGHZ}}} = \sigma_{0,1|1,0}^{C^{\text{GGHZ}}} = \sigma_{1,0|1,0}^{C^{\text{GGHZ}}} = \frac{1}{4} \Big| \theta_+^3 \Big\rangle \Big\langle \theta_+^3 \Big|, \nonumber \\
    \sigma_{0,0|0,2}^{C^{\text{GGHZ}}} &= \sigma_{1,0|0,2}^{C^{\text{GGHZ}}} = \sigma_{0,0|1,2}^{C^{\text{GGHZ}}} = \sigma_{1,0|1,2}^{C^{\text{GGHZ}}} = \sigma_{0,0|2,0}^{C^{\text{GGHZ}}} = \sigma_{0,1|2,0}^{C^{\text{GGHZ}}} = \sigma_{0,0|2,1}^{C^{\text{GGHZ}}} = \sigma_{0,1|2,1}^{C^{\text{GGHZ}}} =  \frac{\cos^2 \theta}{2} | 0 \rangle \langle 0 |, \nonumber \\
    \sigma_{0,1|0,2}^{C^{\text{GGHZ}}} &= \sigma_{1,1|0,2}^{C^{\text{GGHZ}}} = \sigma_{0,1|1,2}^{C^{\text{GGHZ}}} = \sigma_{1,1|1,2}^{C^{\text{GGHZ}}} = \sigma_{1,0|2,0}^{C^{\text{GGHZ}}} = \sigma_{1,1|2,0}^{C^{\text{GGHZ}}} = \sigma_{1,0|2,1}^{C^{\text{GGHZ}}} = \sigma_{1,1|2,1}^{C^{\text{GGHZ}}} =  \frac{\sin^2 \theta}{2} | 1 \rangle \langle 1 |, \nonumber \\
	\sigma_{0,0|2,2}^{C^{\text{GGHZ}}} &= \cos^2 \theta | 0 \rangle \langle 0 |,\quad  \sigma_{1,1|2,2}^{C^{\text{GGHZ}}} = \sin^2 \theta | 1 \rangle \langle 1 |.
	\label{GGHZ2Sassemblage}
\end{align}
where, $\ket{\theta_{\pm}^2} = \cos \theta \ket{0} \pm \sin \theta \ket{1}$ and $\ket{\theta_{\pm}^3} = \cos \theta \ket{0} \pm i \, \sin \theta  \ket{1}$. The assemblage components $\sigma_{0,1|2,2}^{C^{\text{GGHZ}}}$ and  $\sigma_{1,0|2,2}^{C^{\text{GGHZ}}}$ do not exist because the probabilities of their occurrences are  zero. Hence, we need not consider these two components.

\section{Components of the assemblage produced from the GHZ state in our 2SDI scenario}\label{2SAGHZ}

The components of the assemblages $\{\sigma_{a,b|x,y}^{C^{\text{GHZ}}}\}_{a,b,x,y}$ (the target assemblage of our distillation protocol), obtained from the three qubit GHZ state through measurements of the observables $X_0= \sigma_x$, $X_1=\sigma_y$ and $X_2=\sigma_z$ by Alice and $Y_0= \sigma_x$, $Y_1=\sigma_y$ and $Y_2=\sigma_z$ by Bob, are given by,
\begin{eqnarray}
	\sigma_{0,0|0,0}^{C^{\text{GHZ}}} &=& \sigma_{1,1|0,0}^{C^{\text{GHZ}}} = \sigma_{0,1|1,1}^{C^{\text{GHZ}}} = \sigma_{1,0|1,1}^{C^{\text{GHZ}}} = \frac{1}{4} \Bigg| \frac{\pi}{4}_+^2 \Bigg\rangle \Bigg\langle \frac{\pi}{4}_+^2 \Bigg|, \nonumber \\ 
	\sigma_{0,1|0,0}^{C^{\text{GHZ}}} &=& \sigma_{1,0|0,0}^{C^{\text{GHZ}}} = \sigma_{0,0|1,1}^{C^{\text{GHZ}}} = \sigma_{1,1|1,1}^{C^{\text{GHZ}}} = \frac{1}{4} \Bigg| \frac{\pi}{4}_-^2 \Bigg\rangle \Bigg\langle \frac{\pi}{4}_-^2 \Bigg|, \nonumber \\ 
    \sigma_{0,0|0,1}^{C^{\text{GHZ}}} &=& \sigma_{1,1|0,1}^{C^{\text{GHZ}}} = \sigma_{0,0|1,0}^{C^{\text{GHZ}}} = \sigma_{1,1|1,0}^{C^{\text{GHZ}}} = \frac{1}{4} \Bigg| \frac{\pi}{4}_-^3 \Bigg\rangle \Bigg\langle \frac{\pi}{4}_-^3 \Bigg|, \nonumber \\ 
    \sigma_{0,1|0,1}^{C^{\text{GHZ}}} &=& \sigma_{1,0|0,1}^{C^{\text{GHZ}}} = \sigma_{0,1|1,0}^{C^{\text{GHZ}}} = \sigma_{1,0|1,0}^{C^{\text{GHZ}}} = \frac{1}{4} \Bigg| \frac{\pi}{4}_+^3 \Bigg\rangle \Bigg\langle \frac{\pi}{4}_+^3 \Bigg|, \nonumber \\
    \sigma_{0,0|0,2}^{C^{\text{GHZ}}} &=& \sigma_{1,0|0,2}^{C^{\text{GHZ}}} = \sigma_{0,0|1,2}^{C^{\text{GHZ}}} = \sigma_{1,0|1,2}^{C^{\text{GHZ}}} = \sigma_{0,0|2,0}^{C^{\text{GHZ}}} = \sigma_{0,1|2,0}^{C^{\text{GHZ}}} = \sigma_{0,0|2,1}^{C^{\text{GHZ}}} = \sigma_{0,1|2,1}^{C^{\text{GHZ}}} =  \frac{1}{4} | 0\rangle \langle 0 |, \nonumber \\
    \sigma_{0,1|0,2}^{C^{\text{GHZ}}} &=& \sigma_{1,1|0,2}^{C^{\text{GHZ}}} = \sigma_{0,1|1,2}^{C^{\text{GHZ}}} = \sigma_{1,1|1,2}^{C^{\text{GHZ}}} = \sigma_{1,0|2,0}^{C^{\text{GHZ}}} = \sigma_{1,1|2,0}^{C^{\text{GHZ}}} = \sigma_{1,0|2,1}^{C^{\text{GHZ}}} = \sigma_{1,1|2,1}^{C^{\text{GHZ}}} =  \frac{1}{4} | 1 \rangle \langle 1 |, \nonumber \\
	\sigma_{0,0|2,2}^{C^{\text{GHZ}}} &=& \frac{1}{2} | 0 \rangle \langle 0 |,\quad  \sigma_{1,1|2,2}^{C^{\text{GHZ}}} = \frac{1}{2} | 1 \rangle \langle 1 |.
	\label{GHZ2Sassemblage}
\end{eqnarray}
where, $\Big| \dfrac{\pi}{4}_{\pm}^2 \Big\rangle = \frac{1}{\sqrt{2}}(\ket{0} \pm \ket{1})$ and $\Big| \dfrac{\pi}{4}_{\pm}^3 \Big\rangle = \frac{1}{\sqrt{2}}(\ket{0} \pm i \, \ket{1})$. The assemblage components $\sigma_{0,1|2,2}^{C^{\text{GHZ}}}$ and  $\sigma_{1,0|2,2}^{C^{\text{GHZ}}}$ do not exist because the probabilities of their occurrences are  zero. Hence, we need not consider these two components.

\section{Derivation of the assemblage fidelity for GHZ states in 2SDI scenario}\label{app6}

In the regime of finite $N$ copies with $N \geq 2$, a single assemblage can be extracted $\{\sigma_{a,b|x,y}^{\text{dist}^{\text{GHZ}}}\}_{a,b,x,y}$ as a convex combination of the initial assemblage $\{\sigma_{a,b|x,y}^{C^{\text{GGHZ}}}\}_{a,b,x,y}$ with components given in Eq.(\ref{GGHZ2Sassemblage}) and the target assemblage derived from the GHZ state $\{\sigma_{a,b|x,y}^{C^{\text{GHZ}}}\}_{a,b,x,y}$ with components given in Eq.(\ref{GHZ2Sassemblage}). The components of $\{\sigma_{a,b|x,y}^{\text{dist}^{\text{GHZ}}}\}_{a,b,x,y}$ are given by,
\begin{eqnarray}
	\sigma_{0,0|0,0}^{\text{dist}^{\text{GHZ}}} &=& \sigma_{1,1|0,0}^{\text{dist}^{\text{GHZ}}} = \sigma_{0,1|1,1}^{\text{dist}^{\text{GHZ}}} = \sigma_{1,0|1,1}^{\text{dist}^{\text{GHZ}}} = \frac{1}{4} \Bigg(P^{\text{GHZ}_2}_{\text{success}}\Bigg| \frac{\pi}{4}_+^2 \Bigg\rangle \Bigg\langle \frac{\pi}{4}_+^2 \Bigg| + P^{\text{GHZ}_2}_{\text{fail}} \Big| \theta_+^2 \Big\rangle \Big\langle \theta_+^2 \Big| \Bigg), \nonumber \\ 
	\sigma_{0,1|0,0}^{\text{dist}^{\text{GHZ}}} &=& \sigma_{1,0|0,0}^{\text{dist}^{\text{GHZ}}} = \sigma_{0,0|1,1}^{\text{dist}^{\text{GHZ}}} = \sigma_{1,1|1,1}^{\text{dist}^{\text{GHZ}}} = \frac{1}{4} \Bigg(P^{\text{GHZ}_2}_{\text{success}}\Bigg| \frac{\pi}{4}_-^2 \Bigg\rangle \Bigg\langle \frac{\pi}{4}_-^2 \Bigg| + P^{\text{GHZ}_2}_{\text{fail}} \Big| \theta_-^2 \Big\rangle \Big\langle \theta_-^2 \Big|\Bigg), \nonumber \\ 
    \sigma_{0,0|0,1}^{\text{dist}^{\text{GHZ}}} &=& \sigma_{1,1|0,1}^{\text{dist}^{\text{GHZ}}} = \sigma_{0,0|1,0}^{\text{dist}^{\text{GHZ}}} = \sigma_{1,1|1,0}^{\text{dist}^{\text{GHZ}}} = \frac{1}{4} \Bigg(P^{\text{GHZ}_2}_{\text{success}}\Bigg| \frac{\pi}{4}_-^3 \Bigg\rangle \Bigg\langle \frac{\pi}{4}_1^3 \Bigg| + P^{\text{GHZ}_2}_{\text{fail}} \Big| \theta_-^3 \Big\rangle \Big\langle \theta_-^3 \Big|\Bigg), \nonumber \\ 
    \sigma_{0,1|0,1}^{\text{dist}^{\text{GHZ}}} &=& \sigma_{1,0|0,1}^{\text{dist}^{\text{GHZ}}} = \sigma_{0,1|1,0}^{\text{dist}^{\text{GHZ}}} = \sigma_{1,0|1,0}^{\text{dist}^{\text{GHZ}}} = \frac{1}{4} \Bigg(P^{\text{GHZ}_2}_{\text{success}}\Bigg| \frac{\pi}{4}_+^3 \Bigg\rangle \Bigg\langle \frac{\pi}{4}_+^3 \Bigg| + P^{\text{GHZ}_2}_{\text{fail}} \Big| \theta_+^3 \Big\rangle \Big\langle \theta_+^3 \Big| \Bigg), \nonumber \\
    \sigma_{0,0|0,2}^{\text{dist}^{\text{GHZ}}} &=& \sigma_{1,0|0,2}^{\text{dist}^{\text{GHZ}}} = \sigma_{0,0|1,2}^{\text{dist}^{\text{GHZ}}} = \sigma_{1,0|1,2}^{\text{dist}^{\text{GHZ}}} = \sigma_{0,0|2,0}^{\text{dist}^{\text{GHZ}}} = \sigma_{0,1|2,0}^{\text{dist}^{\text{GHZ}}} = \sigma_{0,0|2,1}^{\text{dist}^{\text{GHZ}}} = \sigma_{0,1|2,1}^{\text{dist}^{\text{GHZ}}} = \nonumber \\
    & & \quad \quad \quad \quad \quad \quad \hspace{3cm} =\Big(\frac{1}{4} P^{\text{GHZ}_2}_{\text{success}} | 0 \rangle \langle 0 | + \frac{\cos^2 \theta}{2} P^{\text{GHZ}_2}_{\text{fail}}  | 0 \rangle \langle 0 | \Big), \nonumber \\
    \sigma_{0,1|0,2}^{\text{dist}^{\text{GHZ}}} &=& \sigma_{1,1|0,2}^{\text{dist}^{\text{GHZ}}} = \sigma_{0,1|1,2}^{\text{dist}^{\text{GHZ}}} = \sigma_{1,1|1,2}^{\text{dist}^{\text{GHZ}}} = \sigma_{1,0|2,0}^{\text{dist}^{\text{GHZ}}} = \sigma_{1,1|2,0}^{\text{dist}^{\text{GHZ}}} = \sigma_{1,0|2,1}^{\text{dist}^{\text{GHZ}}} = \sigma_{1,1|2,1}^{\text{dist}^{\text{GHZ}}} = \nonumber \\
    & & \quad \quad \quad \quad \quad \quad \hspace{3cm} = \Big(\frac{1}{4} P^{\text{GHZ}_2}_{\text{success}} | 1 \rangle \langle 1 | + \frac{ \sin^2 \theta}{2} P^{\text{GHZ}_2}_{\text{fail}}  | 1 \rangle \langle 1 | \Big), \nonumber \\
	\sigma_{0,0|2,2}^{\text{dist}^{\text{GHZ}}} &=& \Big(\frac{1}{2} P^{\text{GHZ}_2}_{\text{success}} | 0 \rangle\langle 0 | + \cos^2 \theta \, P^{\text{GHZ}_2}_{\text{fail}}  | 0 \rangle \langle 0 |\Big),\quad  \sigma_{1,1|2,2}^{\text{dist}^{\text{GHZ}}} = \Big(\frac{1}{2} P^{\text{GHZ}_2}_{\text{success}} | 1 \rangle \langle 1 | + \sin^2 \theta \, P^{\text{GHZ}_2}_{\text{fail}}  | 1 \rangle \langle 1 | \Big) \nonumber \\
	\sigma_{0,1|2,2}^{\text{dist}^{\text{GHZ}}}&,& \sigma_{1,0|2,2}^{\text{dist}^{\text{GHZ}}} \text{ do not exist.}
	\label{GGHZ2Sdistassemblage}
\end{eqnarray}

 Since, our target assemblage $\{\sigma_{a,b|x,y}^{C^{\text{GHZ}}} \}_{a,x}$ given by  Eq.(\ref{GHZ2Sassemblage}) is pure, we can write $\sigma_{a,b|x.y}^{C^{\text{GHZ}}}$ = $p(a,b|x,y) \Big| \phi_{C}^{\text{GHZ}} (a,b, x,y) \Big\rangle \Big\langle \phi_{C}^{\text{GHZ}} (a,b, x,y) \Big|$ $\forall$ $a,b,x,y$. Hence, the assemblage fidelity of the assemblage (\ref{GGHZ2Sdistassemblage}) and our target assemblage (\ref{GHZ2Sassemblage}) corresponding to GHZ state in 2SDI scenario can be written as: 
\begin{align}
    &\mathcal{F}_A \Big(\{\sigma_{a,b|x,y}^{\text{dist}^{\text{GHZ}}}\}_{a,b,x,y}, \{\sigma_{a,b|x,y}^{C^{\text{GHZ}}}\}_{a,b,x,y} \Big) = \underset{x,y}{\mathrm{min}} \sum_{a,b} \sqrt{p(a,b|x,y) \, \Big\langle \phi_{C}^{\text{GHZ}} (a,b, x,y) \Big| \sigma_{a,b|x,y}^{\text{dist}^{\text{GHZ}}} \Big| \phi_{C}^{\text{GHZ}} (a,b, x,y) \Big\rangle}.
     \label{assfidghz1app2sdi}
\end{align}
Next, we have,
\begin{eqnarray}
	\sum_{a,b} \sqrt{p(a,b|0,0) \, \Big\langle \phi_{C}^{\text{GHZ}} (a,b, 0,0) \Big| \sigma_{a,b|0,0}^{\text{dist}^{\text{GHZ}}} \Big| \phi_{C}^{\text{GHZ}} (a,b, 0,0) \Big\rangle} &=&  \sqrt{1 - \frac{1}{2} (1 - \sin 2 \theta ) (\cos 2 \theta )^{N-1}}, \label{AF2SG36} \\
	\sum_{a,b} \sqrt{p(a,b|0,1) \, \Big\langle \phi_{C}^{\text{GHZ}} (a,b, 0,1) \Big| \sigma_{a,b|0,1}^{\text{dist}^{\text{GHZ}}} \Big| \phi_{C}^{\text{GHZ}} (a,b, 0,1) \Big\rangle}) &=& \sqrt{1 - \frac{1}{2} (1 - \sin 2 \theta ) (\cos 2 \theta )^{N-1}}, \label{AF2SG37} \\
	\sum_{a,b} \sqrt{p(a,b|0,2) \, \Big\langle \phi_{C}^{\text{GHZ}} (a,b, 0,2) \Big| \sigma_{a,b|0,2}^{\text{dist}^{\text{GHZ}}} \Big| \phi_{C}^{\text{GHZ}} (a,b, 0,2) \Big\rangle} &=& \frac{1}{2} \left(\sqrt{1- (\cos 2 \theta )^N}+\sqrt{ 1+ (\cos 2 \theta )^N }\right), \label{AF2SG38} \\
	\sum_{a,b} \sqrt{p(a,b|1,0) \, \Big\langle \phi_{C}^{\text{GHZ}} (a,b, 1,0) \Big| \sigma_{a,b|1,0}^{\text{dist}^{\text{GHZ}}} \Big| \phi_{C}^{\text{GHZ}} (a,b, 1,0) \Big\rangle} &=& \sqrt{1 - \frac{1}{2} (1 - \sin 2 \theta ) (\cos 2 \theta )^{N-1}}, \label{AF2SG39} \\
	\sum_{a,b} \sqrt{p(a,b|1,1) \, \Big\langle \phi_{C}^{\text{GHZ}} (a,b, 1,1) \Big| \sigma_{a,b|1,1}^{\text{dist}^{\text{GHZ}}} \Big| \phi_{C}^{\text{GHZ}} (a,b, 1,1) \Big\rangle} &=& \sqrt{1 - \frac{1}{2} (1 - \sin 2 \theta ) (\cos 2 \theta )^{N-1}},  \label{AF2SG40}\\
	\sum_{a,b} \sqrt{p(a,b|1,2) \, \Big\langle \phi_{C}^{\text{GHZ}} (a,b, 1,2) \Big| \sigma_{a,b|1,2}^{\text{dist}^{\text{GHZ}}} \Big| \phi_{C}^{\text{GHZ}} (a,b, 1,2) \Big\rangle} &=& \frac{1}{2} \left(\sqrt{1- (\cos 2 \theta )^N}+\sqrt{ 1+ (\cos 2 \theta )^N }\right), \label{AF2SG41} \\
	\sum_{a,b} \sqrt{p(a,b|2,0) \, \Big\langle \phi_{C}^{\text{GHZ}} (a,b, 2,0) \Big| \sigma_{a,b|2,0}^{\text{dist}^{\text{GHZ}}} \Big| \phi_{C}^{\text{GHZ}} (a,b, 2,0) \Big\rangle} &=& \frac{1}{2} \left(\sqrt{1- (\cos 2 \theta )^N}+\sqrt{ 1+ (\cos 2 \theta )^N }\right), \label{AF2SG42} \\
	\sum_{a,b} \sqrt{p(a,b|2,1) \, \Big\langle \phi_{C}^{\text{GHZ}} (a,b, 2,1) \Big| \sigma_{a,b|2,1}^{\text{dist}^{\text{GHZ}}} \Big| \phi_{C}^{\text{GHZ}} (a,b, 2,1) \Big\rangle} &=& \frac{1}{2} \left(\sqrt{1- (\cos 2 \theta )^N}+\sqrt{ 1+ (\cos 2 \theta )^N }\right), \label{AF2SG43} \\
	\sum_{a,b} \sqrt{p(a,b|2,2) \, \Big\langle \phi_{C}^{\text{GHZ}} (a,b, 2,2) \Big| \sigma_{a,b|2,2}^{\text{dist}^{\text{GHZ}}} \Big| \phi_{C}^{\text{GHZ}} (a,b, 2,2) \Big\rangle} &=& \frac{1}{2} \left(\sqrt{1- (\cos 2 \theta )^N}+\sqrt{ 1+ (\cos 2 \theta )^N }\right), \label{AF2SG44}
\end{eqnarray}

Hence, from Eq.(\ref{assfidghz1app2sdi}) and following Appendix \ref{app3}, we can write for $\theta \in ( 0.22, \frac{\pi}{4})$
\begin{align}
    &\mathcal{F}_A \Big(\{\sigma_{a,b|x,y}^{\text{dist}^{\text{GHZ}}}\}_{a,b,x,y}, \{\sigma_{a,b|x,y}^{C^{\text{GHZ}}}\}_{a,b,x,y} \Big) \nonumber \\
    &= \mathrm{min} \Bigg[\sqrt{1 - \frac{1}{2} (1 - \sin 2 \theta ) (\cos 2 \theta )^{N-1}},  \frac{1}{2} \left(\sqrt{1- (\cos 2 \theta )^N}+\sqrt{ 1+ (\cos 2 \theta )^N }\right) \Bigg] \nonumber \\
    &= \sqrt{1 - \frac{1}{2} (1 - \sin 2 \theta ) (\cos 2 \theta )^{N-1}}
    \label{app6eq3new}
\end{align}

\section{Components of assemblages produced from the generalized W states in our 1SDI scenario}\label{W_assemb_1SDI}
The components of the assemblages $\{\sigma_{a|x}^{BC^{GW}}\}$ (the initial assemblage of our distillation protocol), obtained from the three qubit generalized W state (\ref{wclass}) through measurements of the observables $X_0= \sigma_x$, $X_1=\sigma_y$ and $X_2=\sigma_z$ by Alice, are given by,  
\begin{eqnarray}
	\sigma_{0|0}^{BC^{\text{GW}}} &=& \frac{1}{2} \Big|w_+^0 \Big\rangle \Big\langle w_+^0 \Big|, \quad  \hspace{0.87cm}		
	\sigma_{1|0}^{BC^{\text{GW}}} = \frac{1}{2} \Big|w_-^0 \Big\rangle \Big\langle w_-^0 \Big|, \quad 	\hspace{0.87cm}
	\sigma_{0|1}^{BC^{\text{GW}}} = \frac{1}{2} \Big|w_+^1 \Big\rangle \Big\langle w_+^1 \Big|, 	 \nonumber \\
	\sigma_{1|1}^{BC^{\text{GW}}} &=& \frac{1}{2} \Big|w_-^1 \Big\rangle \Big\langle w_-^1 \Big|, \quad 	\hspace{0.87cm}
	\sigma_{0|2}^{BC^{\text{GW}}} =   (c_0^2+c_1^2) \, \Big|w^2 \Big\rangle \Big\langle w^2 \Big|,\quad 	\hspace{0.87cm}
	\sigma_{1|2}^{BC^{\text{GW}}} =  (1-c_0^2-c_1^2) \ket{00}\bra{00},
	\label{wassemblage}
\end{eqnarray}
where, $\Big|w_{\pm}^0\Big\rangle := \sqrt{1-c_0^2-c_1^2} \ket{00} \pm c_0 \ket{01} \pm c_1 \ket{10}$, $\Big|w_{\pm}^1\Big\rangle := \sqrt{1-c_0^2-c_1^2} \ket{00} \pm i \, c_0 \ket{01} \pm i \, c_1 \ket{10}$ and $\Big|w^2\Big\rangle :=  \dfrac{1}{\sqrt{c_0^2 + c_1^2}}(c_0 \ket{01} + c_1 \ket{10})$.\\

The components of the assemblages $\{\sigma_{a|x}^{BC^{W}}\}$ (the target assemblage of our distillation protocol), obtained from the three qubit W state through measurements of the observables $X_0= \sigma_x$, $X_1=\sigma_y$ and $X_2=\sigma_z$ by Alice, are given by,  
\begin{eqnarray}
	\sigma_{0|0}^{BC^{\text{W}}} &=& \frac{1}{2} \Big|w_{m_+}^0 \Big\rangle \Big\langle w_{m_+}^0 \Big|, \quad  		
	\sigma_{1|0}^{BC^{\text{W}}} = \frac{1}{2} \Big|w_{m_-}^0 \Big\rangle \Big\langle w_{m_-}^0 \Big|, \quad 
	\sigma_{0|1}^{BC^{\text{W}}} = \frac{1}{2} \Big|w_{m_+}^1 \Big\rangle \Big\langle w_{m_+}^1 \Big|,  \nonumber \\ 		
	\sigma_{1|1}^{BC^{\text{W}}} &=& \frac{1}{2} \Big|w_{m_-}^1 \Big\rangle \Big\langle w_{m_-}^1 \Big|, \quad
	\sigma_{0|2}^{BC^{\text{W}}} = \dfrac{2}{3} \, \Big|w_m^2 \Big\rangle \Big\langle w_m^2 \Big|, \quad
	\sigma_{1|2}^{BC^{\text{W}}} = \frac{1}{3} \ket{00}\bra{00} 
	\label{wpassemblage}
\end{eqnarray}
where, $\Big|w_{m_{\pm}}^0\Big\rangle := \frac{1}{\sqrt{3}}( \ket{00} \pm \ket{01} \pm  \ket{10} )$, $\Big|w_{m_{\pm}}^1\Big\rangle := \frac{1}{\sqrt{3}}( \ket{00} \pm i  \ket{01} \pm i  \ket{10})$ and $\Big|w_m^2\Big\rangle := \frac{1}{\sqrt{2}}( \ket{01} +  \ket{10}) $.

\section{Derivation of the assemblage fidelity for generalized W states in 1SDI scenario}\label{appnew}

 Let $\{\sigma_{a|x}^{\text{dist}^{\text{W}}}\}_{a,x}$ denote the assemblage obtained, on average, after the distillation protocol in the regime of finite N copies with $N \geq 2$. The assemblage can be written as a convex combination of the initial assemblage $\{\sigma_{a|x}^{BC^{\text{GW}}}\}_{a,x}$ and the target assemblage of W state $\{\sigma_{a|x}^{BC^{\text{W}}}\}_{a,x}$. The components of  $\{\sigma_{a|x}^{\text{dist}^{\text{W}}}\}_{a,x}$  are given by,
\begin{eqnarray}
	\sigma_{0|0}^{\text{dist}^{\text{W}}} &=& \frac{1}{2} \Bigg( P^{\text{W}_1}_{\text{success}}\Big| w_{m_+}^0 \Big\rangle \Big\langle w_{m_+}^0 \Big| + P^{\text{W}_1}_{\text{fail}} \Big| w_+^0 \Big\rangle \Big\langle w_+^0 \Big| \Bigg),\nonumber \\ \sigma_{1|0}^{\text{dist}^{\text{W}}} &=& \frac{1}{2} \Bigg( P^{\text{W}_1}_{\text{success}} \Big| w_{m_-}^0 \Big\rangle \Big\langle w_{m_-}^0 \Big| + P^{\text{W}_1}_{\text{fail}} \Big| w_-^0 \Big\rangle \Big\langle w_-^0 \Big| \Bigg), \nonumber \\
	\sigma_{0|1}^{\text{dist}^{\text{W}}} &=& \frac{1}{2} \Bigg( P^{\text{W}_1}_{\text{success}} \Big| w_{m_-}^1 \Big\rangle \Big\langle w_{m_-}^1 \Big| + P^{\text{W}_1}_{\text{fail}} \Big| w_-^1 \Big\rangle \Big\langle w_-^1 \Big| \Bigg),\nonumber \\ \sigma_{1|1}^{\text{dist}^{\text{W}}} &=& \frac{1}{2} \Bigg( P^{\text{W}_1}_{\text{success}} \Big| w_{m_+}^1 \Big\rangle \Big\langle w_{m_+}^1 \Big| + P^{\text{W}_1}_{\text{fail}}\Big| w_+^1 \Big\rangle \Big\langle w_+^1 \Big| \Bigg), \nonumber \\
	\sigma_{0|2}^{\text{dist}^{\text{W}}} &=& \Bigg(  \frac{2}{3} P^{\text{W}_1}_{\text{success}} \Big| w_{m}^2 \Big\rangle \Big\langle w_m^2 \Big| +  (c_0^2+c_1^2) P^{\text{W}_1}_{\text{fail}}  \ket{w^2}\bra{w^2}\Bigg), \nonumber \\ \sigma_{1|2}^{\text{dist}^{\text{W}}} &=& \Bigg(\frac{1}{3} P^{\text{W}_1}_{\text{success}} + (1-c_0^2-c_1^2) P^{\text{W}_1}_{\text{fail}} \Bigg)\ket{00}\bra{00}.
	\label{Wdistill}
\end{eqnarray}

 In this case also our target assemblage $\{\sigma_{a|x}^{BC^{\text{W}}} \}_{a,x}$ given by  Eq.(\ref{wassemblage}) is pure. We can, therefore, write $\sigma_{a|x}^{BC^{\text{W}}}$ = $p(a|x) \, \Big| \phi_{BC}^{\text{w}} (a, x) \Big\rangle \Big\langle \phi_{BC}^{\text{W}} (a, x) \Big|$ $\forall$ $a,x$. Hence, the assemblage fidelity of the assemblage (\ref{Wdistill}) and our target assemblage (\ref{wassemblage}) corresponding to W state  is written as,
\begin{align}
    &\mathcal{F}_{A} \Big( \{\sigma_{a|x}^{\text{dist}^{\text{W}}} \}_{a,x}, \{\sigma_{a|x}^{BC^{\text{W}}}\}_{a,x} \Big) \nonumber \\
    &= \underset{x}{\mathrm{min}} \sum_a \sqrt{p(a|x) \, \Big\langle \phi_{BC}^{\text{W}} (a, x) \Big| \sigma_{a|x}^{\text{dist}^{\text{W}}} \Big| \phi_{BC}^{\text{W}} (a, x)} \Big\rangle.
     \label{assfidghz1appap}
\end{align}
Next, we have
\begin{align}
 \sum_a \sqrt{p(a|0) \, \Big\langle \phi_{BC}^{\text{W}} (a, 0) \Big| \sigma_{a|0}^{\text{dist}^{\text{W}}} \Big| \phi_{BC}^{\text{W}} (a, 0)} \Big\rangle  
 & = \frac{1}{\sqrt{3}} \sqrt{X^{N-1} \Big[ (c_0 + c_1 + \sqrt{1- c_0^2 - c_1^2})^2 -3 \Big] + 3} = g_1 (c_0, c_1, N),
 \label{g11}
\end{align}
\begin{align}
 \sum_a \sqrt{p(a|1) \, \Big\langle \phi_{BC}^{\text{W}} (a, 1) \Big| \sigma_{a|1}^{\text{dist}^{\text{W}}} \Big| \phi_{BC}^{\text{W}} (a, 1)} \Big\rangle  &= \frac{1}{\sqrt{3}} \sqrt{X^{N-1} \Big[ (c_0 + c_1 + \sqrt{1- c_0^2 - c_1^2})^2 -3 \Big] + 3} = g_1 (c_0, c_1, N),
 \label{g12}
\end{align}
and 
\begin{align}
 \sum_a \sqrt{p(a|2) \, \Big\langle \phi_{BC}^{\text{W}} (a, 2) \Big| \sigma_{a|2}^{\text{dist}^{\text{W}}} \Big| \phi_{BC}^{\text{W}} (a, 2)} \Big\rangle  & = \frac{1}{3}  \Bigg[ \sqrt{4 + \Big( 3 (c_0 + c_1)^2 - 4\Big) X^{N-1}}  + \sqrt{3 (1- c_0^2 - c_1^2)X^{N-1} - X^{N-1}+1 }\Bigg] \nonumber \\
 &= g_2 (c_0, c_1, N),
 \label{g2}
\end{align}
where $X$ = $1- \frac{3 c_0^2 c_1^2}{1- c_0^2-c_1^2}$. 

Hence, we have 
\begin{align}
    &\mathcal{F}_{A} \Big( \{\sigma_{a|x}^{\text{dist}^{\text{W}}} \}_{a,x}, \{\sigma_{a|x}^{BC^{\text{W}}}\}_{a,x} \Big) = \mathrm{min} \big[ g_1 (c_0, c_1, N), g_2 (c_0, c_1, N) \big].
     \label{assfidghz1app22}
\end{align}

Next, we define the following,
\begin{align}
   g_3 (c_0, c_1, N) = 5 + (6 c_0 c_1 - 2) X^{N-1}.
\end{align}

Henceforth, $g_i (c_0, c_1, N)$ (with $i=1,2,3$) will be written as $g_i$ for simplicity. 

Next, it can be checked that $\dfrac{1}{\sqrt{3}} < \sqrt{1-c_0^2 -c_1^2} < 1$ when $0 < c_0 \leq \dfrac{1}{\sqrt{3}}$, $0 < c_1 < \dfrac{1}{\sqrt{3}}$, i.e., when $c_0$, $c_1$ $\in$ $\mathcal{C}_{0,1}$. Since, as mentioned earlier, $\mathcal{C}^{\text{GW}_1}_{0,1}$ $\subset$ $\mathcal{C}_{0,1}$, we have $\dfrac{1}{\sqrt{3}} < \sqrt{1-c_0^2 -c_1^2} < 1$ for all $c_0$, $c_1$ $\in$ $\mathcal{C}^{\text{GW}_1}_{0,1}$. Also, from the definition of fidelity, we have $\sqrt{p(0|2) \, \Big\langle \phi_{BC}^{\text{W}} (0, 2) \Big| \sigma_{0|2}^{\text{dist}^{\text{W}}} \Big| \phi_{BC}^{\text{W}} (0, 2)} \Big\rangle$ = $ \dfrac{1}{3} \sqrt{4 + \Big( 3 (c_0 + c_1)^2 - 4\Big) X^{N-1}}$ $\geq$ $0$, and $\sqrt{p(1|2) \, \Big\langle \phi_{BC}^{\text{W}} (1, 2) \Big| \sigma_{1|2}^{\text{dist}^{\text{W}}} \Big| \phi_{BC}^{\text{W}} (1, 2)} \Big\rangle$ = $ \dfrac{1}{3} \sqrt{3 (1- c_0^2 - c_1^2)X^{N-1} - X^{N-1}+1 }$ $\geq$ $0$. Moreover, we have $0 < X^{N-1} < 1$ $\forall$ $c_0$, 
 $c_1$ $\in$ $\mathcal{C}^{\text{GW}_1}_{0,1}$ and $\forall$ $N \geq 2$ $\in$ $\mathbb{Z}$ as $X^{N-1}$ denotes the failure probability $P^{\text{W}_1}_{\text{fail}}$ given by Eq.(\ref{failurew1sdi}). Hence, one can write the following,
\begin{align}
&9 g_1^2 - g_3 = 4 + \Big[-4 + 6(c_0 + c_1)\sqrt{1-c_0^2 -c_1^2}\Big] X^{N-1} > 4 \Big(1-  X^{N-1} \Big) > 0 \quad \forall \quad c_0, c_1 \in \mathcal{C}^{\text{GW}_1}_{0,1}, \, N \geq 2 \, \in \, \mathbb{Z}, \nonumber \\
& 9 g_2^2 - g_3 = 2 \, \sqrt{4 + \Big( 3 (c_0 + c_1)^2 - 4\Big) X^{N-1}} \, \sqrt{3 (1- c_0^2 - c_1^2)X^{N-1} - X^{N-1}+1 } \geq 0 \quad \forall \quad c_0, c_1 \in \mathcal{C}^{\text{GW}_1}_{0,1}, \, N \geq 2 \, \in \, \mathbb{Z}.
\label{app7eq10}
\end{align}

 Next, an algebraic manipulation leads to the expression,
\begin{align}
&(9 g_1^2 - g_3)^2 - (9 g_2^2 - g_3)^2 \nonumber \\
&= - 12 \, X^{N-1} \, \Big(1-X^{N-1} \Big) \Bigg[2 \Big(\sqrt{1 - c_0^2 - c_1^2} - c_0 \Big) \Big(\sqrt{1 - c_0^2 - c_1^2} - c_1 \Big) + \Big(\sqrt{1 - c_0^2 - c_1^2} - c_0 \Big)^2 + \Big(\sqrt{1 - c_0^2 - c_1^2} - c_1 \Big)^2 \Bigg] 
\label{app7eq2}
\end{align}
Since, we have $\sqrt{1 - c_0^2 - c_1^2} > c_0$ and $\sqrt{1 - c_0^2 - c_1^2} > c_1$ for all $c_0$, $c_1$ $\in$ $\mathcal{C}^{\text{GW}_1}_{0,1}$, it can be easily checked that
\begin{align}
  \Bigg[2 \Big(\sqrt{1 - c_0^2 - c_1^2} - c_0 \Big) \Big(\sqrt{1 - c_0^2 - c_1^2} - c_1 \Big) &+ \Big(\sqrt{1 - c_0^2 - c_1^2} - c_0 \Big)^2 + \Big(\sqrt{1 - c_0^2 - c_1^2} - c_1 \Big)^2 \Bigg]    > 0 \quad \forall \quad c_0, c_1 \in \mathcal{C}^{\text{GW}_1}_{0,1}, \nonumber 
 \\
  X^{N-1} >& 0 \quad \forall \quad c_0, c_1 \in \mathcal{C}^{\text{GW}_1}_{0,1}, \, N \geq 2 \, \in \, \mathbb{Z}, \nonumber \\
 \Big(1-X^{N-1} \Big) >& 0 \quad \forall \quad c_0, c_1 \in \mathcal{C}^{\text{GW}_1}_{0,1}, \, N \geq 2 \, \in \, \mathbb{Z}.
\end{align}

Hence, from Eq.(\ref{app7eq2}), we get 
\begin{align}
(9 g_1^2 - g_3)^2 - (9 g_2^2 - g_3)^2 < 0 \quad \forall \quad c_0, c_1 \in \mathcal{C}^{\text{GW}_1}_{0,1}, \, N \geq 2 \, \in \, \mathbb{Z}.
\label{app7eq3}
\end{align}
Therefore, Eqs.(\ref{app7eq10}) and (\ref{app7eq3}) lead to 
\begin{equation}
   g_1^2 < g_2^2 \quad \forall \quad c_0, c_1 \in \mathcal{C}^{\text{GW}_1}_{0,1}, \, N \geq 2 \, \in \, \mathbb{Z}.
\end{equation}
Next, from  the definition of fidelity \cite{Nery20}, $g_1 \geq 0$ and $g_2 \geq 0$ $\forall$ $c_0$, $c_1$ $\in$ $\mathcal{C}^{\text{GW}_1}_{0,1}$ and $\forall$ $N \geq 2$ $\in$ $\mathbb{Z}$. Hence, we have
\begin{equation}
   g_1 < g_2 \quad \forall \quad c_0, c_1 \in \mathcal{C}^{\text{GW}_1}_{0,1}, \, N \geq 2 \, \in \, \mathbb{Z}.
   \label{app7e11}
\end{equation}

Therefore, from Eqs.(\ref{assfidghz1app22}) and (\ref{app7e11}), we can write
\begin{align}
    \mathcal{F}_{A} \Big( \{\sigma_{a|x}^{\text{dist}^{\text{W}}} \}_{a,x}, \{\sigma_{a|x}^{BC^{\text{W}}}\}_{a,x} \Big) = \frac{1}{\sqrt{3}} \sqrt{X^{N-1} \Big[ (c_0 + c_1 + \sqrt{1- c_0^2 - c_1^2})^2 -3 \Big] + 3},
     \label{assfinal7}
\end{align}
where $X$ = $1- \frac{3 c_0^2 c_1^2}{1- c_0^2-c_1^2}$.

\section{Components of the assemblages produced from the one parameter generalized W pure states in our 2SDI scenario}\label{OW2Sassemblage}
The components of the assemblages $\{\sigma_{a,b|x,y}^{C^{\text{GW}}}\}_{a,b,x,y}$ (the initial assemblages in our distillation protocol), obtained from the three qubit one parameter generalized W pure states (\ref{owclass}) through the measurements of the observables  $X_0= \sigma_x$, $X_1=\sigma_y$ and $X_2=\sigma_z$ by Alice and $Y_0= \sigma_x$, $Y_1=\sigma_y$ and $Y_2=\sigma_z$ by Bob, are given by,
\begin{align}
	\sigma_{0,0|0,0}^{C^{\text{GW}}} &= \frac{2-d_0^2}{4}\Big| w_{+}^3 \Big\rangle \Big\langle w_{+}^3 \Big|, \nonumber \\
	\sigma_{0,1|0,0}^{C^{\text{GW}}} &=\sigma_{1,0|0,0}^{C^{\text{GW}}} = \sigma_{0,1|1,1}^{C^{\text{GW}}} = \sigma_{1,0|1,1}^{C^{\text{GW}}} = \frac{d_0^2}{4}\Big| 1 \Big\rangle \Big\langle 1 \Big|, \nonumber \\
	\sigma_{1,1|0,0}^{C^{\text{GW}}} & = \frac{2-d_0^2}{4}\Big| w_{-}^3 \Big\rangle \Big\langle w_{-}^3 \Big|, \nonumber \\
	  \sigma_{0,0|1,0}^{C^{\text{GW}}} &=  \sigma_{0,0|0,1}^{C^{\text{GW}}} = \frac{1}{4}\Big| w_{0,1,0}^4 \Big\rangle \Big\langle w_{0,1,0}^4 \Big|, \quad \sigma_{0,1|1,0}^{C^{\text{GW}}} = \sigma_{1,0|0,1} ^{C^{\text{GW}}}  = \frac{1}{4}\Big| w_{0,0,1}^4 \Big\rangle \Big\langle w_{0,0,1}^4 \Big|, \, \nonumber \\
	\sigma_{1,0|1,0}^{C^{\text{GW}}} &= \sigma_{0,1|0,1}^{C^{\text{GW}}} = \frac{1}{4}\Big| w_{0,0,0}^4 \Big\rangle \Big\langle w_{0,0,0}^4 \Big|, \quad  \sigma_{1,1|1,0}^{C^{\text{GW}}} = \sigma_{1,1|0,1}^{C^{\text{GW}}} = \frac{1}{4} \Big| w_{0,1,1}^4 \Big\rangle \Big\langle w_{0,1,1}^4 \Big|,\nonumber \\ 
	\sigma_{0,1|0,2}^{C^{\text{GW}}}&=\sigma_{1,1|0,2}^{C^{\text{GW}}}=\sigma_{0,1|1,2}^{C^{\text{GW}}} = \sigma_{1,1|1,2}^{C^{\text{GW}}} = \sigma_{1,0|2,0}^{C^{\text{GW}}} = \sigma_{1,1|2,0}^{C^{\text{GW}}} = \sigma_{1,0|2,1}^{C^{\text{GW}}}=\sigma_{1,1|2,1}^{C^{\text{GW}}}= \frac{1-d_0^2}{4} \Big| 0 \Big\rangle \Big\langle 0 \Big|, \nonumber \\
\sigma_{0,0|1,1}^{C^{\text{GW}}} &=  \frac{2-d_0^2}{4} \Big| w_{+}^5 \Big\rangle \Big\langle w_{+}^5 \Big| , \quad 
     \sigma_{1,1|1,1}^{C^{\text{GW}}} =  \frac{2-d_0^2}{4} \Big| w_{-}^5 \Big\rangle \Big\langle w_{-}^5 \Big| , \nonumber \\
     \sigma_{0,0|0,2}^{C^{\text{GW}}} &= \sigma_{0,0|2,0}^{C^{\text{GW}}} = \frac{1+d_0^2}{4}\Big| w_{+}^6 \Big\rangle \Big\langle w_{+}^6 \Big|,  \quad
     \sigma_{1,0|0,2}^{C^{\text{GW}}} = \sigma_{0,1|2,0}^{C^{\text{GW}}} = \frac{1+d_0^2}{4}\Big| w_{-}^6 \Big\rangle \Big\langle w_{-}^6 \Big|, \nonumber \\
     \sigma_{0,0|1,2}^{C^{\text{GW}}} &= \sigma_{0,0|2,1}^{C^{\text{GW}}} = \frac{1+d_0^2}{4}\Big| w_{+}^7 \Big\rangle \Big\langle w_{+}^7 \Big|, \quad 
     \sigma_{1,0|1,2}^{C^{\text{GW}}} = \sigma_{0,1|2,1}^{C^{\text{GW}}} = \frac{1+d_0^2}{4}\Big| w_{-}^7 \Big\rangle \Big\langle w_{-}^7 \Big|, \nonumber \\
      \sigma_{0,1|2,2}^{C^{\text{GW}}} &= \sigma_{1,0|2,2}^{C^{\text{GW}}} =  \frac{1-d_0^2}{2} \Big| 0 \Big\rangle \Big\langle 0 \Big| , \quad  \sigma_{0,0|2,2}^{C^{\text{GW}}} = d_0^2 \Big| 1 \Big\rangle \Big\langle 1 \Big| 
     \label{GW2SA}
\end{align}

where, $\Big| w_{\pm}^3 \Big\rangle = \sqrt{\dfrac{2}{2-d_0^2}} \Bigg( \sqrt{1-d_0^2} \ket{0} \pm \dfrac{d_0}{\sqrt{2}} \ket{1} \Bigg)$,

$\Big| w_{x,y,z}^4 \Big\rangle = \Big[ (-1)^x  \, + (-1)^y i\, \Big] \sqrt{\dfrac{1-d_0^2}{2}}  \ket{0} + (-1)^z d_0 \ket{1}$, 

$\Big| w_{\pm}^5 \Big\rangle = \sqrt{\dfrac{2}{2-d_0^2}} \Bigg( \sqrt{1-d_0^2} \ket{0} \pm \, i \, \dfrac{d_0}{\sqrt{2}} \ket{1} \Bigg)$,

$\Big| w_{\pm}^{6} \Big\rangle = \sqrt{\dfrac{2}{1+d^2}} \Bigg( \sqrt{\dfrac{1-d_0^2}{2}} \ket{0} \pm d_0 \ket{1} \Bigg)$,

$\Big| w_{\pm}^7 \Big\rangle = \sqrt{\dfrac{2}{1+d^2}} \Bigg( \sqrt{\dfrac{1-d_0^2}{2}} \ket{0} \pm \, i \, d_0 \ket{1} \Bigg)$.




The assemblage component $\sigma_{1,1|2,2}^{C^{\text{GW}}}$ does not exist because the probability of its occurrence is zero. Hence, we need not consider this component.

\section{Components of the assemblage produced from the W state in our 2SDI scenario}\label{W2Sassemblage}

 The components of the assemblage $\{\sigma_{a,b|x,y}^{C^{\text{W}}}\}_{a,b,x,y}$ (the target assemblage of our distillation protocol), obtained from the three qubit W state through measurements of the observables $X_0= \sigma_x$, $X_1=\sigma_y$ and $X_2=\sigma_z$ by Alice and $Y_0= \sigma_x$, $Y_1=\sigma_y$ and $Y_2=\sigma_z$ by Bob, are given by,
\begin{align}
	\sigma_{0,0|0,0}^{C^{\text{W}}} &= \frac{5}{12}\Big| w_{m_{+}}^3 \Big\rangle \Big\langle w_{m_{+}}^3 \Big|, \nonumber \\
	\sigma_{0,1|0,0}^{C^{\text{W}}} &=\sigma_{1,0|0,0}^{C^{\text{W}}} = \sigma_{0,1|1,1}^{C^{\text{W}}} = \sigma_{1,0|1,1}^{C^{\text{W}}} = \frac{1}{12}\Big| 1 \Big\rangle \Big\langle 1 \Big|, \nonumber \\
	\sigma_{1,1|0,0}^{C^{\text{W}}} &= \frac{5}{12}\Big| w_{m_{-}}^3 \Big\rangle \Big\langle w_{m_{-}}^3 \Big|, \nonumber \\
	  \sigma_{0,0|1,0}^{C^{\text{W}}} &=  \sigma_{0,0|0,1}^{C^{\text{W}}} = \frac{1}{4}\Big| w_{m_{0,1,0}}^4 \Big\rangle \Big\langle w_{m_{0,1,0}}^4 \Big|, \quad \sigma_{0,1|1,0}^{C^{\text{W}}} = \sigma_{1,0|0,1} ^{C^{\text{W}}}  = \frac{1}{4}\Big| w_{m_{0,0,1}}^4 \Big\rangle \Big\langle w_{m_{0,0,1}}^4 \Big|, \, \nonumber \\
	\sigma_{1,0|1,0}^{C^{\text{W}}} &= \sigma_{0,1|0,1}^{C^{\text{W}}} = \frac{1}{4}\Big| w_{m_{0,0,0}}^4 \Big\rangle \Big\langle w_{m_{0,0,0}}^4 \Big|, \quad \sigma_{1,1|1,0}^{C^{\text{W}}} = \sigma_{1,1|0,1}^{C^{\text{W}}} = \frac{1}{4} \Big| w_{m_{0,1,1}}^4 \Big\rangle \Big\langle w_{m_{0,1,1}}^4 \Big|,\nonumber \\ 
	\sigma_{0,1|0,2}^{C^{\text{W}}}&=\sigma_{1,1|0,2}^{C^{\text{W}}}=\sigma_{0,1|1,2}^{C^{\text{W}}} = \sigma_{1,1|1,2}^{C^{\text{W}}} = \sigma_{1,0|2,0}^{C^{\text{W}}} = \sigma_{1,1|2,0}^{C^{\text{W}}} = \sigma_{1,0|2,1}^{C^{\text{W}}}=\sigma_{1,1|2,1}^{C^{\text{W}}}= \frac{1}{6} \Big| 0 \Big\rangle \Big\langle 0 \Big|, \nonumber \\
\sigma_{0,0|1,1}^{C^{\text{W}}} &=  \frac{5}{12} \Big| w_{m_{+}}^5 \Big\rangle \Big\langle w_{m_{+}}^5 \Big|, \quad  \sigma_{1,1|1,1}^{C^{\text{W}}} =  \frac{5}{12} \Big| w_{m_{-}}^5 \Big\rangle \Big\langle w_{m_{-}}^5 \Big| , \nonumber \\
     \sigma_{0,0|0,2}^{C^{\text{W}}} &= \sigma_{0,0|2,0}^{C^{\text{W}}} = \frac{1}{3} \Big| w_{m_{+}}^6 \Big\rangle \Big\langle w_{m_{+}}^6 \Big| , \quad \sigma_{1,0|0,2}^{C^{\text{W}}} = \sigma_{0,1|2,0}^{C^{\text{W}}} = \frac{1}{3} \Big| w_{m_{-}}^6 \Big\rangle \Big\langle w_{m_{-}}^6 \Big| , \nonumber \\
     \sigma_{0,0|1,2}^{C^{\text{W}}} &= \sigma_{0,0|2,1}^{C^{\text{W}}} = \frac{1}{3} \Big| w_{m_{+}}^7 \Big\rangle \Big\langle w_{m_{+}}^7 \Big| , \quad \sigma_{1,0|1,2}^{C^{\text{W}}} = \sigma_{0,1|2,1}^{C^{\text{W}}} = \frac{1}{3} \Big| w_{m_{-}}^7 \Big\rangle \Big\langle w_{m_{-}}^7 \Big| , \nonumber \\
      \sigma_{0,1|2,2}^{C^{\text{W}}} &= \sigma_{1,0|2,2}^{C^{\text{W}}} =  \frac{1}{3} \Big| 0 \Big\rangle \Big\langle 0 \Big| , \quad  \sigma_{0,0|2,2}^{C^{\text{W}}} = \frac{1}{3} \Big| 1 \Big\rangle \Big\langle 1 \Big| 
	\label{W2SA}
\end{align}

where, $ \Big| w_{m_{\pm}}^{3} \Bigg\rangle   = \sqrt{\dfrac{4}{5}} \Bigg(  \Big| 0 \Big\rangle \pm \dfrac{1}{2} \Big| 1 \Big\rangle \Bigg)$,

$\Big|w_{m_{x,y,z}}^4\Big\rangle = \Big[ (-1)^x  \, + (-1)^y i\, \Big] \sqrt{\dfrac{1}{3}} \Big|0\Big\rangle + (-1)^z \dfrac{1}{\sqrt{3}} \Big|1\Big\rangle$, 

$\Big|w_{m_{\pm}}^5\Big\rangle = \sqrt{\dfrac{4}{5}} \Bigg(  \Big| 0 \Big\rangle \pm \, i \, \dfrac{1}{2} \Big| 1 \Big\rangle \Bigg)$,

$ \Big| w_{m_{\pm}}^{6} \Big\rangle  = \dfrac{1}{\sqrt{2}} \Big(  \Big| 0 \Big\rangle \pm  \Big| 1 \Big\rangle \Big)$,

$ \Big| w_{m_{\pm}}^{7} \Big\rangle  = \dfrac{1}{\sqrt{2}} \Big(  \Big| 0 \Big\rangle \pm \, i \,  \Big| 1 \Big\rangle \Big)$.

The assemblage component $\sigma_{1,1|2,2}^{C^{\text{W}}}$ does not exist because the probability of occurrence is zero. Hence, we need not consider this component.


\section{Derivation of the assemblage fidelity for generalized W states in 2SDI scenario }\label{app10}

In the regime of finite $N$ copies with $N \geq 2$, a single assemblage can be extracted $\{\sigma_{a,b|x,y}^{\text{dist}^{\text{W}}}\}_{a,b,x,y}$ as a convex combination of the initial assemblage $\{\sigma_{a,b|x,y}^{C^{\text{GW}}}\}_{a,b,x,y}$ with components mentioned in Appendix \ref{OW2Sassemblage} and the target assemblage $\{\sigma_{a,b|x,y}^{C^{\text{W}}}\}_{a,b,x,y}$ with components mentioned in Appendix \ref{W2Sassemblage}. The components of $\{\sigma_{a,b|x,y}^{\text{dist}^{\text{W}}}\}_{a,b,x,y}$ are given by,
\begin{align}
	\sigma_{0,0|0,0}^{\text{dist}^{\text{W}}} &= \dfrac{1}{4} \Bigg(   P_{\text{success}}^{\text{W}_2} \dfrac{5 }{3} \Big| w_{m_{+}}^3 \Big\rangle \Big\langle w_{m_{+}}^3 \Big| + P_{\text{fail}}^{\text{W}_2} (2-d_0^2) \Big| w_{+}^3 \Big\rangle \Big\langle w_{+}^3 \Big| \Bigg), \nonumber \\
	\sigma_{0,1|0,0}^{\text{dist}^{\text{W}}} &=\sigma_{1,0|0,0}^{\text{dist}^{\text{W}}} = \sigma_{0,1|1,1}^{\text{dist}^{\text{W}}} = \sigma_{1,0|1,1}^{\text{dist}^{\text{W}}} = \dfrac{1}{4} \Bigg( P_{\text{success}}^{\text{W}_2} \dfrac{1}{3} \Big| 1 \Big\rangle \Big\langle 1 \Big| + P_{\text{fail}}^{\text{W}_2} \, d_0^2 \Big| 1 \Big\rangle \Big\langle 1 \Big| \Bigg), \nonumber \\
	\sigma_{1,1|0,0}^{\text{dist}^{\text{W}}} &= \dfrac{1}{4} \Bigg( P_{\text{success}}^{\text{W}_2} \dfrac{5}{3} \Big| w_{m_{-}}^3 \Big\rangle \Big\langle w_{m_{-}}^3 \Big| + P_{\text{fail}}^{\text{W}_2} (2-d_0^2) \Big| w_{-}^3 \Big\rangle \Big\langle w_{-}^3 \Big| \Bigg), \nonumber \\
	  \sigma_{0,0|1,0}^{\text{dist}^{\text{W}}} &=  \sigma_{0,0|0,1}^{\text{dist}^{\text{W}}} = \dfrac{1}{4} \Bigg( P_{\text{success}}^{\text{W}_2} \Big| w_{m_{0,1,0}}^4 \Big\rangle \Big\langle w_{m_{0,1,0}}^4 \Big| + P_{\text{fail}}^{\text{W}_2} \Big| w_{0,1,0}^4 \Big\rangle \Big\langle w_{0,1,0}^4 \Big| \Bigg), \nonumber \\
	   \sigma_{0,1|1,0}^{\text{dist}^{\text{W}}} &= \sigma_{1,0|0,1} ^{\text{dist}^{\text{W}}}  = \dfrac{1}{4} \Bigg( P_{\text{success}}^{\text{W}_2} \Big| w_{m_{0,0,1}}^4 \Big\rangle \Big\langle w_{m_{0,0,1}}^4 \Big| + P_{\text{fail}}^{\text{W}_2} \Big| w_{0,0,1}^4 \Big\rangle \Big\langle w_{0,0,1}^4 \Big| \Bigg),  \nonumber \\
	\sigma_{1,0|1,0}^{\text{dist}^{\text{W}}} &= \sigma_{0,1|0,1}^{\text{dist}^{\text{W}}} = \dfrac{1}{4} \Bigg( P_{\text{success}}^{\text{W}_2} \Big| w_{m_{0,0,0}}^4 \Big\rangle \Big\langle w_{m_{0,0,0}}^4 \Big| + P_{\text{fail}}^{\text{W}_2} \Big| w_{0,0,0}^4 \Big\rangle \Big\langle w_{0,0,0}^4 \Big| \Bigg),  \nonumber \\
	 \sigma_{1,1|1,0}^{\text{dist}^{\text{W}}} &= \sigma_{1,1|0,1}^{\text{dist}^{\text{W}}} = \dfrac{1}{4} \Bigg( P_{\text{success}}^{\text{W}_2} \Big| w_{m_{0,1,1}}^4 \Big\rangle \Big\langle w_{m_{0,1,1}}^4 \Big| + P_{\text{fail}}^{\text{W}_2} \Big| w_{0,1,1}^4 \Big\rangle \Big\langle w_{0,1,1}^4 \Big| \Bigg), \nonumber \\
	\sigma_{0,1|0,2}^{\text{dist}^{\text{W}}} &=\sigma_{1,1|0,2}^{\text{dist}^{\text{W}}} =\sigma_{0,1|1,2}^{\text{dist}^{\text{W}}} = \sigma_{1,1|1,2}^{\text{dist}^{\text{W}}} = \sigma_{1,0|2,0}^{\text{dist}^{\text{W}}} = \sigma_{1,1|2,0}^{\text{dist}^{\text{W}}} = \sigma_{1,0|2,1}^{\text{dist}^{\text{W}}} = \sigma_{1,1|2,1}^{\text{dist}^{\text{W}}}=  \nonumber \\
	& \hspace{3.5cm}  \dfrac{1}{2} \Bigg( P_{\text{success}}^{\text{W}_2} \dfrac{1}{3} \Big| 0 \Big\rangle \Big\langle 0 \Big| + P_{\text{fail}}^{\text{W}_2} \frac{1-d_0^2}{2} \Big| 0 \Big\rangle \Big\langle 0 \Big| \Bigg), \nonumber \\
\sigma_{0,0|1,1}^{\text{dist}^{\text{W}}} &=  \dfrac{1}{4} \Bigg( P_{\text{success}}^{\text{W}_2} \dfrac{5}{3} \Big| w_{m_{+}}^5 \Big\rangle \Big\langle w_{m_{+}}^5 \Big| +P_{\text{fail}}^{\text{W}_2} \big(2-d_0^2 \big) \Big| w_{+}^5 \Big\rangle \Big\langle w_{+}^5 \Big| \Bigg), \nonumber \\
     \sigma_{1,1|1,1}^{\text{dist}^{\text{W}}} &=  \dfrac{1}{4} \Bigg( P_{\text{success}}^{\text{W}_2} \dfrac{5}{3} \Big| w_{m_{-}}^5 \Big\rangle \Big\langle w_{m_{-}}^5 \Big| +P_{\text{fail}}^{\text{W}_2} \big(2-d_0^2 \big) \Big| w_{-}^5 \Big\rangle \Big\langle w_{-}^5 \Big| \Bigg), \nonumber \\
     \sigma_{0,0|0,2}^{\text{dist}^{\text{W}}} &= \sigma_{0,0|2,0}^{\text{dist}^{\text{W}}} = \dfrac{1}{4} \Bigg( P_{\text{success}}^{\text{W}_2} \dfrac{4}{3} \Big| w_{m_{+}}^6 \Big\rangle \Big\langle w_{m_{+}}^6 \Big| + P_{\text{fail}}^{\text{W}_2} (1+d_0^2) \Big| w_{+}^6 \Big\rangle \Big\langle w_{+}^6 \Big| \Bigg), \nonumber \\
     \sigma_{1,0|0,2}^{\text{dist}^{\text{W}}} &= \sigma_{0,1|2,0}^{\text{dist}^{\text{W}}} = \dfrac{1}{4} \Bigg( P_{\text{success}}^{\text{W}_2} \dfrac{4}{3} \Big| w_{m_{-}}^6 \Big\rangle \Big\langle w_{m_{-}}^6 \Bigg| + P_{\text{fail}}^{\text{W}_2} (1+d_0^2) \Big| w_{-}^6 \Big\rangle \Big\langle w_{-}^6 \Big| \Bigg), \nonumber \\
     \sigma_{0,0|1,2}^{\text{dist}^{\text{W}}} &= \sigma_{0,0|2,1}^{\text{dist}^{\text{W}}} = \dfrac{1}{4} \Bigg( P_{\text{success}}^{\text{W}_2} \dfrac{4}{3} \Big| w_{m_{+}}^7 \Big\rangle \Big\langle w_{m_{+}}^7 \Bigg| + P_{\text{fail}}^{\text{W}_2} (1+d_0^2) \Big| w_{+}^7 \Big\rangle \Big\langle w_{+}^7 \Big| \Bigg), \nonumber \\
     \sigma_{1,0|1,2}^{\text{dist}^{\text{W}}} &= \sigma_{0,1|2,1}^{\text{dist}^{\text{W}}} = \dfrac{1}{4} \Bigg( P_{\text{success}}^{\text{W}_2} \dfrac{4}{3} \Big| w_{m_{-}}^7 \Big\rangle \Big\langle w_{m_{-}}^7 \Bigg| + P_{\text{fail}}^{\text{W}_2} (1+d_0^2) \Big| w_{-}^7 \Big\rangle \Big\langle w_{-}^7 \Big| \Bigg), \nonumber \\
      \sigma_{0,1|2,2}^{\text{dist}^{\text{W}}} &= \sigma_{1,0|2,2}^{\text{dist}^{\text{W}}} = P_{\text{success}}^{\text{W}_2} \dfrac{1}{3} \Big| 0 \Big\rangle \Big\langle 0 \Big|  + P_{\text{fail}}^{\text{W}_2} \dfrac{1-d_0^2}{2} \Big| 0 \Big\rangle \Big\langle 0 \Big| , \quad  \sigma_{0,0|2,2}^{\text{dist}^{\text{W}}} = P_{\text{success}}^{\text{W}_2} \dfrac{1}{3} \Big| 1 \Big\rangle \Big\langle 1 \Big|  + P_{\text{fail}}^{\text{W}_2} \, d_0^2 \Big| 1 \Big\rangle \Big\langle 1 \Big|, \nonumber \\
       \sigma_{1,1|2,2}^{\text{dist}^{\text{W}}}  & \, \, \text{does not exist because the probability of its occurrence is zero.}
	\label{W2Sdistassemblage}
\end{align}

Since, our target assemblage $\{\sigma_{a,b|x,y}^{C^{\text{W}}} \}_{a,b,x,y}$ given in Appendix \ref{W2Sassemblage} is pure, we can write, $\sigma_{a,b|x.y}^{C^{\text{W}}}$ = $p(a,b|x,y) \Big| \phi_{C}^{\text{W}} (a,b, x,y) \Big\rangle \Big\langle \phi_{C}^{\text{W}} (a,b, x,y) \Big|$ $\forall$ $a,b,x,y$. Hence, the assemblage fidelity of the assemblage (\ref{W2Sdistassemblage}) and our target assemblage (\ref{W2SA}) corresponding to W state in our 2SDI scenario can be written as,
\begin{align}
    &\mathcal{F}_A \Big(\{\sigma_{a,b|x,y}^{\text{dist}^{\text{W}}}\}_{a,b,x,y}, \{\sigma_{a,b|x,y}^{C^{\text{W}}}\}_{a,b,x,y} \Big) = \underset{x,y}{\mathrm{min}} \sum_{a,b} \sqrt{p(a,b|x,y) \, \Big\langle \phi_{C}^{\text{W}} (a,b, x,y) \Big| \sigma_{a,b|x,y}^{\text{dist}^{\text{W}}} \Big| \phi_{C}^{\text{W}} (a,b, x,y) \Big\rangle}.
     \label{assfidw1app2sdi}
\end{align}
Next, we have,
\begin{align}
	\sum_{a,b} \sqrt{p(a,b|0,0) \, \Big\langle \phi_{C}^{\text{W}} (a,b, 0,0) \Big| \sigma_{a,b|0,0}^{\text{dist}^{\text{W}}} \Big| \phi_{C}^{\text{W}} (a,b, 0,0) \Big\rangle} &= h_0(d_0,N) \label{www1} \\
	\sum_{a,b} \sqrt{p(a,b|1,1) \, \Big\langle \phi_{C}^{\text{W}} (a,b, 1,1) \Big| \sigma_{a,b|1,1}^{\text{dist}^{\text{W}}} \Big| \phi_{C}^{\text{W}} (a,b, 1,1) \Big\rangle} &= h_0(d_0,N)  \label{www2} \\
	 \sum_{a,b} \sqrt{p(a,b|0,1) \, \Big\langle \phi_{C}^{\text{W}} (a,b, 0,1) \Big| \sigma_{a,b|0,1}^{\text{dist}^{\text{W}}} \Big| \phi_{C}^{\text{W}} (a,b, 0,1) \Big\rangle}) &= h_1(d_0,N) \label{www3} \\ 
	 \sum_{a,b} \sqrt{p(a,b|1,0) \, \Big\langle \phi_{C}^{\text{W}} (a,b, 1,0) \Big| \sigma_{a,b|1,0}^{\text{dist}^{\text{W}}} \Big| \phi_{C}^{\text{W}} (a,b, 1,0) \Big\rangle} 	&= h_1(d_0,N) \label{www4}\\
	\sum_{a,b} \sqrt{p(a,b|0,2) \, \Big\langle \phi_{C}^{\text{W}} (a,b, 0,2) \Big| \sigma_{a,b|0,2}^{\text{dist}^{\text{W}}} \Big| \phi_{C}^{\text{W}} (a,b, 0,2) \Big\rangle}&= h_2(d_0,N) \label{www5} \\
	\sum_{a,b} \sqrt{p(a,b|2,0) \, \Big\langle \phi_{C}^{\text{W}} (a,b, 2,0) \Big| \sigma_{a,b|2,0}^{\text{dist}^{\text{W}}} \Big| \phi_{C}^{\text{W}} (a,b, 2,0) \Big\rangle} &= h_2(d_0,N) \label{www6} \\
	\sum_{a,b} \sqrt{p(a,b|1,2) \, \Big\langle \phi_{C}^{\text{W}} (a,b, 1,2) \Big| \sigma_{a,b|1,2}^{\text{dist}^{\text{W}}} \Big| \phi_{C}^{\text{W}} (a,b, 1,2) \Big\rangle} &= h_2(d_0,N) \label{www7} \\
	\sum_{a,b} \sqrt{p(a,b|2,1) \, \Big\langle \phi_{C}^{\text{W}} (a,b, 2,1) \Big| \sigma_{a,b|2,1}^{\text{dist}^{\text{W}}} \Big| \phi_{C}^{\text{W}} (a,b, 2,1) \Big\rangle} 
	&= h_2(d_0,N) \label{www8} \\
	\sum_{a,b} \sqrt{p(a,b|2,2) \, \Big\langle \phi_{C}^{\text{W}} (a,b, 2,2) \Big| \sigma_{a,b|2,2}^{\text{dist}^{\text{W}}} \Big| \phi_{C}^{\text{W}} (a,b, 2,2) \Big\rangle} &= h_3(d_0,N),
	\label{www9}
\end{align}
where
\begin{align}
   h_0(d_0, N) &= \frac{1}{6} \left(\sqrt{25 - \Big(1 - 12 d_0 \sqrt{2 - 2 d_0^2} + 21 d_0^2 \Big) \Big(1 - 3 d_0^2 \Big)^{N-1}}+\sqrt{1- \Big(1-3d_0^2 \Big)^{\text{N}}}\right) \nonumber \\
   h_1(d_0, N) &= \dfrac{1}{\sqrt{3}}\sqrt{3 - 2 \Bigg(\sqrt{\frac{1-d_0^2}{2}} - d_0 \Bigg)^2 \Big(1-3d_0^2 \Big)^{N-1}} \nonumber \\
 h_2(d_0, N) &= \frac{1}{3 \sqrt{2}} \left( \sqrt{8 + \Big( 6 d_0  \sqrt{2 - 2 d_0^2} + 3 d_0^2 -5  \Big) \Big(1-3d_0^2 \Big)^{N-1} }+\sqrt{2+\Big(1-3d_0^2 \Big)^{N}} \right) \nonumber \\
h_3(d_0, N) &= \frac{1}{3} \left(\sqrt{1-\Big(1-3d_0^2 \Big)^{N} }+\sqrt{2} \sqrt{2+ \Big(1-3d_0^2 \Big)^{N}}\right)
\label{expressionw2}
\end{align}.

Hence, we have 
	\begin{equation}
		\mathcal{F}_A \Big(\{\sigma_{a,b|x,y}^{\text{dist}^{\text{W}}}\}_{a,b,x,y}, \{\sigma_{a,b|x,y}^{C^{\text{W}}}\}_{a,b,x,y} \Big) = \text{min} \Big[ h_0(d_0,N),h_1(d_0,N),h_2(d_0,N),h_3(d_0,N)\Big]
		\label{www2cc7}
	\end{equation}
Let us first  define the following,
 \begin{equation}
 	h_4(d_0,N) = \dfrac{1}{36} \Big[ 26 - \Big(1 - 3 d_0^2\Big)^N  - \Big(1 + 21 d_0^2 - 12 d_0  \sqrt{2 - 2 d_0^2} \Big) \Big(1 - 3 d_0^2\Big)^{N-1} \Big].
 \end{equation}
 Henceforth, $h_i(d_0,N)$ will be written as $h_i$ for simplicity. Next, one can write the following,
 \begin{align}
 	h_0^2 - h_4 &= \frac{1}{18} \sqrt{25 - \Big(1 - 12 d_0 \sqrt{2 - 2 d_0^2} + 21 d_0^2 \Big) \Big(1 - 3 d_0^2 \Big)^{N-1}} \, \sqrt{1- \Big(1-3d_0^2 \Big)^{\text{N}}} \geq 0 \quad \forall \quad d_0 \in \Bigg( \dfrac{3}{25}, \dfrac{1}{\sqrt{3}} \Bigg), N\geq 2  \, \in \, \mathbb{Z}, \nonumber \\
	h_1^2 - h_4 &= \frac{1}{18}\Bigg[ 5 \Bigg(1 - \Big(1 - 3 d_0^2 \Big)^{N - 1} \Bigg) + \Big( 6 d_0 \sqrt{2 - 2 d_0^2} + 
     3 d_0^2 \Big) \Big( 1 - 3 d_0^2 \Big)^{N - 1} \Big] > 0 \quad \forall \quad d_0 \in \Bigg( \dfrac{3}{25}, \dfrac{1}{\sqrt{3}} \Bigg), N\geq 2  \, \in \, \mathbb{Z}.
     \label{www2c1}
 \end{align}
 where the first inequality is obtained from the definition of fidelity \cite{Nery20}, $\sqrt{p(0,0|0,0) \, \Big\langle \phi_{C}^{\text{W}} (0,0, 0,0) \Big| \sigma_{0,0|0,0}^{\text{dist}^{\text{W}}} \Big| \phi_{C}^{\text{W}} (0,0, 0,0) \Big\rangle}$ = $\frac{1}{12} \sqrt{25 - \Big(1 - 12 d_0 \sqrt{2 - 2 d_0^2} + 21 d_0^2 \Big) \Big(1 - 3 d_0^2 \Big)^{N-1}}$ $\geq 0$, $\sqrt{p(0,1|0,0) \, \Big\langle \phi_{C}^{\text{W}} (0,1, 0,0) \Big| \sigma_{0,1|0,0}^{\text{dist}^{\text{W}}} \Big| \phi_{C}^{\text{W}} (0,1, 0,0) \Big\rangle}$ = $\frac{1}{12} \sqrt{1- \Big(1-3d_0^2 \Big)^{\text{N}}}$ $\geq 0$ $\forall$ $d_0$ $\in$ $\Bigg( \dfrac{3}{25}, \dfrac{1}{\sqrt{3}} \Bigg)$, and $\forall$ $N\geq 2$ $\in$ $\mathbb{Z}$.
 
 Next, an algebraic manipulation leads to the expression,
 \begin{equation}
 	(h_1^2 - h_4)^2 - (h_0^2 - h_4)^2 = -\, \frac{4}{27} \Big( 1 - 3 d_0^2 \Big)^{N - 1} \Big[1-\Big( 1 - 3 d_0^2 \Big)^{N - 1}\Big] \Bigg(\sqrt{\frac{1-d_0^2}{2}} - d_0 \Bigg)^2 < 0 \quad \forall \quad d_0 \in \Bigg( \dfrac{3}{25}, \dfrac{1}{\sqrt{3}} \Bigg), N\geq 2  \, \in \, \mathbb{Z}.
 	\label{www2c2}
 \end{equation}
 
 Hence, From (\ref{www2c1}) and (\ref{www2c2}), we can write, $h_1^2 < h_0^2$ $\forall$ $d_0$ $\in$ $\Bigg( \dfrac{3}{25}, \dfrac{1}{\sqrt{3}} \Bigg)$, and $\forall$ $N\geq 2$ $\in$ $\mathbb{Z}$. Now from the definition of fidelity, $h_0 \geq 0$ and $h_1 \geq 0$. Hence, we can conclude that $h_1 < h_0$ $\forall$ $d_0$ $\in$ $\Bigg( \dfrac{3}{25}, \dfrac{1}{\sqrt{3}} \Bigg)$, and $\forall$ $N\geq 2$ $\in$ $\mathbb{Z}$.

Next, we again define another function as below,
 \begin{equation}
 	h_5(d_0,N) = \frac{1}{18}\left[ 10 + \Big(1 - 3 d_0^2 \Big)^N
     + \Big(3 d_0^2 + 6 d_0 \sqrt{2 - 2 d_0^2} -5\Big) \Big(1 - 3 d_0^2\Big)^{N - 1}
     \right].
 \end{equation}
 We can write the following,
 \begin{align}
 	h_1^2 - h_5 &= \frac{1}{9}\Bigg[ 4 \Bigg( 1 - \Big(1 - 3 d_0^2 \Big)^{N - 1} \Bigg) + 
   3 \Bigg( d_0  \sqrt{2 - 2 d_0^2} + \Big(1 - d_0^2\Big) \Bigg) \Big(1 - 3 d_0^2 \Big)^{N - 1} \Bigg] > 0 \quad \forall \quad d_0 \in \Bigg( \dfrac{3}{25}, \dfrac{1}{\sqrt{3}} \Bigg), N\geq 2  \, \in \, \mathbb{Z}, \nonumber \\
	h_2^2 - h_5 &= \frac{1}{9} \sqrt{8 + \Big( 6 d_0  \sqrt{2 - 2 d_0^2} + 3 d_0^2 -5  \Big) \Big(1-3d_0^2 \Big)^{N-1} } \, \sqrt{2+ \Big(1-3d_0^2 \Big)^{N}} \geq 0 \quad \forall \quad d_0 \in \Bigg( \dfrac{3}{25}, \dfrac{1}{\sqrt{3}} \Bigg), N\geq 2  \, \in \, \mathbb{Z},
	\label{www2c3}
 \end{align}
 where the second inequality is obtained from the definition of fidelity: $\sqrt{p(0,0|0,2) \, \Big\langle \phi_{C}^{\text{W}} (0,0, 0,2) \Big| \sigma_{0,0|0,2}^{\text{dist}^{\text{W}}} \Big| \phi_{C}^{\text{W}} (0,0, 0,2) \Big\rangle}$ = $\frac{1}{6 \sqrt{2}} \sqrt{8 + \Big( 6 d_0  \sqrt{2 - 2 d_0^2} + 3 d_0^2 -5  \Big) \Big(1-3d_0^2 \Big)^{N-1} }$ $\geq 0$, $\sqrt{p(0,1|0,2) \, \Big\langle \phi_{C}^{\text{W}} (0,1, 0,2) \Big| \sigma_{0,1|0,2}^{\text{dist}^{\text{W}}} \Big| \phi_{C}^{\text{W}} (0,1, 0,2) \Big\rangle}$ = $\frac{1}{6 \sqrt{2}} \sqrt{2+\Big(1-3d_0^2 \Big)^{N}}$ $\geq 0$ $\forall$ $d_0$ $\in$ $\Bigg( \dfrac{3}{25}, \dfrac{1}{\sqrt{3}} \Bigg)$, and $\forall$ $N\geq 2$ $\in$ $\mathbb{Z}$.
 
 Next, we have,
 \begin{equation}
 	(h_1^2 - h_5)^2 - (h_2^2 - h_5)^2 = -\, \frac{4}{27} \Big( 1 - 3 d_0^2 \Big)^{N - 1} \Big[1-\Big( 1 - 3 d_0^2 \Big)^{N - 1}\Big] \Bigg(\sqrt{\frac{1-d_0^2}{2}} - d_0 \Bigg)^2 < 0 \quad \forall \quad d_0 \in \Bigg( \dfrac{3}{25}, \dfrac{1}{\sqrt{3}} \Bigg), N\geq 2  \, \in \, \mathbb{Z}.
 	\label{www2c4}
 \end{equation}
  Hence, from (\ref{www2c3}) and (\ref{www2c4}), we can write, $h_1^2 < h_2^2$ $\forall$ $d_0$ $\in$ $\Bigg( \dfrac{3}{25}, \dfrac{1}{\sqrt{3}} \Bigg)$, and $\forall$ $N\geq 2$ $\in$ $\mathbb{Z}$. Now from the definition of fidelity \cite{Nery20}, $h_0 \geq 0$ and $h_2 \geq 0$. Hence, we can conclude that $h_1 < h_2$ $\forall$ $d_0$ $\in$ $\Bigg( \dfrac{3}{25}, \dfrac{1}{\sqrt{3}} \Bigg)$, and $\forall$ $N\geq 2$ $\in$ $\mathbb{Z}$.

Finally, we consider the following function,
 \begin{equation}
 	h_6 = \frac{1}{9}\Bigg[5+ \Big( 1 - 3 d_0^2 \Big)^{N} \Bigg]
 \end{equation}
 One can write,
 \begin{align}
 	h_1^2 - h_6 &= \frac{1}{9}\Bigg[ 4 \Bigg( 1 - \Big(1 - 3 d_0^2 \Big)^{N - 1} \Bigg) + 
   6 d_0  \sqrt{2 - 2 d_0^2}  \Big(1 - 3 d_0^2 \Big)^{N - 1} \Bigg] > 0 \quad \forall \quad d_0 \in \Bigg( \dfrac{3}{25}, \dfrac{1}{\sqrt{3}} \Bigg), N\geq 2  \, \in \, \mathbb{Z}, \nonumber \\
	h_3^2 - h_6 &= \frac{2\sqrt{2}}{9} \sqrt{1-\Big(1-3d_0^2 \Big)^{N} } \, \sqrt{2+ \Big(1-3d_0^2 \Big)^{N}} \geq  0 \quad \forall \quad d_0 \in \Bigg( \dfrac{3}{25}, \dfrac{1}{\sqrt{3}} \Bigg), N\geq 2  \, \in \, \mathbb{Z}.
	\label{www2c5}
 \end{align}
 Here the second inequality is obtained from the definition of fidelity: $\sqrt{p(0,0|2,2) \, \Big\langle \phi_{C}^{\text{W}} (0,0, 2,2) \Big| \sigma_{0,0|2,2}^{\text{dist}^{\text{W}}} \Big| \phi_{C}^{\text{W}} (0,0, 2,2) \Big\rangle}$ = $\frac{1}{3} \sqrt{1-\Big(1-3d_0^2 \Big)^{N} }$ $\geq 0$, $\sqrt{p(0,1|2,2) \, \Big\langle \phi_{C}^{\text{W}} (0,1, 2,2) \Big| \sigma_{0,1|2,2}^{\text{dist}^{\text{W}}} \Big| \phi_{C}^{\text{W}} (0,1, 2,2) \Big\rangle}$ = $\frac{1}{3 \sqrt{2}} \sqrt{2+ \Big(1-3d_0^2 \Big)^{N}}$ $\geq 0$ $\forall$ $d_0$ $\in$ $\Bigg( \dfrac{3}{25}, \dfrac{1}{\sqrt{3}} \Bigg)$, and $\forall$ $N\geq 2$ $\in$ $\mathbb{Z}$.
 
 Next, after performing an algebraic manipulation, we get the following,
 \begin{equation}
 	(h_1^2 - h_6)^2 - (h_3^2 - h_6)^2 = -\, \frac{16}{27} \Big( 1 - 3 d_0^2 \Big)^{N - 1} \Big[1-\Big( 1 - 3 d_0^2 \Big)^{N - 1}\Big] \Bigg(\sqrt{\frac{1-d_0^2}{2}} - d_0 \Bigg)^2 < 0 \quad \forall \quad d_0 \in \Bigg( \dfrac{3}{25}, \dfrac{1}{\sqrt{3}} \Bigg), N\geq 2  \, \in \, \mathbb{Z}.
 	\label{www2c6}
 \end{equation}
 Therefore, from (\ref{www2c5}) and (\ref{www2c6}), we can write, $h_1^2 < h_3^2$ $\forall$ $d_0$ $\in$ $\Bigg( \dfrac{3}{25}, \dfrac{1}{\sqrt{3}} \Bigg)$, and $\forall$ $N\geq 2$ $\in$ $\mathbb{Z}$. Now from the definition of fidelity \cite{Nery20}, $h_0 \geq 0$ and $h_3 \geq 0$. Hence, we can conclude that $h_1 < h_3$ $\forall$ $d_0$ $\in$ $\Bigg( \dfrac{3}{25}, \dfrac{1}{\sqrt{3}} \Bigg)$, and $\forall$ $N\geq 2$ $\in$ $\mathbb{Z}$.
 
 To summarize, we get $h_1 < h_k$ $\forall$ $k$ $\in$ $\{0,2,3\}$, $\forall$ $d_0$ $\in$ $\Bigg( \dfrac{3}{25}, \dfrac{1}{\sqrt{3}} \Bigg)$, and $\forall$ $N\geq 2$ $\in$ $\mathbb{Z}$.
 
Hence, from Eq.(\ref{www2cc7}), we can conclude the following,
 \begin{equation}
		\mathcal{F}_A \Big(\{\sigma_{a,b|x,y}^{\text{dist}^{\text{W}}}\}_{a,b,x,y}, \{\sigma_{a,b|x,y}^{C^{\text{W}}}\}_{a,b,x,y} \Big) = \dfrac{1}{\sqrt{3}}\sqrt{3 - 2 \Bigg(\sqrt{\frac{1-d_0^2}{2}} - d_0 \Bigg)^2 \Big(1-3d_0^2 \Big)^{N-1}}.
	\end{equation}


\end{widetext}

\end{document}